\documentclass[apj]{emulateapj}

\newcommand{\kms}{\ifmmode {\rm km\,s}^{-1} \else km\,s$^{-1}$\fi}
\newcommand{\ergs}{\ifmmode {\rm erg\,s}^{-1} \else erg\,s$^{-1}$\fi}
\newcommand{\lsim}{\stackrel{\scriptscriptstyle <}{\scriptstyle {}_\sim}}
\newcommand{\gsim}{\stackrel{\scriptscriptstyle >}{\scriptstyle {}_\sim}}
\newcommand{\et}{\mbox{et~al.}\ }
\newcommand{\eg}{\mbox{e.g.,}\ }
\newcommand{\ie}{\mbox{i.e.,}\ }
\newcommand{\lya}{Ly $\alpha$}
\newcommand{\hb}{H$\beta$}
\newcommand{\Hbeta}{\ifmmode {\rm H}\beta \else H$\beta$\fi}

\newcommand{\civ}{C\,{\sc iv}}

\newcommand{\mgii}{Mg\,{\sc ii}}
\newcommand{\feii}{Fe\,{\sc ii}}

\newcommand{\lam}{$\lambda$}
\newcommand{\Msigma}{\ifmmode M_{\rm BH} - \sigma \else $M_{\rm BH} - \sigma$\fi}
\newcommand{\Mbh}{\ifmmode M_{\rm BH} \else $M_{\rm BH}$\fi}
\newcommand{\mbh}{\ifmmode M_{\rm BH} \else $M_{\rm BH}$\fi}
\newcommand{\lbol}{$L_{\rm bol}$}
\newcommand{\lol}{$L_{\rm bol}/L_{\rm Edd}$}
\newcommand{\Msol}{\ifmmode M_{\odot} \else $M_{\odot}$\fi}

\slugcomment{Accepted by The Astrophysical Journal, April 18, 2009}

\shorttitle{Mass Functions of Active Black Holes}
\shortauthors{Vestergaard and Osmer}

\begin{document}


\title{Mass Functions of the Active Black Holes in Distant
   Quasars from the Large Bright Quasar Survey, the Bright Quasar
   Survey, and the Color-Selected Sample of the SDSS Fall Equatorial 
   Stripe.}


\author{M. Vestergaard\altaffilmark{1} and
Patrick S. Osmer\altaffilmark{2}}
\email{m.vestergaard@tufts.edu, osmer.1@osu.edu}

\altaffiltext{1}{Dept. of Physics and Astronomy,
Robinson Hall, Tufts University, Medford, MA 02155}
\altaffiltext{2}{Graduate School and Department of Astronomy, 
The Ohio State University, 230 N. Oval Mall, Columbus, OH, 43210}

\begin{abstract}
We present mass functions of distant actively accreting supermassive 
black holes residing in luminous quasars discovered in the Large Bright
Quasar Survey (LBQS), the Bright Quasar Survey (BQS), and the Fall 
Equatorial Stripe of the Sloan Digital Sky Survey (SDSS). The quasars 
cover a wide range of redshifts from the local universe to $z$ = 5 and 
were subject to different selection criteria and flux density limits.  
This makes these samples complementary and can help us gain additional 
insight on the true underlying black hole mass distribution free from 
selection effects and mass estimation errors through future studies. 
By comparing these quasar samples, we see evidence that the active black 
hole population at redshift four is somewhat different than that at lower 
redshifts, including that in the nearby universe. In particular, there is 
a sharp increase in the space density of the detected active black holes
(\mbh $\gsim 10^8$\Msol) between redshifts $\sim$4 and $\sim$2.5.  
Also, the mass function of the SDSS quasars at $3.6 \leq z \leq 5$
has a somewhat flatter high mass-end slope of $\beta = -1.75 \pm 0.56$,
compared to the mass functions based on quasars below $z$ of 3 (BQS and
LBQS quasars), which display typical slopes of $\beta \approx -3.3$; the
latter are consistent with the mass functions at similar redshifts based 
on the SDSS Data Release 3 quasar catalog presented by Vestergaard et al.
We see clear evidence of cosmic downsizing in the comoving space 
density distribution of active black holes in the LBQS sample alone.
In forthcoming papers, further analysis, comparison, and discussion of 
these mass functions will be made with other existing black hole mass 
functions, notably that based on the SDSS DR3 quasar catalog.
We present the relationships used to estimate the black hole mass based 
on the \mgii{} emission line; the relations are calibrated to the \hb{}
and \civ{} relations by means of several thousand high quality SDSS
spectra. Mass estimates of the individual black holes of these 
samples are also presented.
\end{abstract}

\keywords{cosmology: observations -- galaxies: active -- galaxies: luminosity function, mass function -- quasars: emission lines -- quasars: general -- surveys
}

\section{Introduction}

Black hole demographics has become a common and important ingredient 
of cosmological studies in recent years. One reason is the indication 
that black holes and their activity play a crucial role in 
the formation and evolution of galaxies (\eg Granato \et 2004; 
Springel, Di Matteo, \& Hernquist 2005; Somerville \et 2008) and 
galaxy clusters (\eg McNamara \& Nulsen 2007). 
To fully understand the impact of black holes on mass structures and 
their evolution we need to understand much better how they form, grow, 
and interact with their surroundings. A first step toward this goal 
is to take inventory of the population of supermassive black holes 
across the history of the universe. Unfortunately, it is not possible
to study supermassive black holes in all types of galaxies with the 
same (mass estimation) method (\eg cf.\ Magorrian \et 1998; 
Vestergaard 2004b, 2009; Peterson \et 2004),
due to the varying physical conditions of their host galaxies, the 
varying activity level of the black holes themselves, and the large 
range of physical distances to the black hole host galaxies. 
Due to their faintness and the small angular extent of the central 
region that needs to be studied, quiescent or weakly active black 
holes can typically not be well studied beyond the local neighborhood 
of a few hundred Mpc (\eg Ferrarese 2003). 
Since actively accreting black holes power the luminous quasars that are 
observable across the universe (\eg Fan 2006), quasars 
offers a convenient way to trace the black hole population in the 
distant universe.  The goal is that the combination of studies of active 
black holes with empirical and theoretical insight on the relationship 
between active and quiescent black holes and between obscured and 
unobscured black holes will eventually lead to realistic 
representations of the true and complete black hole population. 

We can study the population of actively accreting supermassive black 
holes by studying the large catalogs of quasars and active galaxies 
obtained through the numerous large, extensive quasar and AGN UV and 
optical surveys that have been made since the discovery of quasars [\eg 
The multi-color survey of stellar objects by Koo \& Kron (1982),
Palomar-Green Survey (Schmidt \& Green 1983), 
UK-Schmidt Telescope Survey (\eg Kibblewhite \et 1984; Hewett \et 2001), 
Canada-France-Hawaii Telescope Survey (Crampton, Cowley, \& Hartwick 1987), 
CfA Redshift Survey (Huchra \& Burg 1992), 
Palomar Transit Grism Survey (Schmidt, Schneider, \& Gunn 1995), 
2-degree-field survey (2dF; Smith \et 2005), and
Sloan Digital Sky Survey (SDSS; York \et 2000)].
Notably, any survey with its own set of selection criteria will tend to 
preferentially select either for or against objects with particular 
spectral or broad band properties. Therefore, by studying only quasars 
studied by a single selection method there is a risk that we may limit 
ourselves to only part of the actual underlying black hole population. 
We aim to minimize this potential issue by including in our ongoing study 
of the black hole population and the black hole mass functions several 
quasar samples.
The current work focuses on the mass functions based on the following
samples: the Bright Quasar Survey (BQS; \eg Schmidt \& Green 1983), 
the Large Bright Quasar Survey (LBQS; \eg Hewett, Foltz, \& Chaffee 1995), 
and the color-selected sample from the Fall Equatorial Stripe (Fan \et 2001a) 
of the SDSS. 

The BQS, LBQS, and the SDSS color-selected samples are each based on
different selection criteria, namely UV excess, spectral shape on objective 
prism plates, and broad-band colors, respectively. Especially, the BQS and
the LBQS are each complementary to the SDSS DR3 quasar selection (Richards \et
2002). For example, the SDSS selection probability decreases significantly 
for quasars at redshifts of about 2.8 to 3.2 because the colors of quasars 
at these redshifts coincide with the stellar locus in the selected color 
spaces (Richards \et 2006). This limits the usefulness of the DR3 quasar 
black hole mass function (Vestergaard \et 2008) at and near this redshift 
range which is particularly important since this is the epoch at which the 
quasar space density peaks (\eg Osmer 1982; Warren, Hewett, \& Osmer 1994; 
Schmidt \et 1995; Fan \et 2001a, 2001b). The different selection criterion
of the LBQS can instead help shed light on the mass distribution at the 
affected redshift range, especially given the large size of this quasar sample 
(1067 quasars) and its wide redshift range.  
The favorable properties of the somewhat smaller BQS sample are that it is 
selected over a very large sky area (10700 square degrees) and includes the 
brightest quasars in the nearby universe. Therefore, the BQS is suitable for 
anchoring the bright end of the luminosity function and the high mass end of
the black hole mass function in the nearby universe. Also, the colors of the 
BQS quasars are not typical of 
SDSS quasars (Jester \et 2005) and the BQS thus makes a different contribution 
to our insight on black hole mass distribution than the SDSS alone can provide.
The value of the SDSS color-selected sample studied here is that it is a 
well-defined, complete (to within the color-selection criteria, the survey area, 
and the flux density limits) and 
homogeneous quasar sample at redshift 3.6\,$\leq z \leq$\,5.0 that is highly 
suitable for statistical studies.  This redshift range is beyond that of the 
LBQS ($z \leq$3.0 for the mass functions) and is where the number statistics of 
the SDSS DR3 mass function quasar sample is lower. 
The availability of a well-defined selection function for this color-selected 
sample (Fan \et 2001a) therefore gives us an excellent opportunity to determine 
the black hole mass function for a homogeneous and well-defined sample at high 
redshift which thus renders an additional opportunity for constraining the 
black hole mass distribution at earlier epochs.

The added values of the samples studied here are that luminosity functions have
been determined for both the LBQS (\eg Boyle \et 2000) and the SDSS color-selected 
samples (Fan \et 2001a). Potentially, by combining the luminosity and black hole
mass functions for the same quasar samples we can break the degeneracies (related 
in part to the unknown mass dependency of the radiative efficiency and mass 
accretion rate) that limit the use of the luminosity functions alone (\eg Wyithe 
\& Radmanabhan 2006).  By combining the three quasar samples studied here with 
the SDSS DR3 quasar sample for which the luminosity and mass functions have 
already been determined (Richards \et 2006; Vestergaard \et 2008) we anticipate 
to gain a better understanding of the true underlying black hole mass 
distribution than is possible by studying either of these samples alone.

In this work we present the black hole mass functions of the BQS, LBQS, and 
the SDSS color-selected quasar samples which collectively cover the entire 
redshift range up to $z = 5$ and contain nearly 1200 sources. 
We adopt the most recently calibrated mass scaling relations utilizing 
broad line widths and continuum luminosities (\ie Vestergaard \& Peterson
2006; Vestergaard \et 2008 and references therein) in cases where the more
robust reverberation mapping mass determinations (Peterson \et 2004) are not 
available. Although the mass estimates based on scaling relations are less 
robust than the reverberation mapping masses and are potentially 
prone to systematic uncertainties (\eg Krolik 2001; Richards \et 2002;
Vestergaard 2004a, 2004b, 2009) 
they perform surprisingly well considering the circumstances (\eg Vestergaard
2004b, 2009; Vestergaard \& Peterson 2006; Marconi \et 2008). In fact, there
is suggestive evidence based on \hb{} data that scaling relations can be 
improved to yield mass estimates that are within a factor 1.6 (or 0.2\,dex) 
of the reverberation masses when radiation pressure on the broad line gas is 
taken into account (Marconi \et 2008). This uncertainty rivals the scatter of
the quiescent black holes in the \mbh $- \sigma$ relationship 
($0.25 - 0.3$\,dex; Tremaine \et 2002).

Source inclination is known to affect the velocity dispersion of the broad 
line region as measured using the widths of broad optical and UV emission lines 
(\eg Wills \& Browne 1986; Vestergaard, Wilkes, \& Barthel 2000). In fact, 
source inclination is expected to be one of the factors that account for at
least part of the scatter in the \mbh{} values around the \mbh $- \sigma$
relationship (although the inclination is not trivially connected to object
location relative to the \mbh $- \sigma$ relationship; \eg Collin \et 2006) 
and which at present limits the accuracy 
of mass determinations of active black holes to some degree. Considerations 
have also been made on the relative usefulness of the various line
widths (\eg the line dispersion versus the FWHM width) (\eg Peterson
\et 2004; Denney \et 2008) and a correction scheme was suggested (Collin 
\et 2006) to correct mass estimates based on FWHM measurements of \hb{} to 
mass estimates based on the line dispersion values which are preferred for 
the high-quality reverberation mapping database (Peterson \et 2004).
Unfortunately, for neither of these effects (radiation pressure, inclination,
and the potential inadequacies of the FWHM parameter) are the influences on 
our determination of black hole masses understood or parametrized well 
enough that corrections can be applied to our mass estimates for all 
three emission lines, \hb, \mgii, and \civ, used here.  Also, for this 
study we need our entire data base to be coherent and homogeneously 
analyzed and all mass estimates to be based on similar assumptions. 
For these reasons, we make no attempt to correct the mass estimates
at this time. 

The mass functions presented
here will be analyzed further in relation to the black hole mass 
functions of the SDSS DR3 quasar sample (Vestergaard \et 2008) for the
purpose of extracting the true underlying mass distribution using the
statistical methods introduced and discussed for the luminosity function
by Kelly, Fan, \& Vestergaard (2008a) and for the mass function by Kelly, 
Vestergaard \& Fan (2009).  If, at that time, we are capable of making
corrections for any of the effects outlined earlier such that the mass
estimates or the associated error distribution change significantly, the
mass functions will be updated accordingly. 

A cosmology of H$_0$ = 70 ${\rm km~ s^{-1} Mpc^{-1}}$,
$\Omega_{\Lambda}$ = 0.7, and $\Omega_m$ = 0.3 is used throughout.

\begin{figure}
\plotone{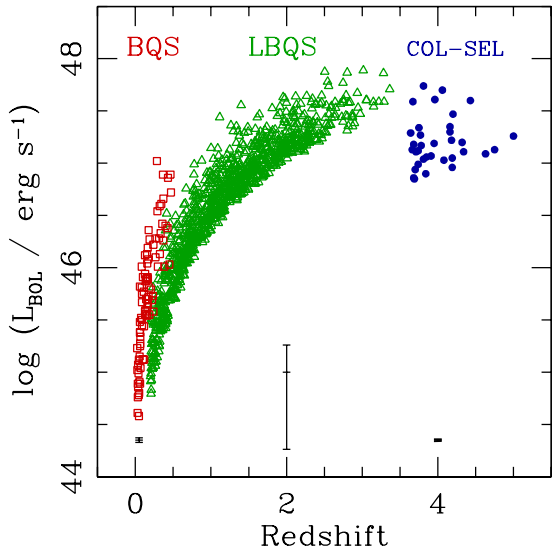}
\caption{Distributions of bolometric luminosities, \lbol, as a function of
redshift for the LBQS (triangles), BQS (squares), and SDSS color-Selected
(filled circles) samples. Typical measurement errors are shown in the lower
portion of the diagram. For the LBQS the \lbol{} values are based on the
survey $B_J$ magnitudes, causing large errors due to the necessary
extrapolation across the spectrum.
\label{lzfig}}
\end{figure}

\begin{figure}
\epsscale{1.0}
\plotone{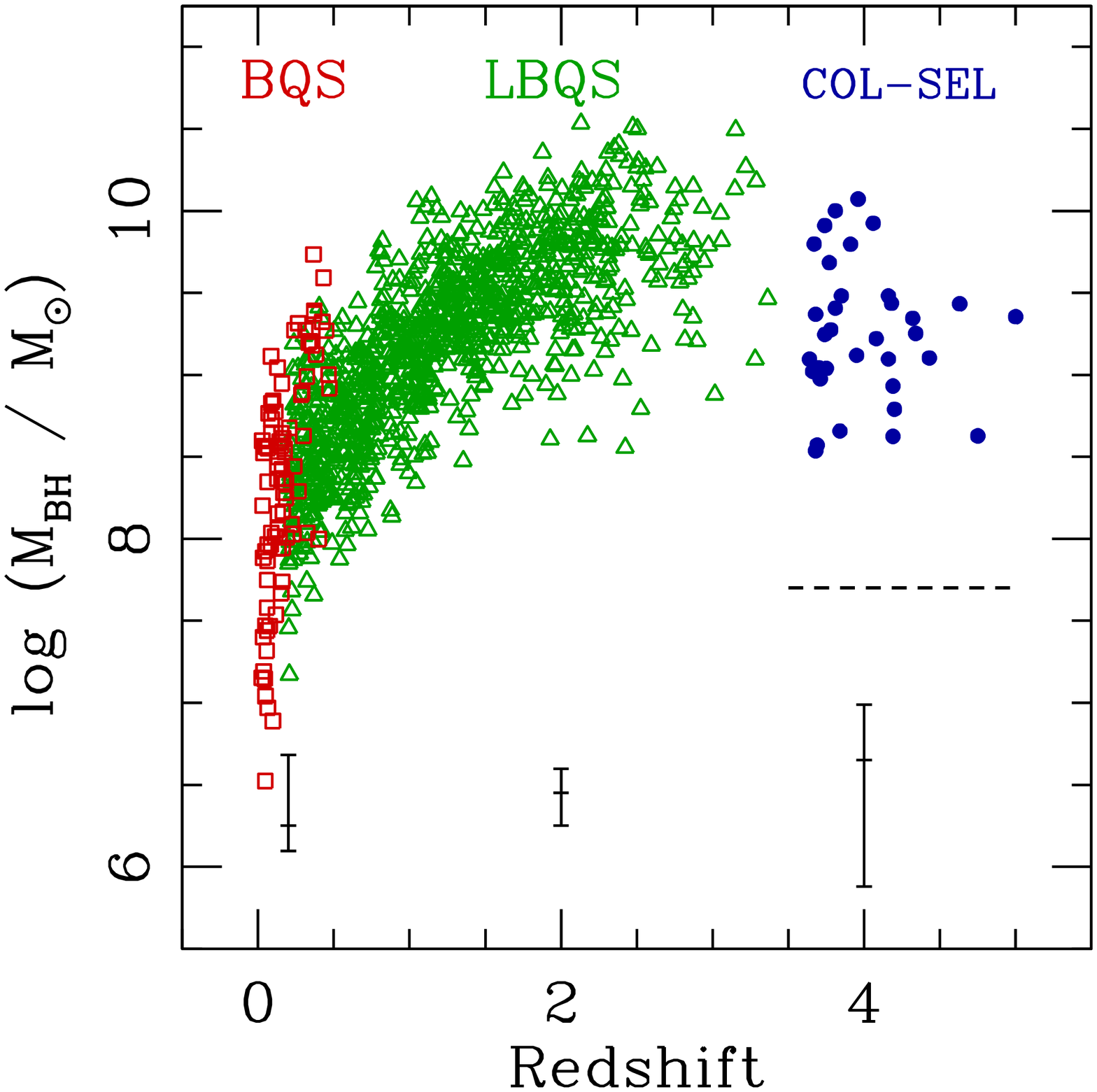}
\vspace{0.3cm}
\plotone{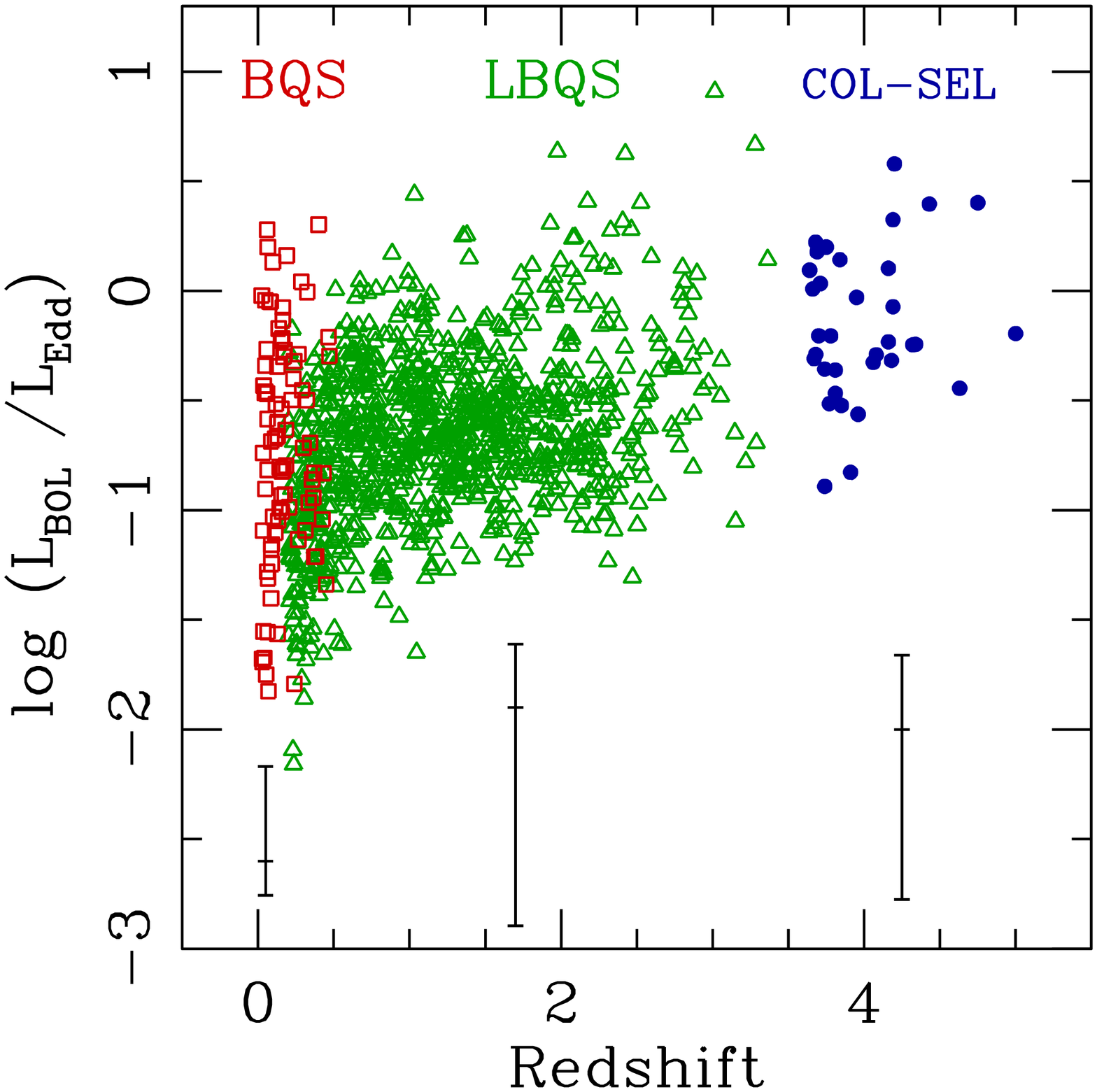}
\caption{Distributions of black hole mass, \mbh, (top panel) and Eddington
luminosity ratios, \lol, (bottom panel) as a function of redshift for the LBQS,
BQS, and SDSS color-Selected samples. Symbols are as in Figure~\ref{lzfig}.
The dashed line in the top panel shows the SDSS flux density limit folded with
the line width cut-off of 1000\,\kms{} adopted for SDSS quasars. See the discussion
for details.
\label{mlolzfig}}
\end{figure}

\section{Data}

The three quasar samples for which we determine the black hole mass
functions are summarized in the following 
sections, in order of their redshift coverage. As outlined in 
section~3, 
the black hole mass estimates are based on
measurements of the widths of the \hb, \mgii, and \civ{} profiles and
nuclear continuum luminosities. These spectral measurements are 
therefore also summarized in the following.

\subsection{The Bright Quasar Survey \label{bqsdata.sec}}

The Bright Quasar Survey (BQS; Schmidt \& Green 1983) is a subset of 
the quasars discovered in the Palomar-Green Survey of UV excess sources 
($U - B < - 0.46$; Green, Schmidt, \& Liebert 1986) undertaken in 1973 
$-$ 1974 using the 18 inch Palomar Schmidt telescope with classification 
spectroscopy obtained with the Hale 5\,m telescope. This survey was 
done using photographic $B_J$ plates that were later digitized.  It has 
an impressive area coverage of 10714 deg$^2$ and a flux limit of about 
$B_J \approx 16.1$ mag.  Here, we include the 87 objects at $z \leq$ 0.5 
for which Boroson \& Green (1992) present spectroscopic data.
We use the spectral measurements (\ie line widths and continuum luminosities)
adopted by Vestergaard \& Peterson (2006) where the details of the data
are also described.

The mass distribution of the BQS sample was previously studied by
Vestergaard (2004a). Since that work, the mass scaling relationships
have been updated owing to improvements made to the reverberation 
mapping database (Peterson \et 2004). Therefore, for the few BQS quasars 
for which robust black hole mass measurements have been obtained using 
reverberation mapping (Peterson \et 2004), we adopt these mass 
determinations in our analysis with exception of PG2130$+$099 for which 
the mass has recently been further improved (Grier \et 2008).
For the remaining sources we adopt the mass estimates based on the 
most recently updated mass scaling relationships. These mass values
are listed in Table~7 of Vestergaard \& Peterson (2006).

\begin{figure}
\epsscale{1.2}
\plotone{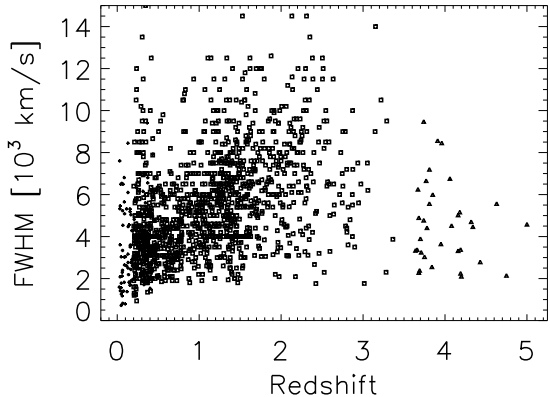}
\caption{Distributions of line widths (FWHM) as a function of redshift for the LBQS
(squares), BQS (asterics), and SDSS color-Selected (triangles) samples.
No significant changes are seen in the distribution with redshift, except the SDSS
color selected sample does not have as extreme wide lines as some LBQS quasars.
\label{FWzfig}}
\end{figure}

\subsection{The Large Bright Quasar Survey} \label{lbqsdata.sec}

The Large Bright Quasar Survey (LBQS; \eg Foltz \et 1987; Hewett \et 2001) 
is the largest published spectroscopic survey of optically selected quasars 
at bright ($B_J < 19$ mag) apparent magnitudes as of 1995.
It consists of 1067 quasars 
located at redshifts between 0.2 and 3.4 and it covers an effective 
area of 453.8 deg$^2$ on the sky. The quasar candidates were selected
from the Automated Plate Measuring machine scans of the United Kingdom
Schmidt Telescope direct and objective-prism plates based on their
spectral energy distribution shapes on these plates. The details of the
survey are published in a series of papers and most recently summarized
by Hewett \et (2001) who also present an updated account of the survey
completeness as a function of redshift.

Spectroscopic observations were obtained in the late 1980's at a resolution 
of 6 $-$ 10\,\AA{} using the 4.5\,m Multi Mirror Telescope (MMT) with a 
wavelength range of \lam \lam 3300 $-$ 7500\,\AA{} or with the Las
Campanas 2.5\,m Du Pont Telescope with a wavelength range of \lam \lam
3400 $-$ 7000\,\AA{} (Morris \et 1991). In these configurations the 
\hb{}, \mgii, and \civ{} emission lines are observed in the spectra of 
quasars at redshifts between 0.2 to about 0.4 $-$ 0.5, between 0.2 to 
1.5 $-$ 1.65, and between 1.2 $-$ 1.26 to a redshift of about 3.4, 
respectively. The presence of two different spectral wavelength ranges
should not affect our results since each spectrum will contain at least
one of the \hb, \mgii, and \civ{} emission lines for the redshift range
from 0.2 to 3.0.

Forster \et (2001) performed automated spectral modeling of the LBQS
spectra and, in particular, present line width measurements for 993 of
the 1067 quasars of the full LBQS sample. The remaining quasars were
not modeled because of strong broad and narrow absorption in the \lya{}
and \civ{} emission lines. The spectra were modeled with a powerlaw
continuum component, UV and optical \feii{} emission templates 
(Vestergaard \& Wilkes 2001; Boroson \& Green 1992), and Gaussian 
emission and absorption line functions. 
Most of the broad emission lines were modeled
with a single Gaussian profile from which the FWHM was obtained. A
small fraction of the spectra are of high enough quality to allow two
Gaussian functions to be fitted to the lines. For the single Gaussian
models we adopted the measurements of the FWHM and uncertainties 
tabulated by Forster \et (their Table~5). To determine the FWHM of the 
emission lines modeled with multiple Gaussian components we first 
regenerated the modeled line profile from the tabulated model 
parameters of all the components. Then we measured the FWHM of the 
regenerated profile (i.e., as the full width at the half maximum peak 
value). The measurement uncertainties of the FWHM in this case were 
obtained by a
suitable weighting of the uncertainties in the tabulated measurements
of the individual Gaussian components. These weights were estimated as
the weights to be applied to the FWHM measurements of the individual
Gaussian components so to obtain the FWHM of the sum of the components 
(determined from the regenerated line profile, described above).
For each of the broad emission lines (\hb, \mgii, \civ) a separate 
weight was determined. We note that only a few quasars had emission lines
modeled with multiple Gaussian components: 7 \hb, 118 \mgii, and 79 
\civ{} lines, respectively (out of a total of 148 \hb, 677 \mgii, and
488 \civ{} profiles, respectively). For profiles for which Forster \et
fixed the line width (no uncertainties were estimated) we adopt a 
typical measurement error of 10\% (\eg Brotherton 1996).

Because the LBQS spectra are not flux calibrated, we determined the 
nuclear monochromatic continuum luminosities not from the spectra but
from the survey $B_J$ magnitudes. 

Given the modest quality of the spectra we visually inspected the 
spectra of quasars with FWHM measurements listed by Forster \et to be 
either below 2000 \kms{} or above 12,000 \kms. The reason is that very 
small and very large line widths measurements are more prone to be 
spurious in spectra of modest quality and they can significantly affect
the mass functions owing to the relatively smaller number of such extreme 
objects. The fact that the Forster \et profile measurements above 12,000 
\kms{} tend to have large uncertainties confirm our general suspicion. 
Most of the very broad line profile fits of Forster \et are typically 
listed for profiles of quite poor quality (with a few having strong 
absorption in addition); in a couple of cases, the fitting must have 
gone bad. We found all \civ{} profiles with FWHM $\geq$ 17,000\,\kms{} 
had to be discarded for these reasons and a few profiles of about 
12,500 $-$ 14,000\,\kms{} have too noisy profiles to be useful. In 
addition, most of the \mgii{} FWHM measurements listed to be above 
12,000\,\kms{} are based on very noisy, unreliable data; these 
measurements were discarded from further analysis.
For the profiles listed to have FWHM $\lsim$ 2000\,\kms, we find that 
for some of the \hb{} and \mgii{} profiles the single (or the second) 
Gaussian component was fitted to noise spikes or what appears to be a 
strong contribution from the narrow line region (judging from the 
strength of the [OIII] \lam 5007 line, when available) which gives the 
appearance of intrinsically very narrow broad-line profiles. For these
profiles, the narrow component was discarded and the broad component
alone, if available, was used to characterize the profile. For profiles
fitted with only a single Gaussian component, the emission line was
discarded from further analysis. A total of 6 \hb{} and 7 \mgii{}
narrow profiles were discarded.

After this filtering of the data with the most uncertain measurements,
we are left with measurements of a total of 139 \hb, 654 \mgii, and
480 \civ{} profiles; a total of 134 quasars have measurements of both 
\hb{} and \mgii, while 161 quasars have both \mgii{} and \civ{}
measurements. Of the original 993 quasars analyzed by Forster \et we 
were able to estimate the black hole masses for 978 quasars.
The adopted FWHM values and the continuum luminosities used for the
mass estimates are listed in Table~\ref{LBQS_FWL.tab}. The black hole
masses, bolometric luminosities, and Eddington luminosity ratios for 
the LBQS are listed in Table~\ref{LBQS_ML.tab}. 
Table~\ref{LBQS_Lz_missingQ.tab} lists the basic properties (name, 
redshift, $B_J$ magnitudes, and luminosity) of the LBQS quasars 
without mass estimates. The determination of the \mbh{} values are 
described in Section~3. 

\begin{figure}
\epsscale{1.0}
\plotone{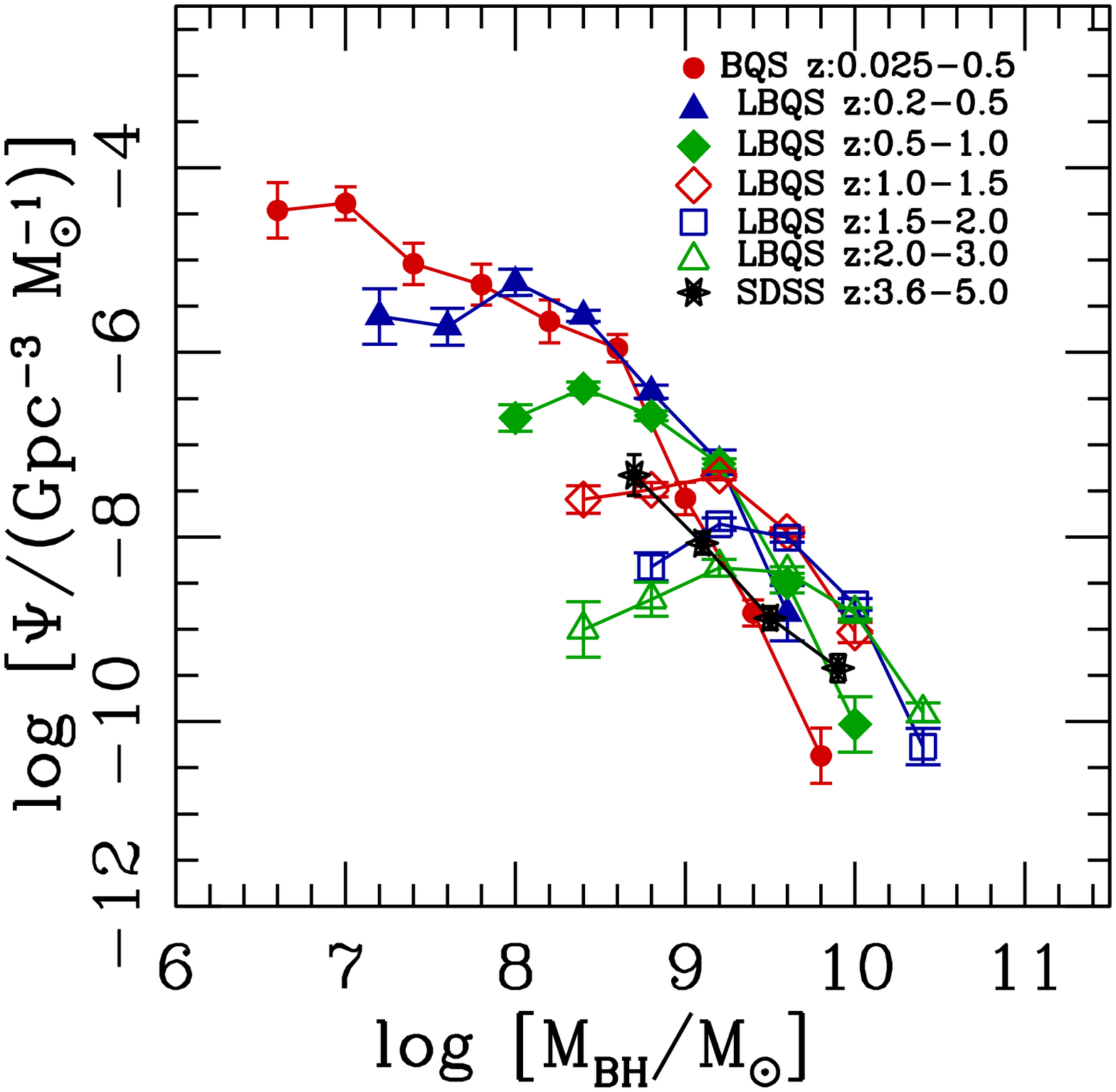}
\vspace{0.2cm}
\plotone{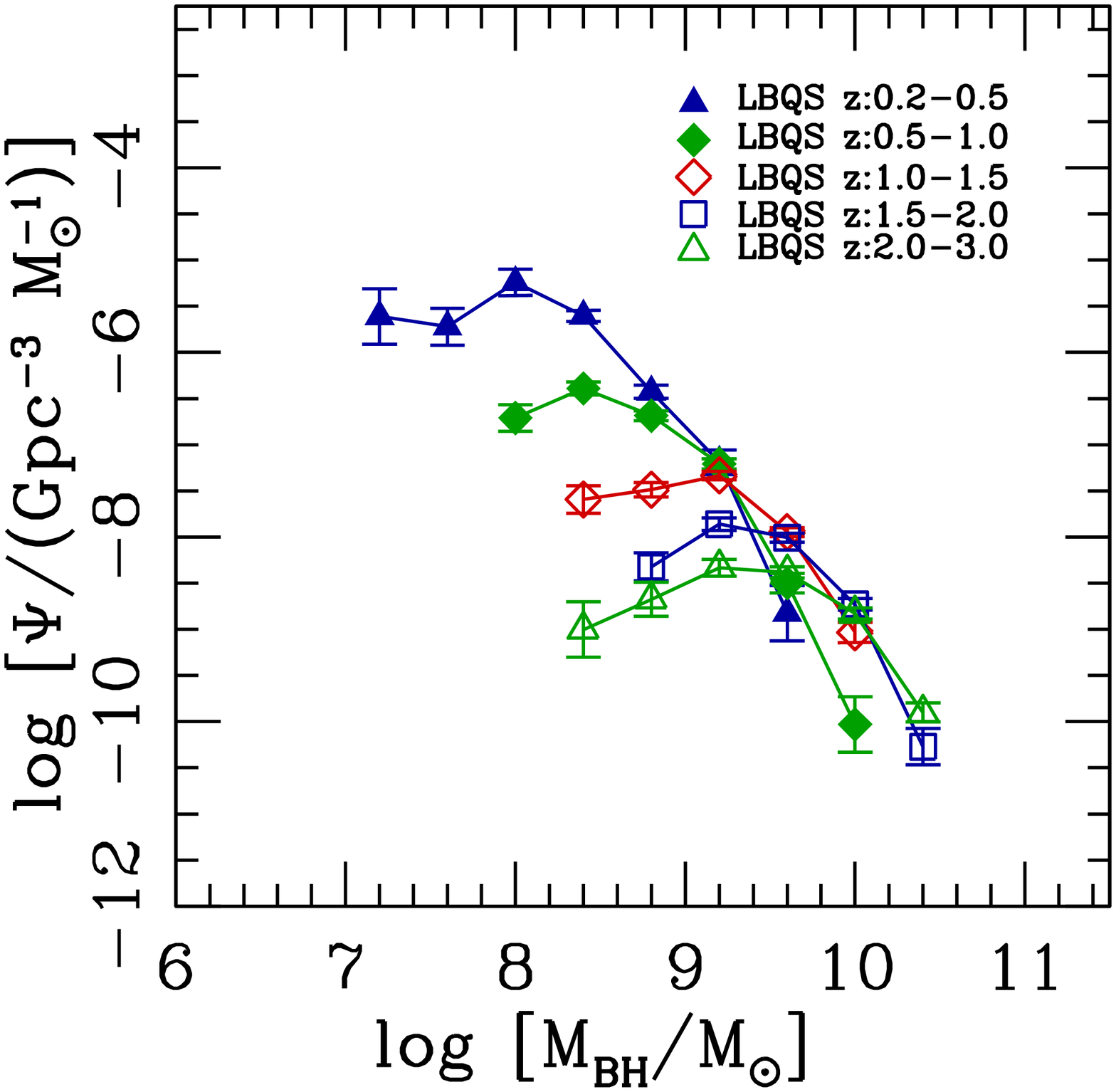}
\vspace{0.2cm}
\plotone{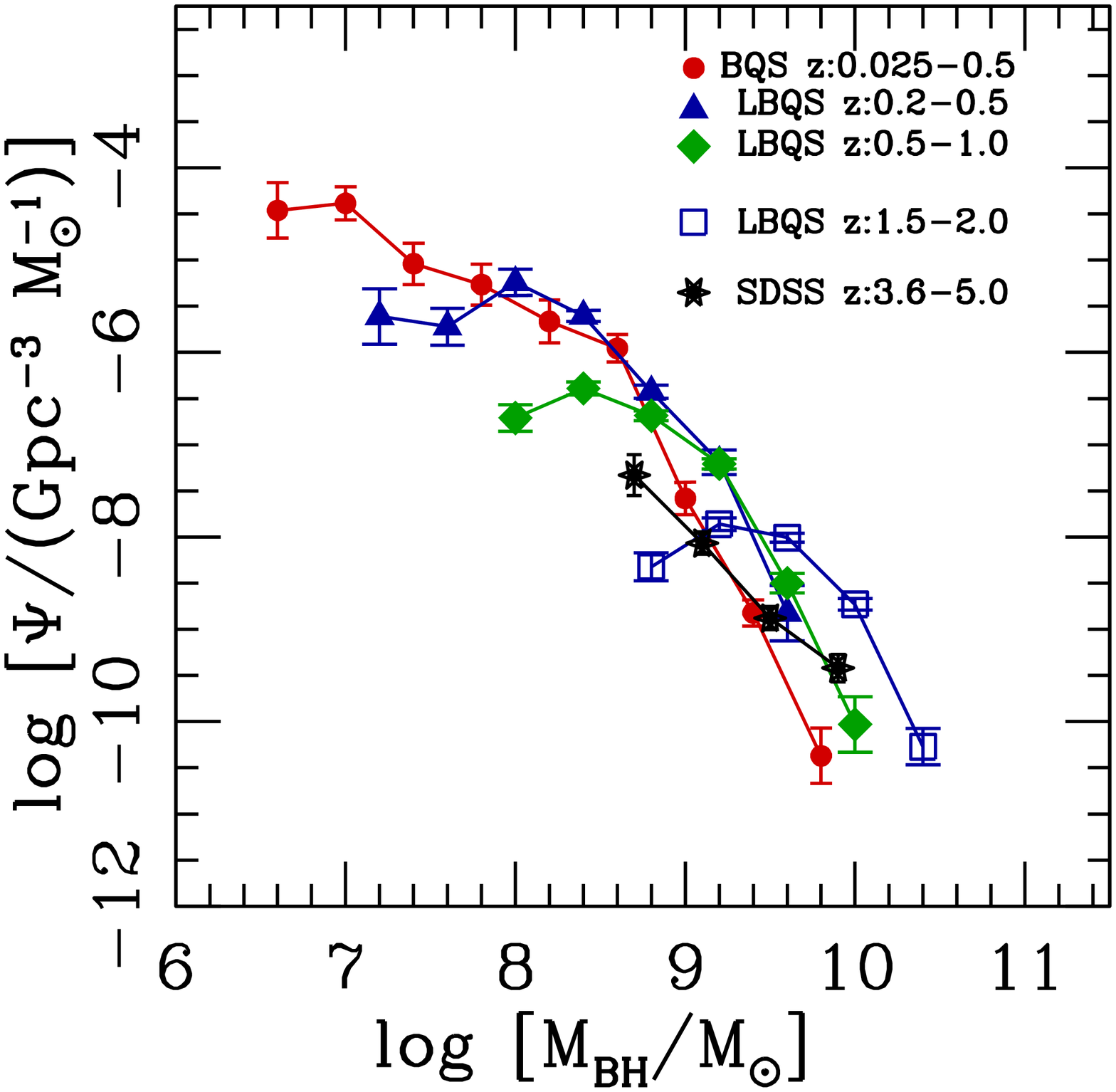}
\caption{Mass functions as a function of mass for the samples analyzed in this work:
{\it Top Panel}$-$ the LBQS (for all five redshift bins), the BQS, and the SDSS
color-selected samples;
{\it Middle Panel} $-$ the LBQS alone (for the five different redshift bins);
{\it Lower Panel} $-$ the BQS and the SDSS color-selected samples are shown with
selected redshift bins of the LBQS for ease of comparison.
The mass functions show turn-overs toward low masses due to incompleteness.
\label{MFallfig}}
\end{figure}

\subsection{The SDSS Color-selected Sample in the Fall Equatorial Stripe}
 \label{ColSeldata.sec}

Fan \et (2001a) present a well-defined color-selected sample of 
38 quasars at 3.6 $< z \leq$ 5.0 from a 182 deg$^2$ field in the SDSS
Fall Equatorial Stripe for which they determine the quasar luminosity
function. The quasars were selected based on $gri$ and $riz$ colors
to be complete in the survey area down to $i^{\ast}$ = 20 mag. 
The continuum luminosities measured from the spectra were recalibrated
to the dereddened AB(1450\AA) magnitudes (see \eg Vestergaard 2004a;
see Fan \et 2001a for details on these data).
For four of the 38 quasars we are unable to obtain black hole mass 
estimates, because the (discovery) spectrum did not include the \civ{} 
emission line (J021043.17$-$001818.4, J021102.72$-$000910.3, 
J025019.78$+$004650.3) or the quality of the spectrum was too poor to 
measure FWHM(\civ) reasonably reliably (J020731.68$+$010348.9).  For 
quasar J021102.72$-$000910.3 another spectrum is available in the SDSS 
archive, but we were unable to reliably measure spectral parameters for 
the mass estimate from that spectrum. 

Fan \et originally presented 39 quasars in this sample, but one 
source (J225529.09$-$003433.4) has since been reclassified as a 
star\footnote{This classification is verified by the NASA/IPAC Extragalactic 
Database (NED): http://nedwww.ipac.caltech.edu.}; the spectrum is
also missing from the discovery papers (Fan \et 2001a; Schneider \et 
2001) and the SDSS Data Release 6 Archives.

This quasar sample was also included in the study by Vestergaard 
(2004a) on the distributions of black hole mass and Eddington 
luminosity ratios for distant quasars. Since improved mass estimation
relations have been published more recently (Vestergaard \& Peterson 
2006), we redetermine the black hole masses for this sample.
We adopt the measurements of FWHM(\civ) and 1350\,\AA{} continuum
luminosities obtained and analyzed by Vestergaard (2004a).
 
The FWHM and continuum luminosity measurements used for the
mass estimates of the SDSS color-selected sample are listed in 
Table~\ref{SDSS_FWL.tab}. The black hole
masses, bolometric luminosities, and Eddington luminosity ratios for 
these quasars are listed in Table~\ref{SDSS_ML.tab}.
The computations are described in the next section.

\begin{figure}
\epsscale{.95}
\plotone{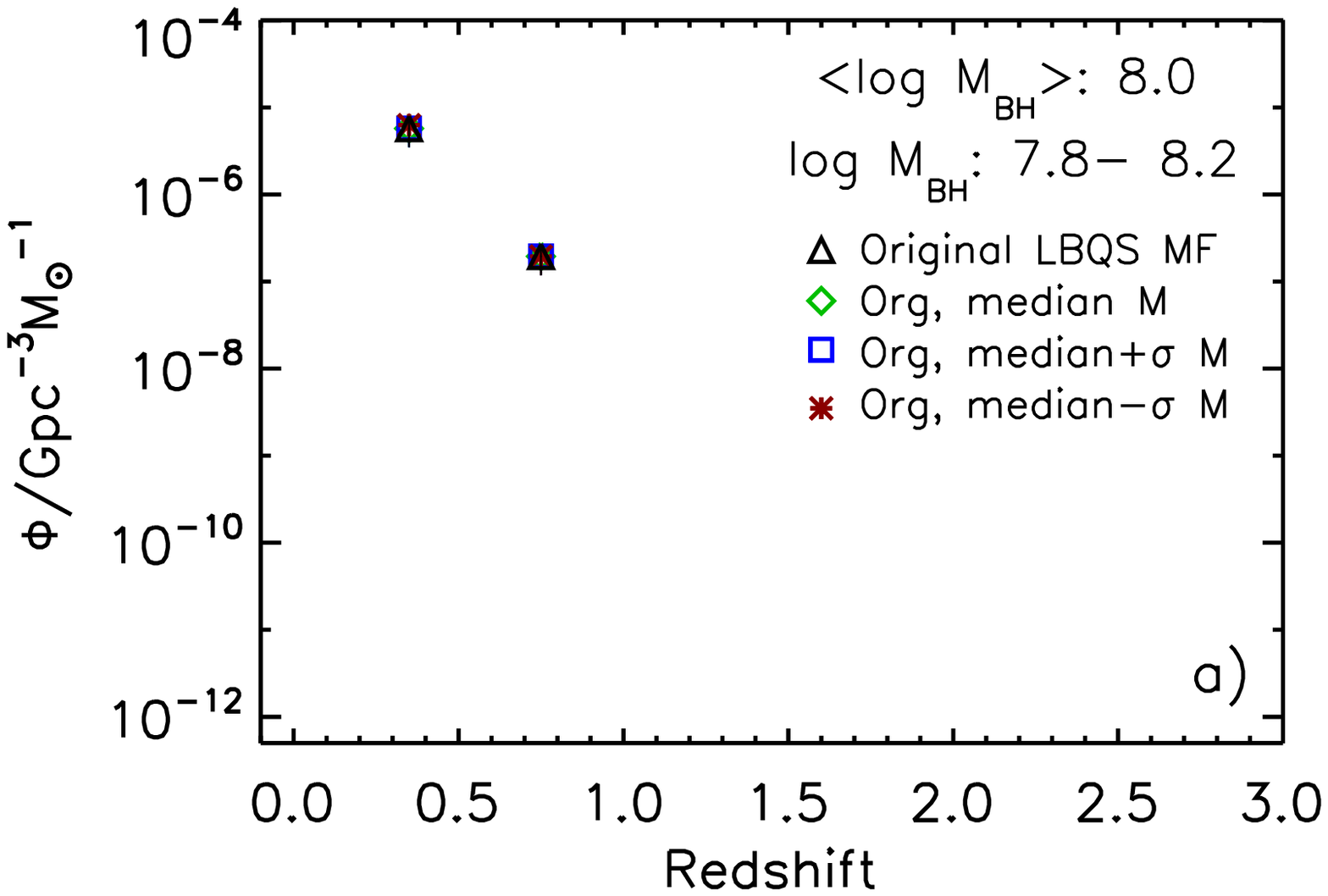}
\vspace{0.1cm}
\plotone{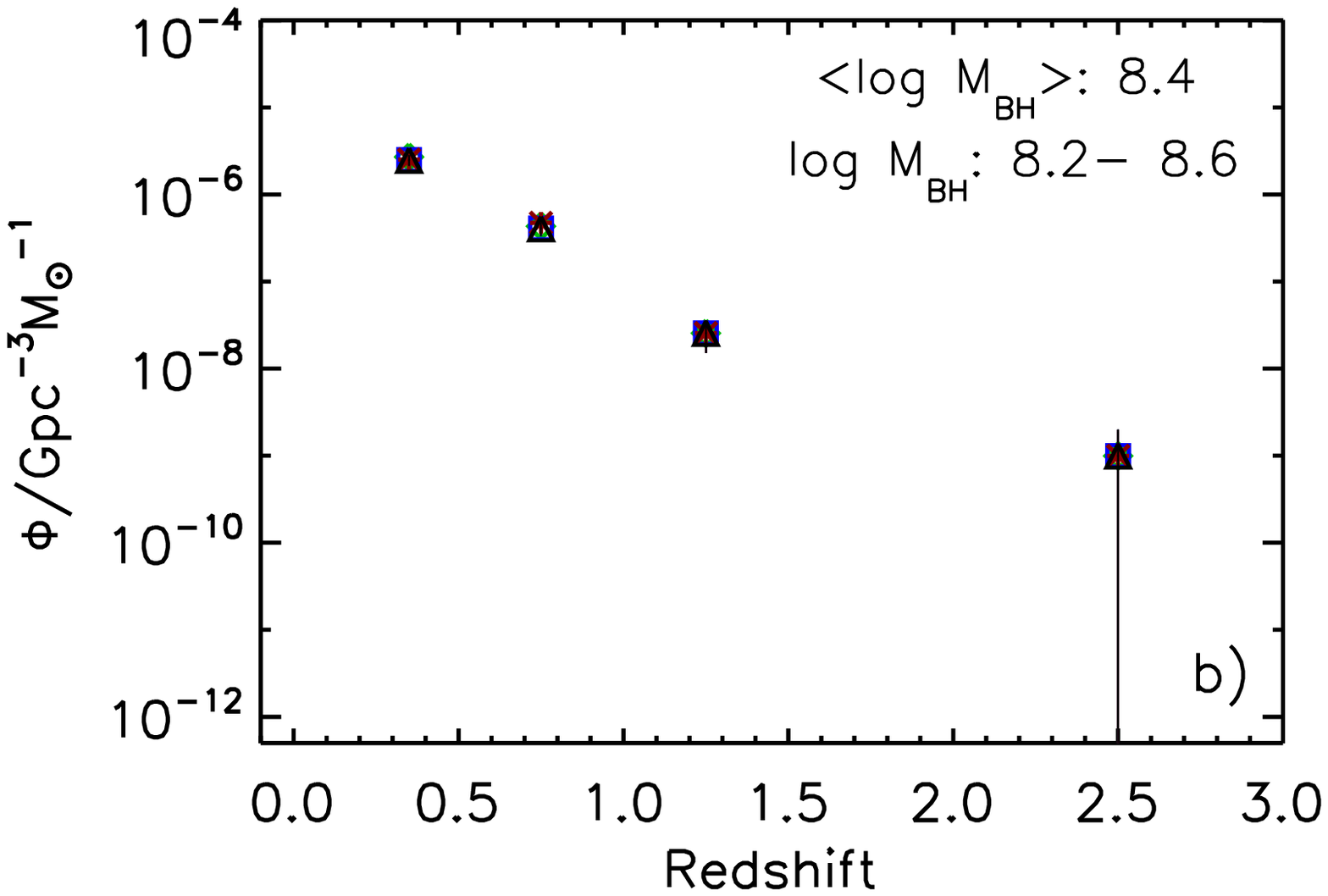}
\vspace{0.1cm}
\plotone{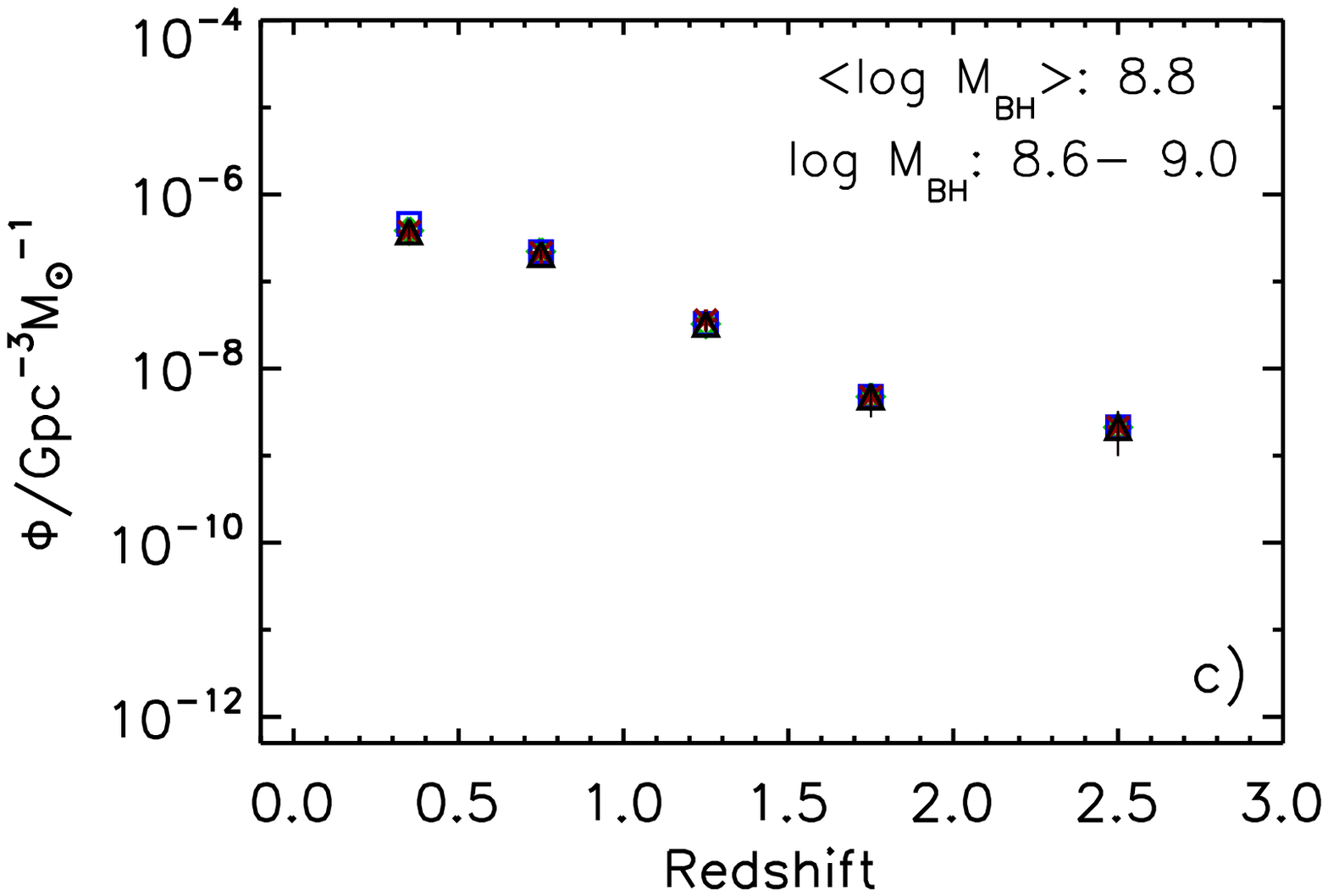}
\vspace{0.1cm}
\plotone{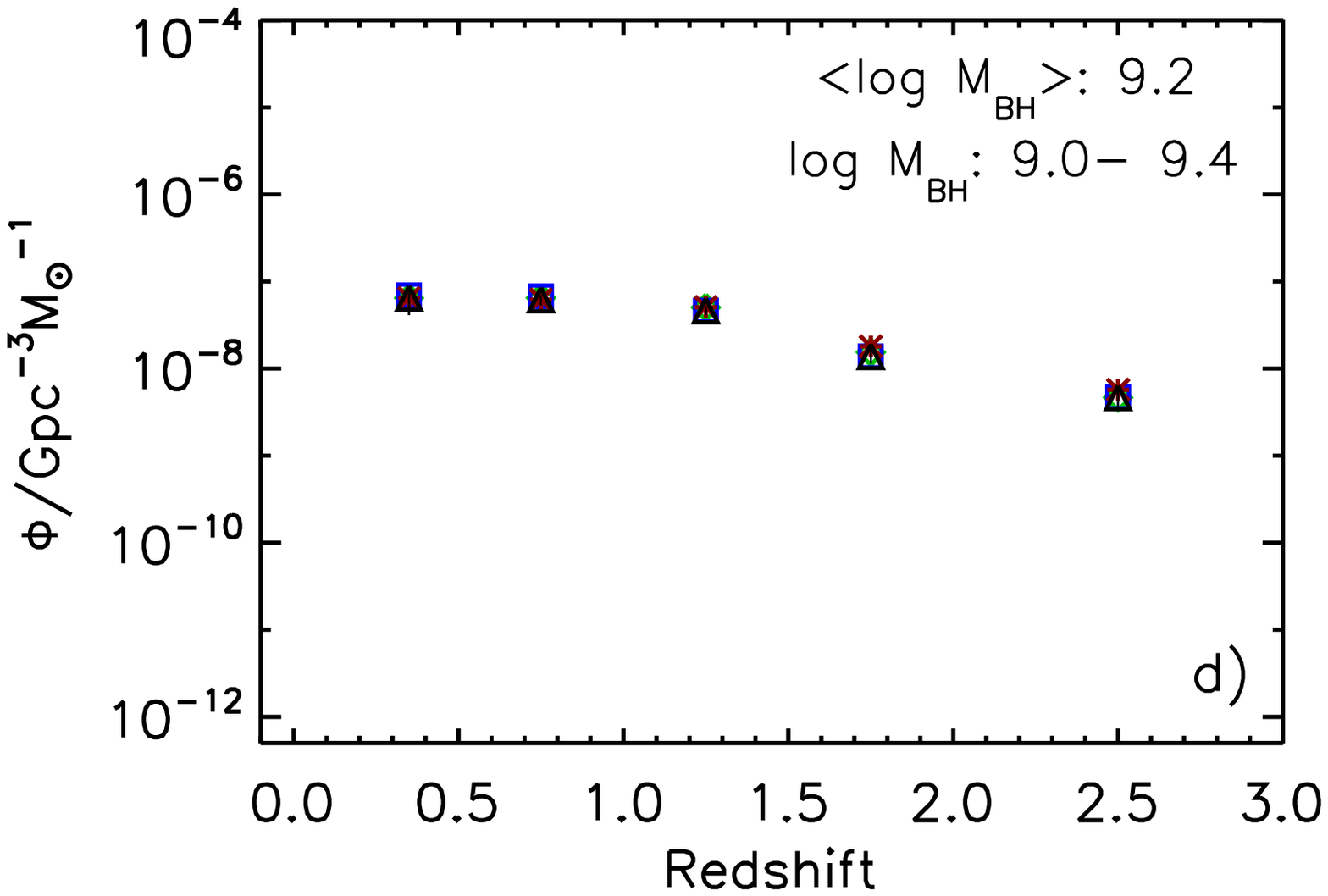}
\caption{...figure is continued in next column...}
\end{figure}

\setcounter{figure}{4}
\begin{figure}
\epsscale{0.95}
\plotone{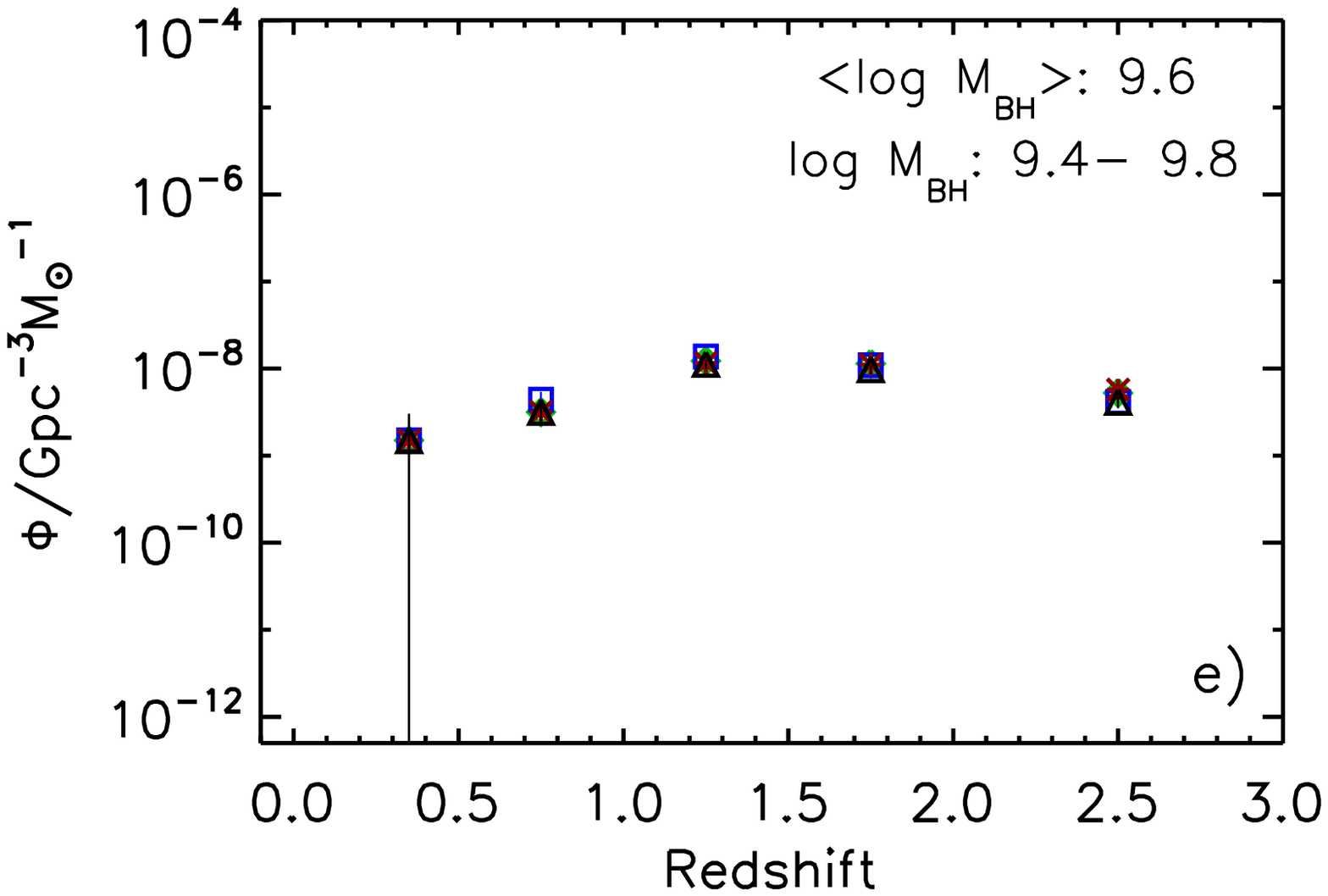}
\vspace{0.1cm}
\plotone{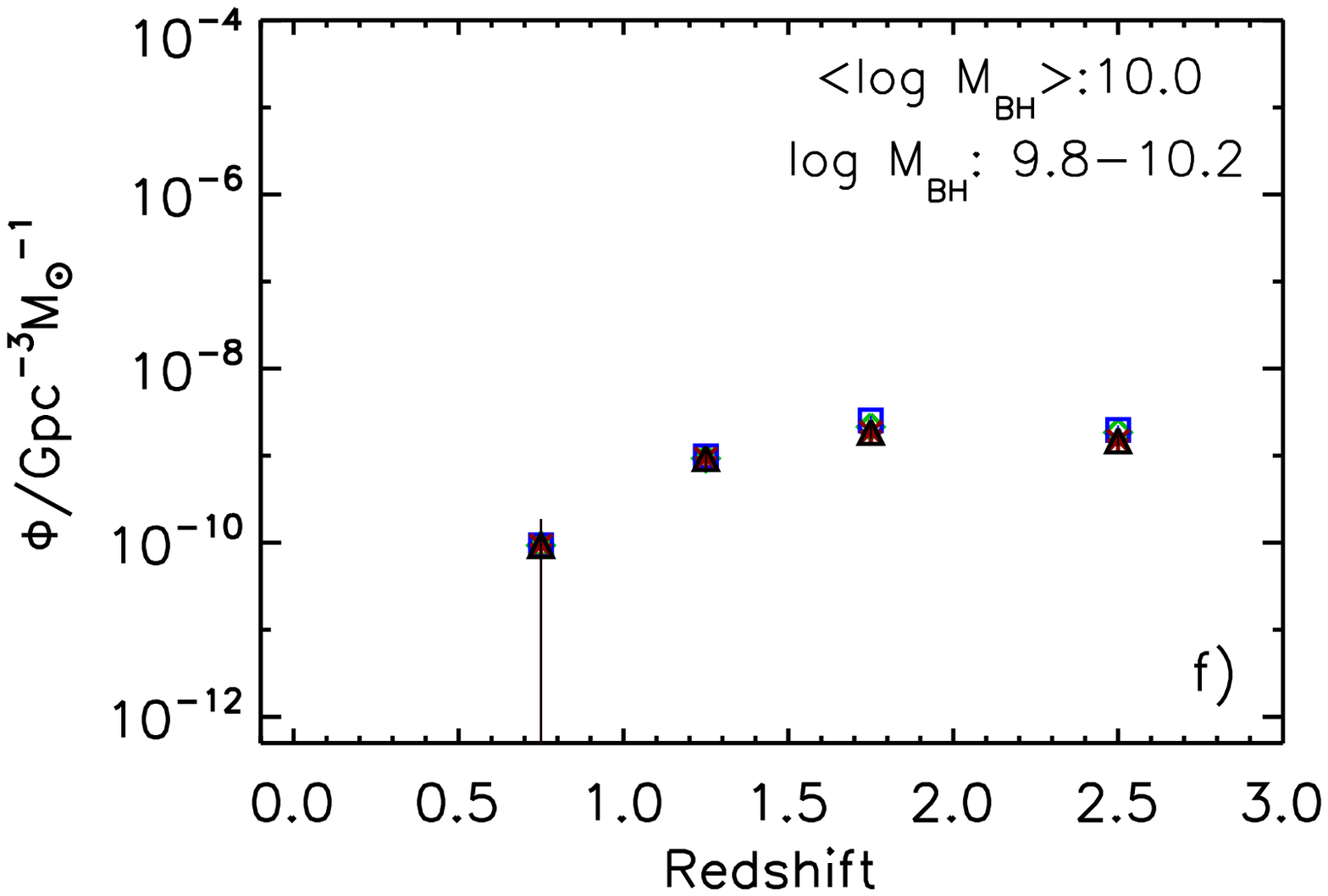}
\vspace{0.1cm}
\plotone{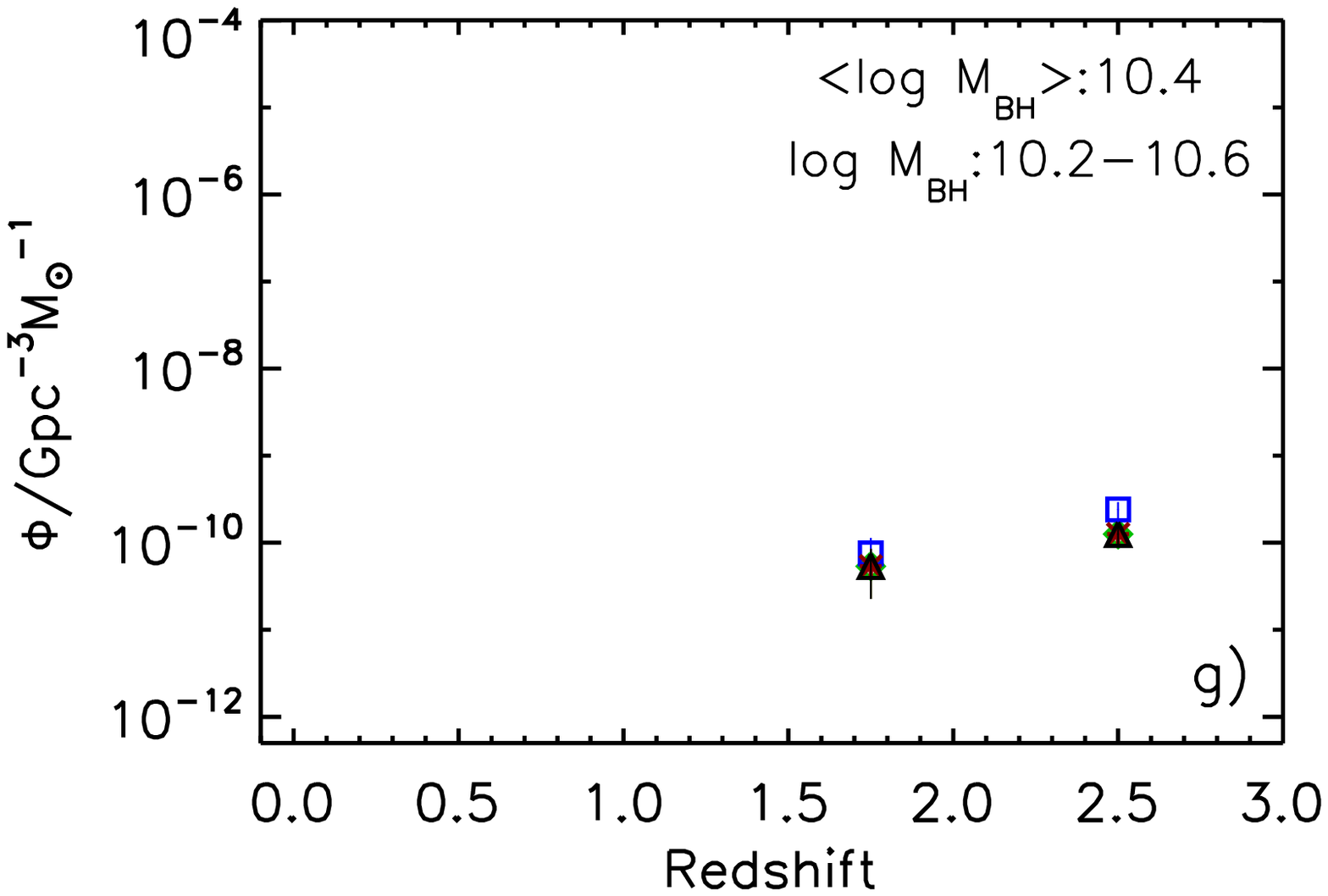}
\caption{Mass function of LBQS as a function of redshift at different mass bins
as labeled in each panel.
The mass function based only on (969) measurements of suitable spectroscopic data is shown
as black open triangles. In addition, the full sample of 1058 quasars is also shown in
three different versions. Sources without suitable spectroscopy (`the missing subset') is
assigned one of three mass values determined according to their individual redshift and
luminosity based on the 978 quasars with mass estimates. Diamonds denote inclusion of
sources assigned the median \mbh{} value; Boxes denote inclusion of sources assigned
the (median $+ 1\sigma$) \mbh{} value; Crosses denote inclusion of sources assigned the
(median $- 1\sigma$) \mbh{} value.
\label{MFzfig}}
\end{figure}

\section{Black Hole Mass Estimates \label{mestim}}

We determine black hole mass estimates using the socalled mass scaling 
relationships which utilize the widths of the broad emission lines and
the nuclear continuum luminosities (\eg Wandel, Peterson, \& Malkan 1999; 
Vestergaard 2002, 2004b, 2009; McLure \& Jarvis 2002; Warner \et 2003; 
see also Dietrich \& Hamann 2004). This method is preferred for several 
reasons (see also Vestergaard 2009). 
Only a couple of methods are applicable both to active
galaxies in the nearby universe as well as the most distant quasars.
Of these, mass scaling relationships have some of the lowest associated
uncertainties (\eg Vestergaard 2004b) which can be further improved upon 
in the future (\eg Marconi \et 2008). And
quite importantly, these relationships are anchored in robust black 
hole mass determinations of low redshift active nuclei based on the 
reverberation mapping method (\eg Blandford \& McKee 1982; Peterson 
1993) which have recently been updated following a homogeneous 
reanalysis of the available reverberation database (Peterson \et 2004; 
Onken \et 2004).
Moreover, the evidence in favor of our application of this method to 
more distant sources is quite strong (see \eg Vestergaard 2004b, 2009,
and references therein).

Black hole mass estimates can be obtained using one or more of the 
\hb, \mgii, or \civ{} emission lines. For mass estimates based on \hb{}
and \civ{} we use equations (5) and (7), respectively, presented by
Vestergaard \& Peterson (2006); these relationships are calibrated to 
the most recently updated robust reverberation mapping mass 
determinations (Peterson \et 2004; Onken \et 2004). At present, there 
is no published relationship for the \mgii{} emission line which is 
(re)calibrated to the improved reverberation masses and also
intercalibrated to the mass estimates based on the \civ{} emission 
line\footnote{
The McGill \et (2008) relations do not satisfy these requirements; they  
are based on a sample of only 19 SDSS quasars at $z \approx$0.36 and do not 
include a relation for \civ. We remind the reader that 
our ultimate intent is to compare and combine the mass functions presented
here with those published by Vestergaard \et (2008) on the SDSS DR3 quasar 
survey. For this particular purpose we need to use the {\it exact same} mass 
estimation relationships for all the individual mass functions.
}. 
We therefore obtained a new relationship for \mgii{} using several thousand 
high-quality spectra from the SDSS DR3 quasar sample (Schneider \et 2005). 
This relationship has been applied to the subset of the DR3 quasar sample 
used to establish the luminosity (Richards \et 2006) and black hole mass 
(Vestergaard \et 2008) functions. 
For completeness, we here present the relationships used. We obtained a 
relation for each of four monochromatic continuum luminosities because the
3000\,\AA{} luminosity may not always be accurately determined, sitting below the
strong \feii{} line emission at those wavelengths. We use the relation 
pertaining to the nuclear 
monochromatic continuum luminosity that can be best and most accurately 
measured in the observed spectrum. We refrain from extrapolating or 
adopting an assumed continuum slope in any of our work. For a given wavelength, 
\lam, the black hole mass based on \mgii{} was obtained according to:

\begin{equation}
M_{\rm BH} = 10^{zp(\lambda)} \left[\frac{{\rm FWHM(Mg{\sc II})}}{1000 \,km/s}\right]^2 \left[\frac{\lambda L_{\lambda}}{10^{44} \,erg/s}\right]^{0.5}
\end{equation}
where $zp(\lambda)$ is 6.72, 6.79, 6.86, and 6.96 for \lam 1350\,\AA, 
\lam 2100\,\AA, \lam 3000\,\AA, and \lam 5100\,\AA, respectively.
The 1\,$\sigma$ scatter in the absolute zero-points, $zp$, is 0.55\,dex
which includes the factor $\sim$2.9 uncertainties of the reverberation
mapping masses to which these mass estimation relations are anchored
[see e.g., Vestergaard \& Peterson (2006) and Onken \et (2004) for details].
On average, these relations are consistent to within about 0.1\,dex of
the \hb{} and \civ{} mass estimates.  The relationships will be discussed 
further in a forthcoming paper (M.\ Vestergaard et al., in preparation).

A few of the BQS sources have been targeted by reverberation mapping. For
those sources, we adopt the reverberation mass listed by Peterson \et 
(2004). Since the objects in the BQS are all located at $z < 0.5$ we use
the \hb{} relationship to estimate the black hole mass for the remaining
sources in this sample. Those mass estimates are listed in Table~7 of 
Vestergaard \& Peterson (2006) and by Grier \et (2008) for PG\,2130$+$099.

The quasars in the LBQS span a large range of redshifts and therefore we 
applied all three emission line relations as follows. For any given 
source for which the FWHM of any of the three emission lines (\hb, 
\mgii, \civ) could be reliably measured we determined an estimate of
the mass based on each of these emission lines. If more than one 
emission line estimate is available for a given quasar, the final mass 
estimate of that source was determined as the weighted average of the 
available individual mass estimates. 
The adopted weights are the inverse variance determined from the 
propagated measurement errors. Therefore, for redshifts between 0.2 
and 0.66 the mass estimate for each object is based on both the \hb{}
and \mgii{} emission lines while for redshifts between 1.04 and 1.71
the mass is based on both the \mgii{} and \civ{} lines. The black hole
mass estimates for the LBQS are listed in Table~\ref{LBQS_ML.tab}.

The quasars in the SDSS color-selected sample all reside at redshifts 
for which only the \civ{} emission line can be observed using optical
spectroscopy. We therefore use equation (7) of Vestergaard \& 
Peterson (2006) to estimate the black hole masses for this sample. The
mass values are tabulated in Table~\ref{SDSS_ML.tab}.

The distribution of (bolometric) luminosities and black hole masses 
with redshift is shown for all three samples in Figures~\ref{lzfig} 
and~\ref{mlolzfig}, respectively. The bolometric luminosities, \lbol,
are obtained for the LBQS quasars by scaling monochromatic continuum
luminosities at 1350\,\AA, 2100\,\AA, 3000\,\AA, and 5100\,\AA{} with 
a constant (average) bolometric correction factor\footnote{
Note, the bolometric correction factors are suggested to depend on both 
black hole mass (Kelly \et 2008b) and 'Eddington luminosity ratio' 
(Vasudevan \& Fabian 2007) which may introduce systematic uncertainties 
in the derived \lol{} values in addition to those stemming from assuming 
a constant bolometric correction given that a range of spectral energy
distributions exist of quasars (\eg Elvis \et 1994; Kuhn \et 2001; 
Richards \et 2006).
Given that a robust adjustment scheme for the bolometric corrections has
not been isolated and that the Eddington luminosity ratios are not analyzed 
further here, adopting a more complex correction factor is unnecessary at
this time.
} (of 4.3$\pm$0.46, 
5.4$\pm$0.26, 5.8$\pm$0.24, and 10.5$\pm$0.24, respectively) extracted 
from Richards \et (2006).
For each 
quasar the \lbol{} value is determined from the monochromatic luminosity
that is as closely centered in its optical observing window.  For the SDSS
color-selected sample, the 1350\,\AA{} continuum luminosity were
similarly corrected for an estimate of the \lbol{} values. The BQS
bolometric luminosities are adopted from Sanders \et (1989) 
with appropriate cosmological corrections applied to conform to the
cosmological model used in this work.

Figures~\ref{lzfig} and~\ref{mlolzfig} clearly demonstrate that since
we can probe to lower luminosity limits in the nearby universe, we
can also probe less massive black holes than at high redshift. While 
the BQS spans a large range in masses, it does not go very deep; 
recall, the limiting magnitude is $B_J \approx$ 16.2 mag. The LBQS is 
also a `bright' survey but has a lower flux limit (Fig.~\ref{lzfig}). 
This is also clear from the relative distributions of black hole masses 
at low redshifts for which there is some redshift overlap between the 
LBQS and the BQS (Fig.~\ref{mlolzfig}). The 
masses of the high-$z$ color-selected sample and of the LBQS sources 
above a redshift of 1 are all of order a billion solar masses. 

For completeness, we display the redshift distributions of the
Eddington luminosity ratios, \lol, in Figure~\ref{mlolzfig}. The
quasars beyond $z \approx$ 1 typically have \lol{} values between 
$\sim$0.1 and $\sim$1.0, while in the nearby universe the surveys can 
probe sources that accrete at rates down to about 1/100th of the
Eddington limit. This distribution is consistent with earlier studies
(\eg Warner \et 2003; Shemmer \et 2004; McLure \& Dunlop 2004; 
Vestergaard 2004; Kollmeier \et 2006; Shen \et 2008; Netzer \&
Traktenbrot 2007) and results showing that
distant quasars are more actively accreting than local quasars and
active nuclei (\eg Peterson \et 2004). It is important to keep in 
mind, here, that distant quasars also tend to be more luminous (\eg
Figure~\ref{lzfig}).
Also, we show the distribution of FWHM of the emission lines with 
redshift in Figure~\ref{FWzfig}. There is no significant change in
FWHM with redshift with the exception that the SDSS quasars do not
have the extreme wide lines that some LBQS quasars do.

\begin{figure}
\epsscale{1.10}
\plotone{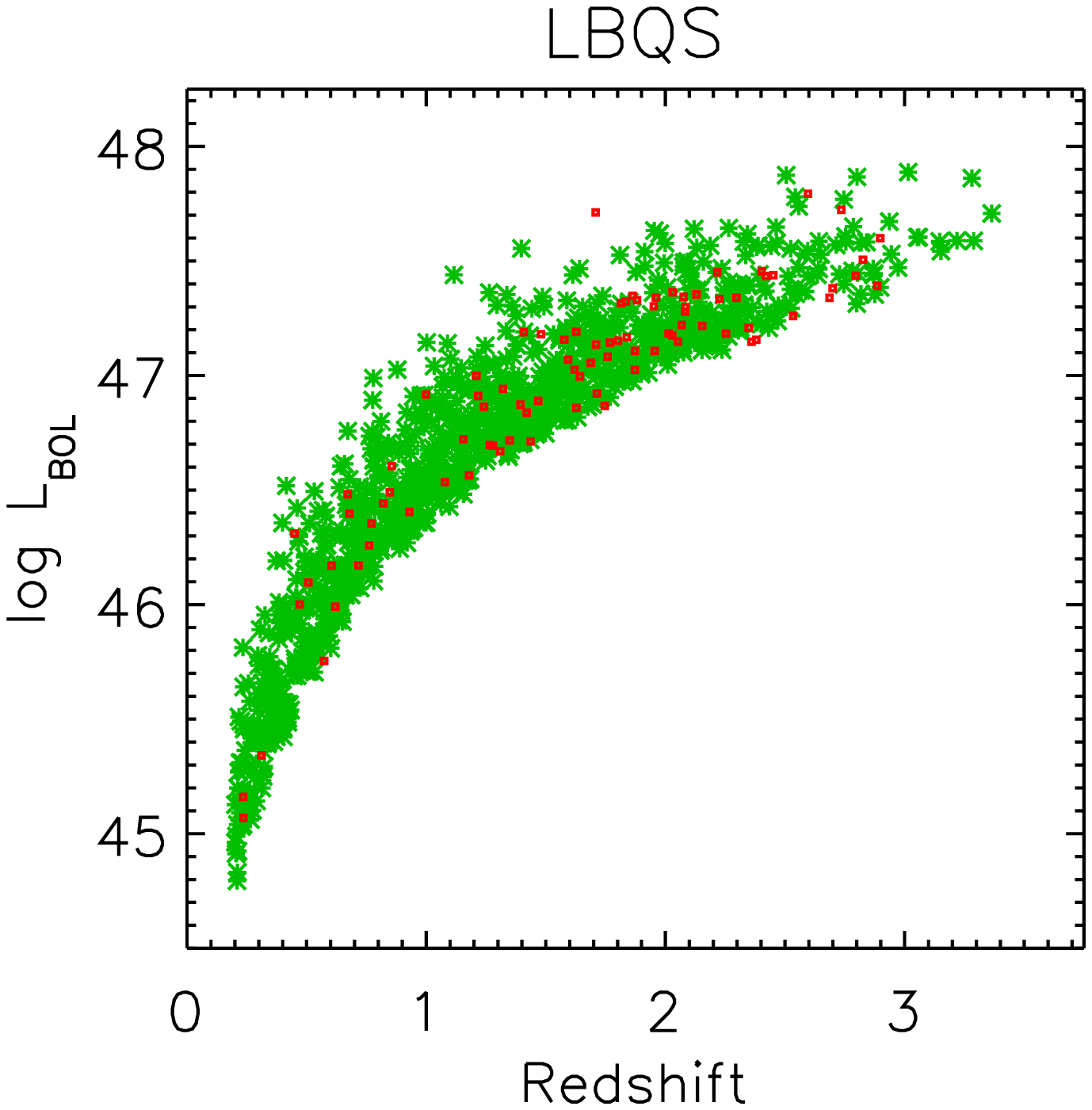}
\plotone{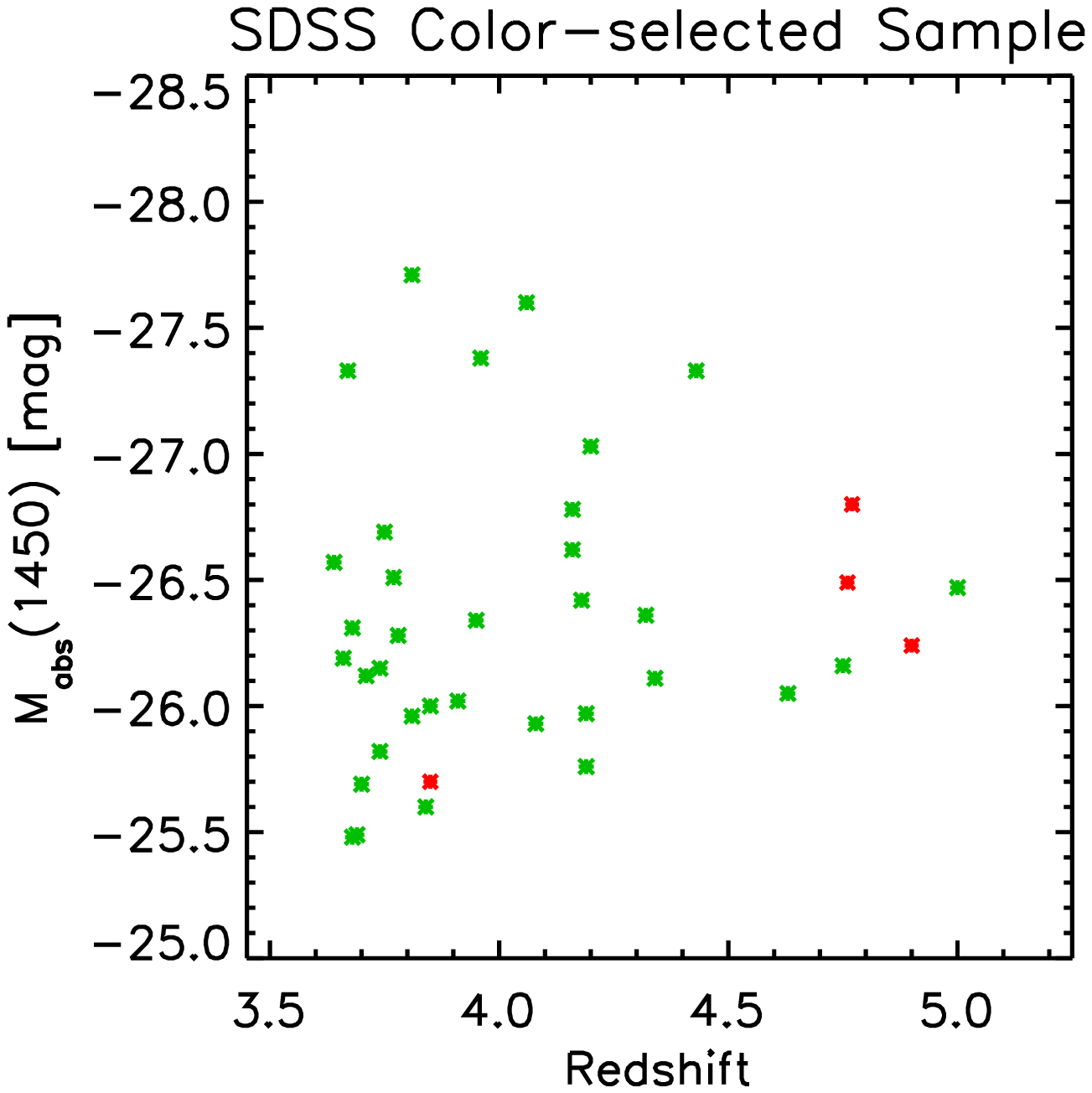}
\caption{Distribution of luminosities with redshift of the LBQS ({\it Upper panel})
and SDSS color-selected ({\it Lower panel}) samples. The sources for which there
are no spectroscopy available suitable for black hole mass estimates are highlighted.
\label{lmissQfig}}
\end{figure}

\section{Black Hole Mass Functions \label{MF.sec}}

The quasar black hole mass function, $\Psi$(\mbh,$z$), is defined
as the comoving space density of black holes per unit black hole 
mass as a function of black hole mass and redshift.  To determine the
space density (\ie the number of black holes per unit comoving volume)
in a given mass and redshift bin, we use the 1/$V_a$ method presented
by Warren, Hewett, \& Osmer (1994), where $V_a$ is the accessible 
volume, defined by Avni \& Bahcall (1980). The mass function and its
statistical uncertainty is described as

\begin{eqnarray}
\Psi(<M_{\rm BH}>,<z>)  =  \sum_{i = 1}\frac{1}{V_{a,i}\, \Delta M_{\rm BH}}, {\rm ~~~and}\\
\sigma(\Psi)  =  \left[\sum_{i = 1}\left(\frac{1}{V_{a,i}\,\Delta M_{\rm BH}} \right)^2\right]^{1/2},
\label{mf.eq}
\end{eqnarray}
\noindent
respectively. The sum is performed over the objects (denoted by $i$) with 
redshift in the range $<z> - \Delta z/2$ and $<z> + \Delta z/2$ and with 
masses in the range 
$<M_{\rm BH}> - \Delta M_{\rm BH}/2$ and $<M_{\rm BH}> + \Delta M_{\rm BH}/2$.

We follow the method of Fan \et (2001a) and Vestergaard \et (2008) of
including the selection function of the quasar survey in the 
computation of the accessible volume:

\begin{equation}
V_{a,i}\,(z_a) = \left({\int_{z_{\rm min}}^{z_a} p(z)\, \frac{dV}{dz} dz} \right)_i
\label{Va.eq}
\end{equation}

\noindent
The ``accessible redshift'', $z_a$, is the minimum of $z_{\rm max}$,
the maximum redshift that object $i$ can have and still be detected by 
the survey,
and the upper redshift limit in the survey or in the adopted redshift 
bin, \ie $z_a = {\rm min} [z_{\rm max},z_{\rm limit}]$.
The volume element, $dV/dz$, is defined by Hogg (1999) for a 
$\Lambda$CDM cosmology.

For the BQS a constant survey completeness of $p$ = 0.88, determined
by Schmidt \& Green (1983), was adopted. For the LBQS we adopt the 
survey completeness as a function of redshift presented by Hewett \et
(2001; their Table~7). The selection function for the SDSS color-selected
sample was presented by Fan \et (2001a) and is a function of both 
luminosity, redshift, and spectral energy distribution, $p(L,z,{\rm SED})$.

We show the black hole mass functions of the three samples as a 
function of black hole mass in Figure~\ref{MFallfig}. We limit the
LBQS mass functions to redshifts below 3 since there are only nine
quasars between redshifts 3 and 3.4; the LBQS mass functions are
thus based on 969 quasars at $z \leq 3$.  The LBQS sample
is large enough to allow the mass functions to be determined for a 
range of redshift bins; they are shown in whole or in part in 
the various panels of Figure~\ref{MFallfig}. This also allows us to 
display the mass functions as a function of redshift for a given black 
hole mass (Figure~\ref{MFzfig}).
The mass functions for the three quasar samples are tabulated in 
Tables~\ref{BQS_MF.tab}, \ref{LBQS_MF.tab}, and~\ref{SDSS_MF.tab}.
Table~\ref{LBQS_MF_zdep.tab} tabulates the redshift dependent mass
function of the LBQS.

For 89 quasars (or about 8\%) of the complete LBQS sample there is no
spectral modeling available and therefore no black hole mass estimates
exist for these sources. We refer henceforth to this subset as the
`missing subset' (Figure~\ref{lmissQfig}). 
This affects the mass functions to some degree as we
will underestimate the space density of certain black hole masses. In
an attempt to estimate the most likely mass function for the entire
LBQS sample we have used the observed distribution of black hole
masses (for those quasars with reliable spectral modeling) 
to determine for each quasar in the missing subset the most likely 
black hole mass and the reasonably expected mass range around this 
value.  From the observed distribution of LBQS black hole masses we
determined the median mass and the standard deviation $\sigma$ around 
this mass value in bins of $z$ and $B_J$ of widths $\Delta z$ = 0.1 
and $\Delta B_J$ = 0.5; each bin has typically between 10 and 45 objects. 
For each of the 89 sources in the missing subset we used the observed 
redshift and $B_J$ magnitude (see Fig.~\ref{lmissQfig}) to identify 
three mass values: the most likely mass (the median mass value of the 
observed distribution in the appropriate $z$ and $B_J$ bin) 
and the $\pm$\,1$\sigma$ mass values relative thereto, respectively. 
We then generated three catalogs of 1058 black hole mass estimates at 
$z \leq 3$, each consisting of the original 969 black hole mass 
estimates based on spectral measurements plus for the missing 
subset either the most likely mass value, the most likely mass 
$+ 1\sigma$(\mbh), or the most likely mass $- 1\sigma$(\mbh), 
respectively. 
For each
of these three mass catalogs we redetermined the mass functions for
LBQS. By adopting the same type of mass estimate for each quasar in
the missing subset we get a handle on the most likely mass functions
and the $1\sigma$ extremes. Notably, this is in practice different
from running Monto Carlo simulations but gives us similar insight on
the possible distributions of the mass functions. We show these
adjusted mass functions as a function of mass in 
Figure~\ref{LBQS_MF_missQ.fig}. For comparison the mass functions 
based on only the 969 quasars (at $z \leq 3$) with spectral measurements 
(shown in Figure~\ref{MFallfig}) are also shown. The adjusted mass
functions are shown as a function of redshift in Figure~\ref{MFzfig}.
Our omission (or inclusion) of the missing subset clearly has no 
significant effect on the LBQS mass function.

Figure~\ref{LBQS_MF_missQ.fig} shows that inclusion of the sources 
without mass estimates does not change the LBQS mass function (to 
within the statistical uncertainties) for redshifts below 2. At higher
redshifts the missing sources will at one extreme 
(\mbh(median) $- \sigma$) tend to increase the peak amplitude of the 
mass function slightly, and at the other extreme 
(\mbh(median) $+ \sigma$) the mass function will broaden slightly 
toward higher masses. The latter has the stronger effect due to the 
lower number of sources with mass estimates above $10^{10}$\Msol.  
However, these variations are all within the statistical uncertainties
with exception of the (\mbh{} $+ \sigma$) mass function which is 
marginally more deviant. 
Notably, in reality the mass function of the full LBQS 
sample, that we could obtain if higher quality data were available 
of the missing subset, is more likely to be between the two extreme
cases; as expected, assigning the median mass of the relevant
$z$ and $B_J$ bin does not change the mass function.  We therefore 
conclude that we are not making a significant error at this point in 
excluding the sources with low quality spectra (\ie the missing subset) 
in determining the LBQS black hole mass function.

We repeated this exercise for the SDSS color-selected sample since
for four of the 38 quasars the black hole mass could not be estimated.
The effects of assigning \mbh(median) $- \sigma$(\mbh) or 
\mbh(median) $+ \sigma$(\mbh) to these few sources are shown in 
Figure~\ref{SDSS_MF_missQ.fig}.  The differences in the mass functions 
are most noticeable when the extreme mass values of \mbh(median) 
$+ \sigma$ are adopted: the mass function flattens slightly toward
higher mass values. Nonetheless, this extreme case mass function is
still consistent to within the statistical uncertainties with the 
``original'' mass function based on the 34 quasars with spectral 
measurements (\ie Figure~\ref{MFallfig}). Therefore, exclusion of 
the four sources without mass estimates does not significantly
affect the mass function of the SDSS color-selected sample.

We will reassess this issue of missing black hole masses in future
work when analyzing the mass functions from this work with respect
to the SDSS DR3 mass functions of Vestergaard \et (2008).

The black hole mass functions for the LBQS and the SDSS color-selected
samples with these adjustments applied are listed in Tables~\ref{LBQS_MFcor.tab}
and~\ref{SDSS_MF.tab}, respectively.

\begin{figure}
\epsscale{0.90}
\plotone{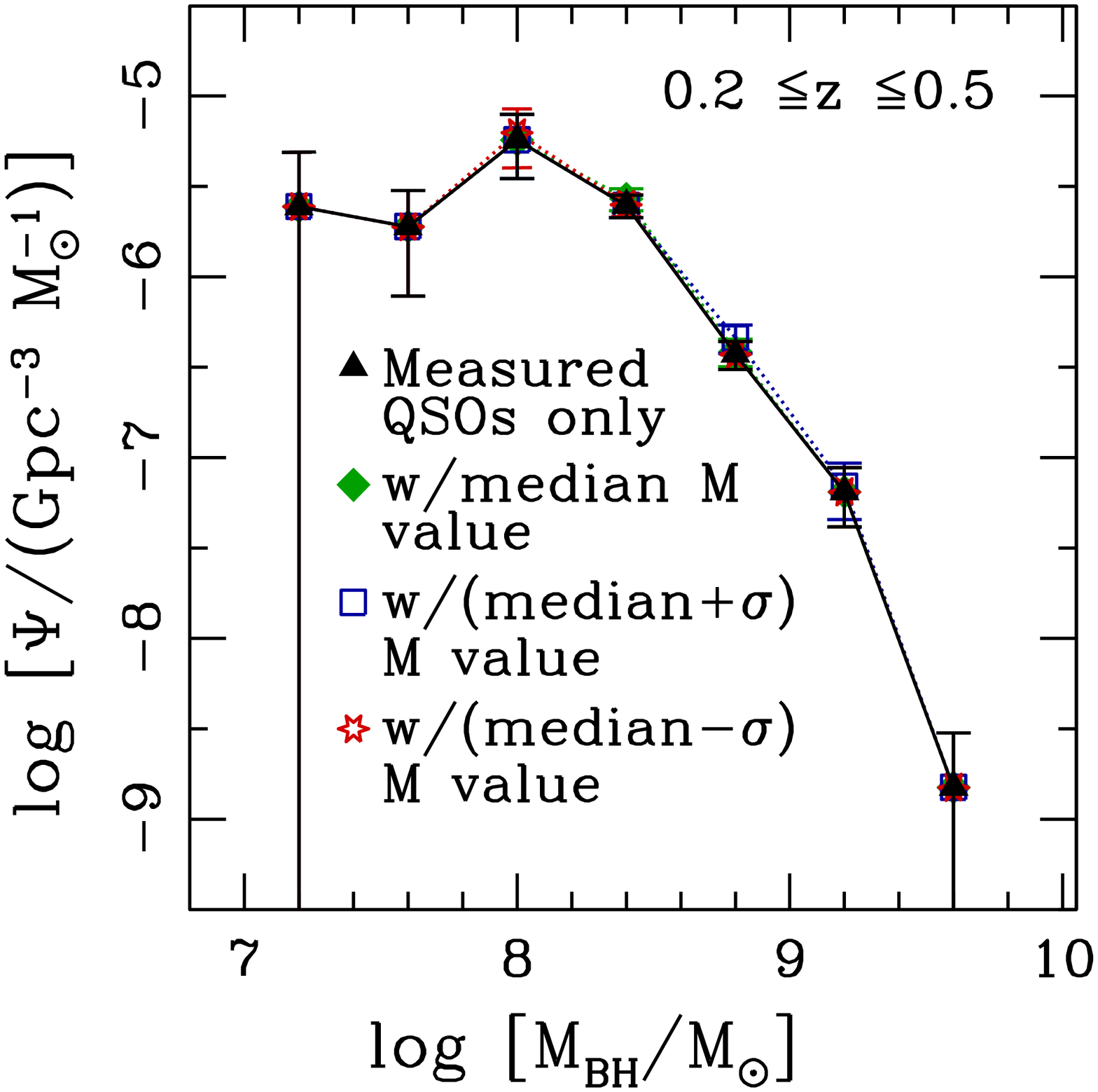}
\vspace{0.2cm}
\plotone{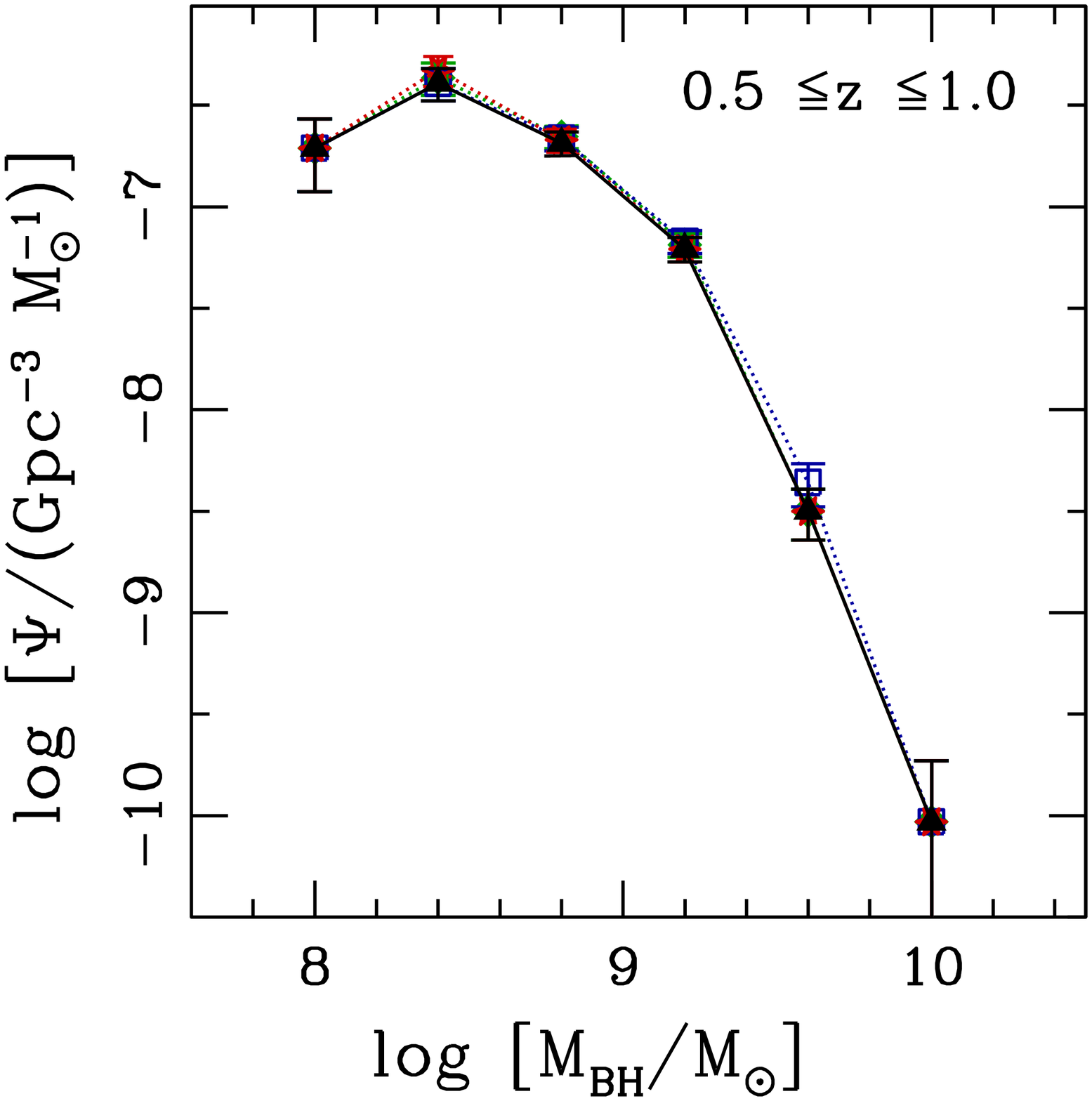}
\vspace{0.2cm}
\plotone{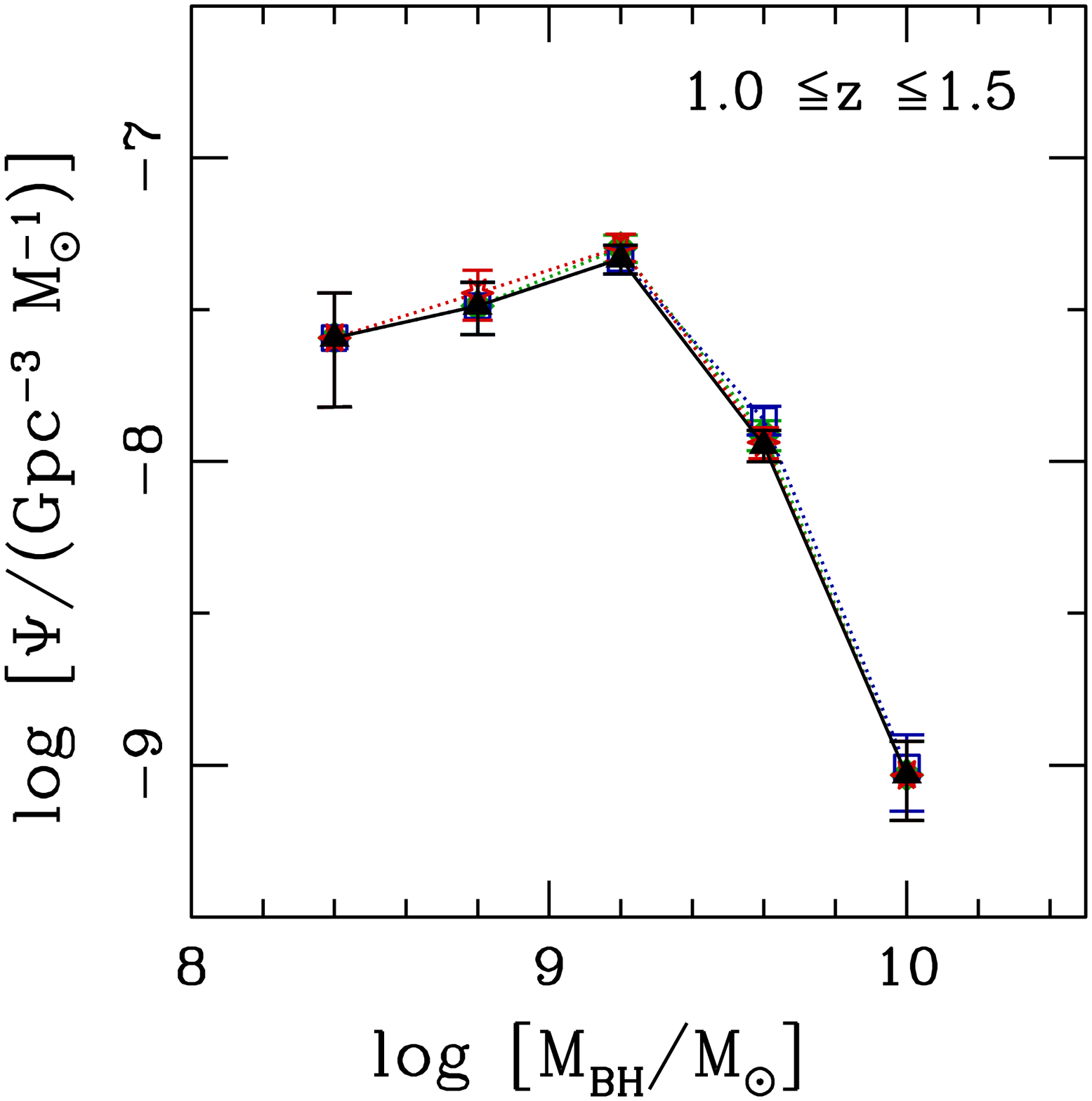}
\caption{  continued in next column }
\end{figure}

\setcounter{figure}{6}
\begin{figure}
\epsscale{.90}
\plotone{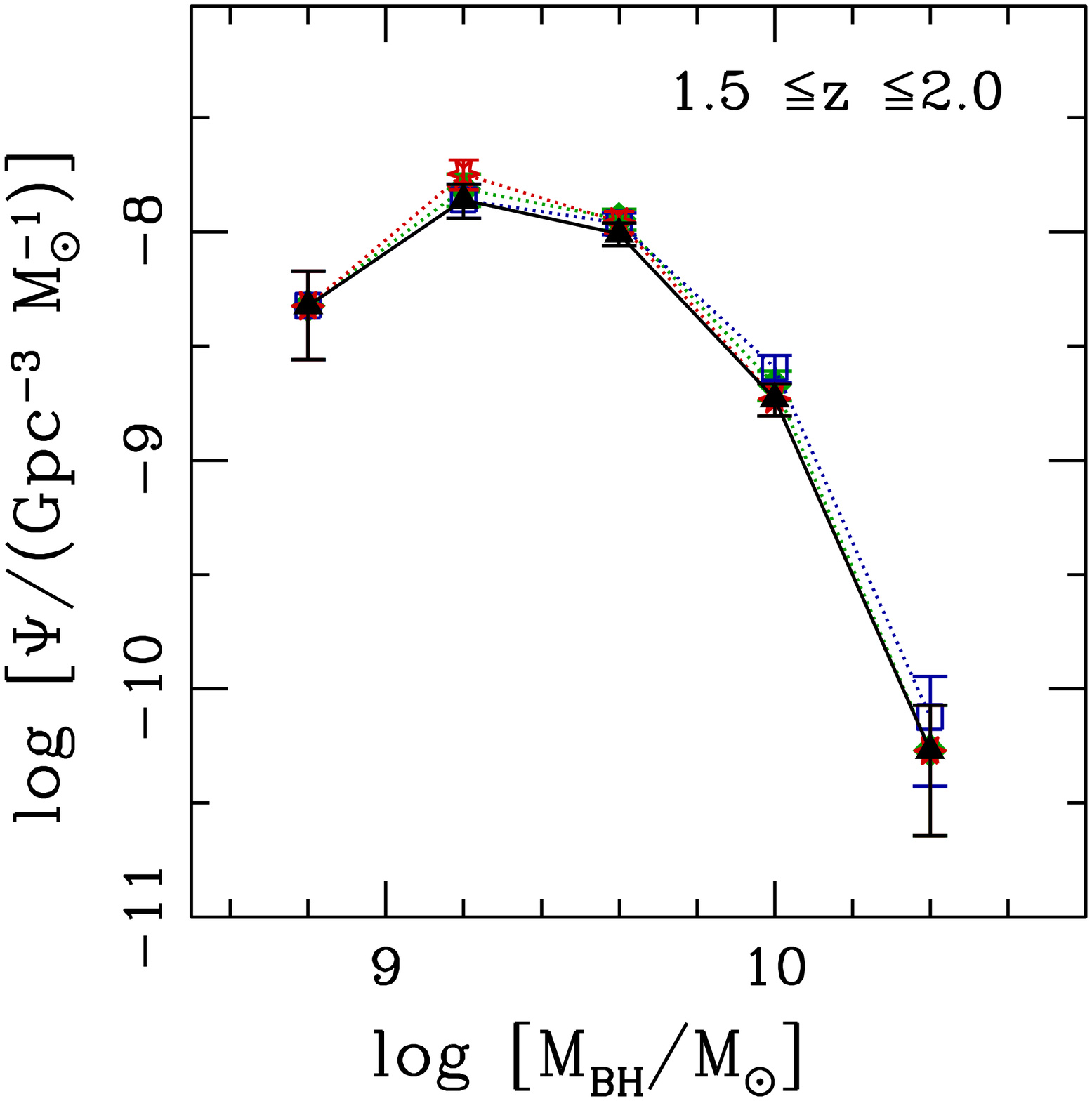}
\vspace{0.2cm}
\plotone{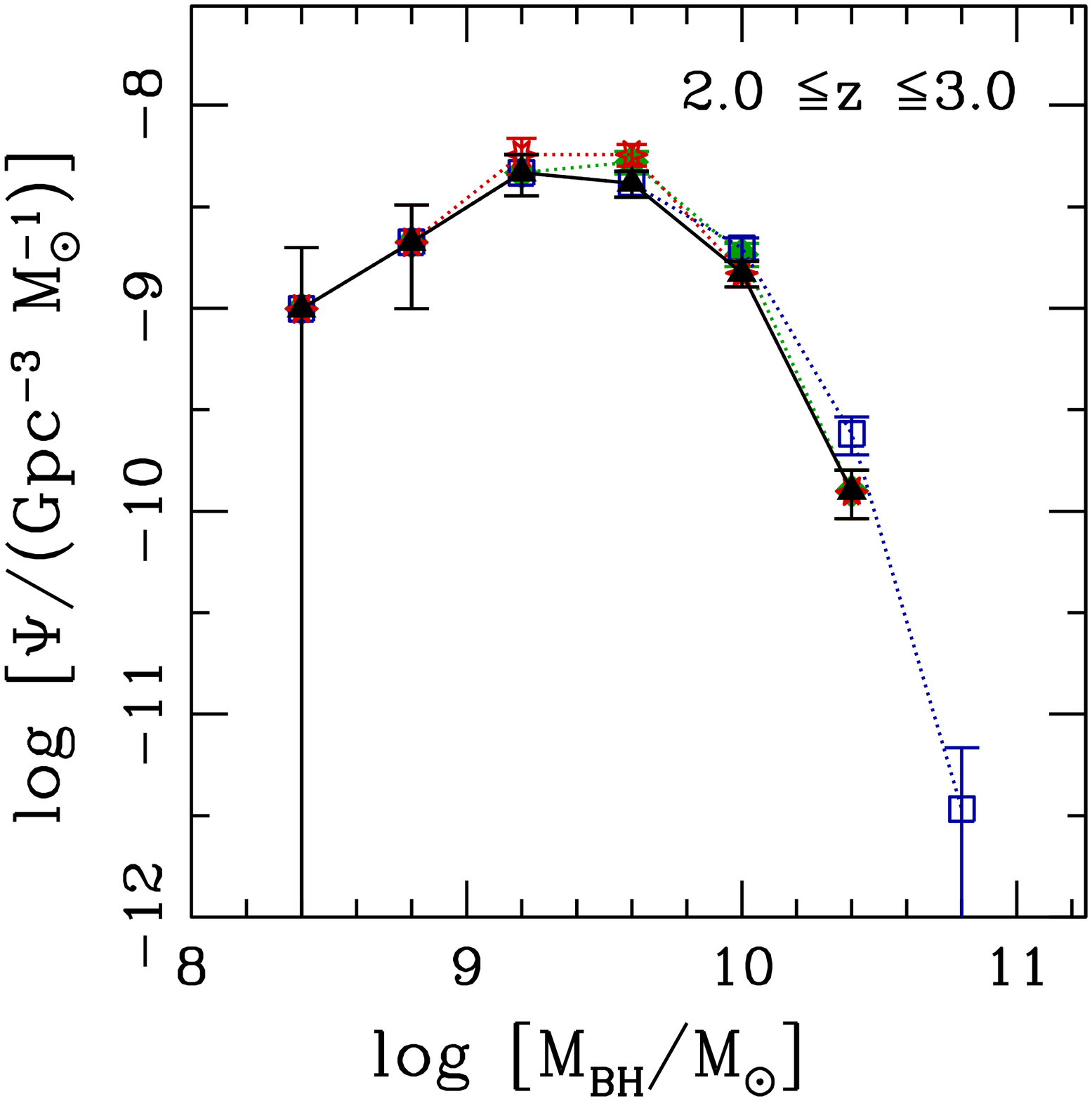}
\caption{Mass functions for LBQS for which conservative
corrections are made for the quasars without black hole mass estimates. For each quasar
without a mass estimate one of three mass values are assumed for each mass function based
on the masses of LBQS quasars at similar redshift and luminosity: (1)
the median black hole mass (of quasars with similar redshift and luminosity), (2) the median
minus one standard deviation mass value, or (3) the median plus one standard deviation mass
value. The mass values within a standard deviation of the median should bracket the most
likely masses of these quasars. The mass functions based on the extreme mass values of the
quasars without other mass estimates thus also brackets the likely amplitudes of the mass
function when all quasars in the sample are included.
\label{LBQS_MF_missQ.fig}}
\end{figure}

\section{Cumulative Mass Densities \label{cumMdens.sec}}
                                                                       
The integrated mass density above a certain mass value in each sample 
is computed by summing the contribution of each individual object with 
central mass above a progressively increasing mass limit, $M_k$:
                                                                       
\begin{equation}
\rho \,(\geq M_k) = 
\int_{M_k}^{\inf}\frac{M_{\rm BH,i}}{V_{a,i}}
\label{cumMdens.eq}
\end{equation}
                                                                       
\begin{figure}
\epsscale{1.00}
\plotone{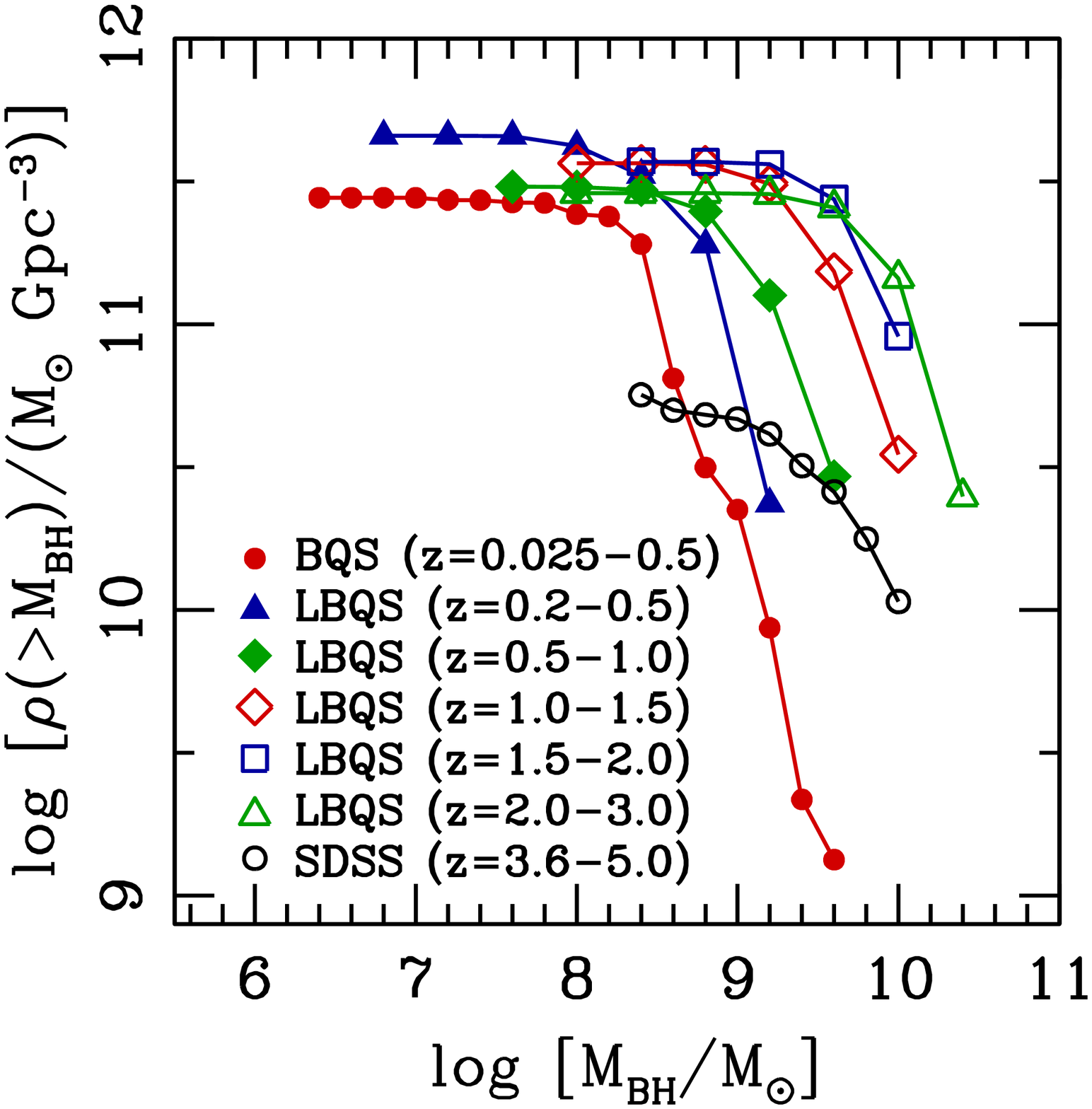}
\vspace{0.2cm}
\plotone{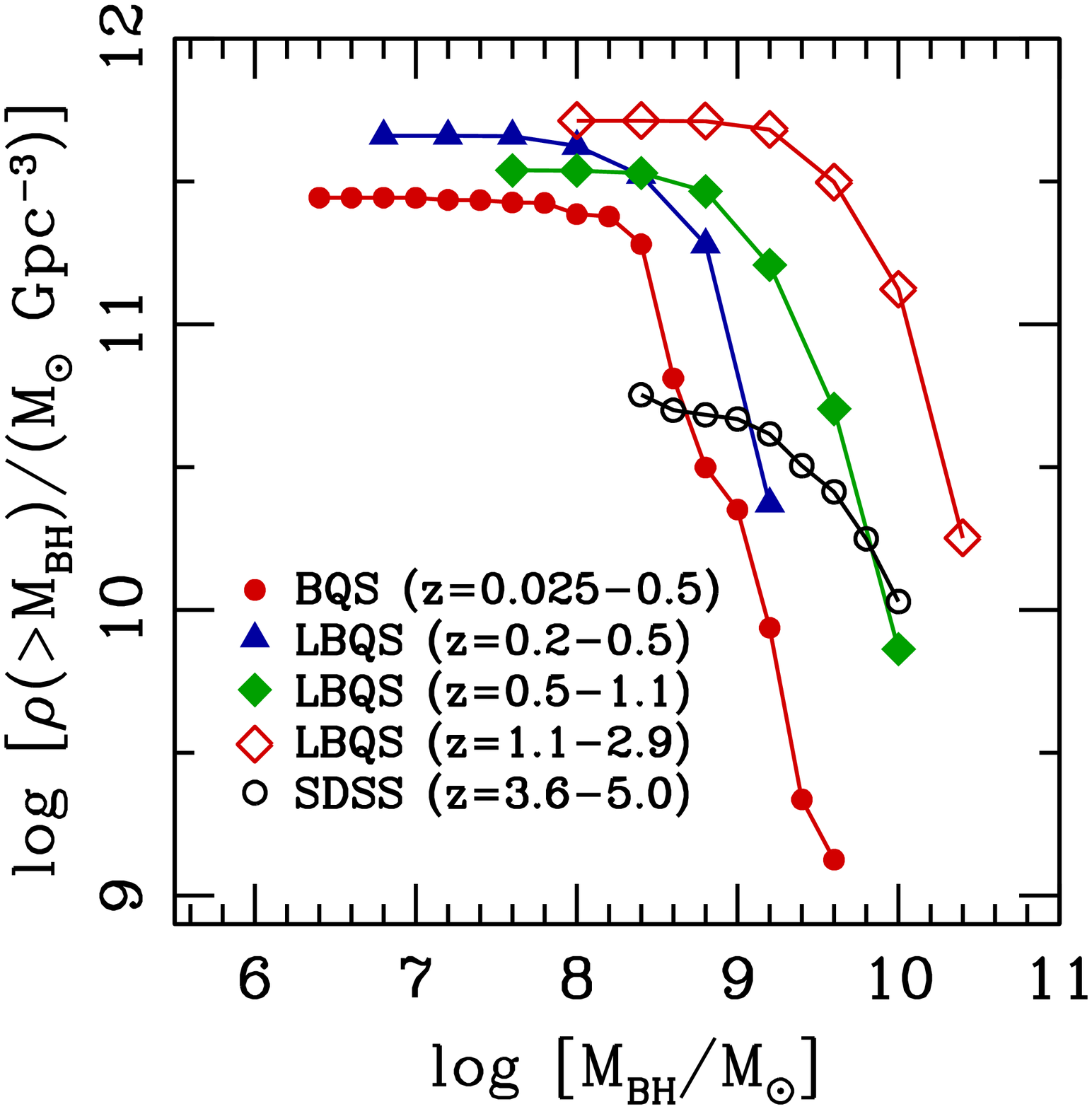}
\caption{Cumulative mass density of active supermassive black holes for different
redshift bins of the LBQS and for the BQS and SDSS color-selected samples as a
function of black hole mass. The top panel shows the BQS and SDSS color-selected
samples with the five redshift bins of the LBQS as are also shown in Figure~\ref{MFallfig}.
The lower panel displays fewer redshift bins for LBQS for a different, less cluttered view.
\label{cumMdens.fig}}
\end{figure}

Figure~\ref{cumMdens.fig} shows the cumulative mass densities
for each of the SDSS, BQS, and LBQS samples. The mass densities are
tabulated in Table~\ref{BQS_SDSS_cumMdens.tab} for the SDSS and BQS
samples and in Tables~\ref{LBQS_MF.tab} and~\ref{LBQS_MFcor.tab} for
the LBQS. 

In Figure~\ref{MFall2fig} we show the distributions of $\Psi \cdot \mbh$ 
for comparison. This representation is closer to the comoving volume
(number) density of active black holes as a function of mass and makes
the space density differences more apparent.

\begin{figure}
\epsscale{1.00}
\plotone{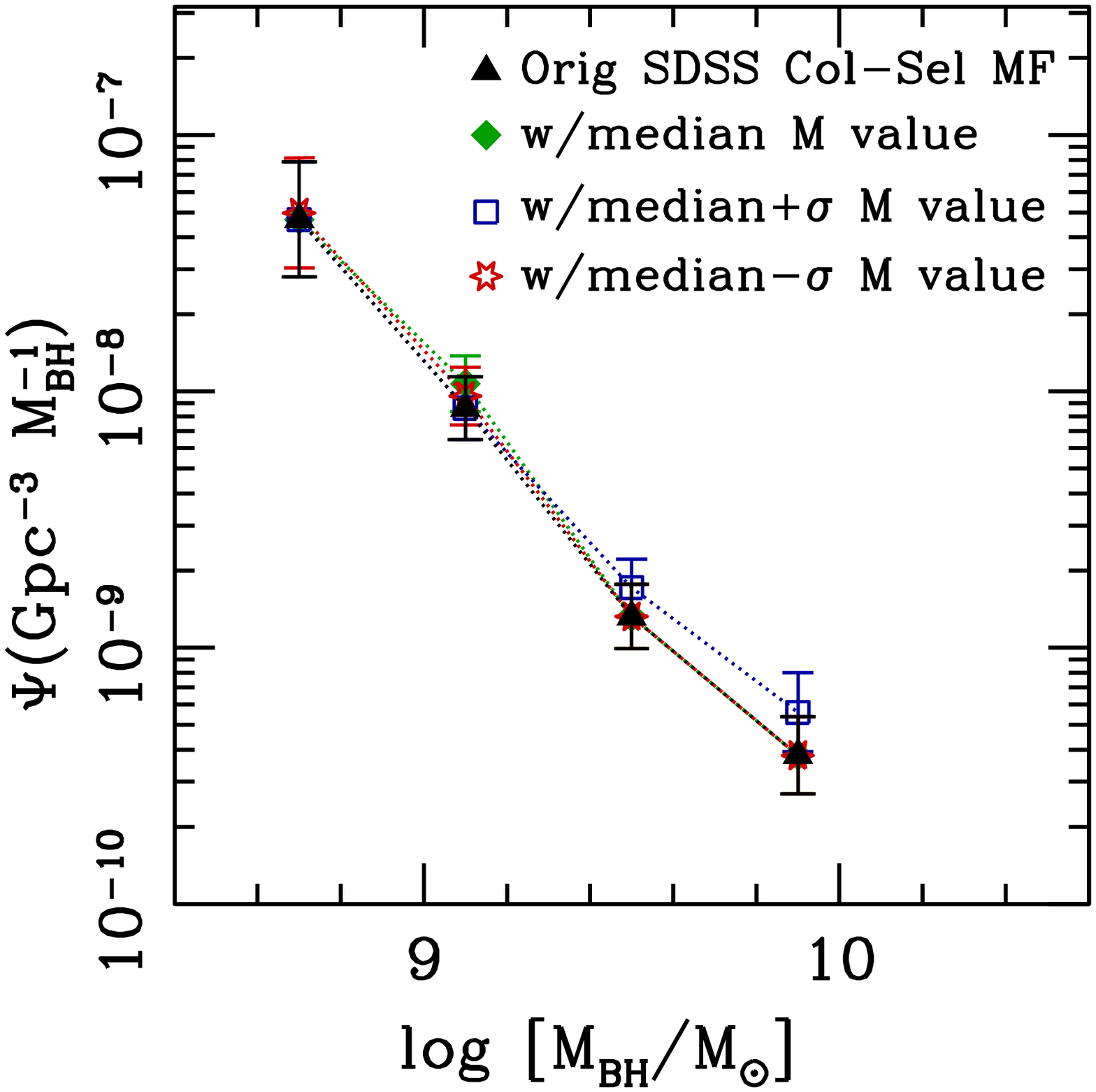}
\caption{Mass function for the SDSS color-selected sample for which conservative
corrections are made for the quasars without black hole mass estimates. For each quasar
without a mass estimate one of three mass values are assumed for each mass function based
on the masses of quasars at similar redshift: (1) the median black hole mass of quasars
with similar redshift, (2) the median minus one standard deviation mass value, or (3) the
median plus one standard deviation mass value.
\label{SDSS_MF_missQ.fig}}
\end{figure}

\section{Discussion \label{discussion.sec}}

The mass functions for the three quasar samples are shown as a function
of black hole mass for different redshift bins in Figure~\ref{MFallfig}. 
Three features are apparent: (1) the
mass functions of the LBQS tend to turn over at the low mass end, (2)
there is a general consistency of the slope of the high mass end between
the mass functions in most of the redshift bins, and (3) the amplitude
of the high mass end increases rapidly between mean redshifts of 4 to
2.5 and then decrease again below a redshift of 1. This is particularly
significant for the cumulative mass density (Fig.~\ref{cumMdens.fig}). 
We briefly comment on each feature in the following.

The turnover at the low mass end tends to occur when the low mass bins are 
incompletely populated [cf.\ with Fig.~\ref{mlolzfig}; see also the 
discussion of the SDSS DR3 mass functions by Vestergaard \et (2008) and
the statistical analysis of Kelly, Vestergaard, \& Fan (2009)]. 
However, the analysis of the BQS mass function by Kelly \et{} indicates 
that part of such turnovers (at least at low redshift) are real, suggesting 
that the number of low mass active black holes does not necessarily stay 
constant or increase with decreasing mass.

\begin{figure}
\epsscale{1.90}
\plottwo{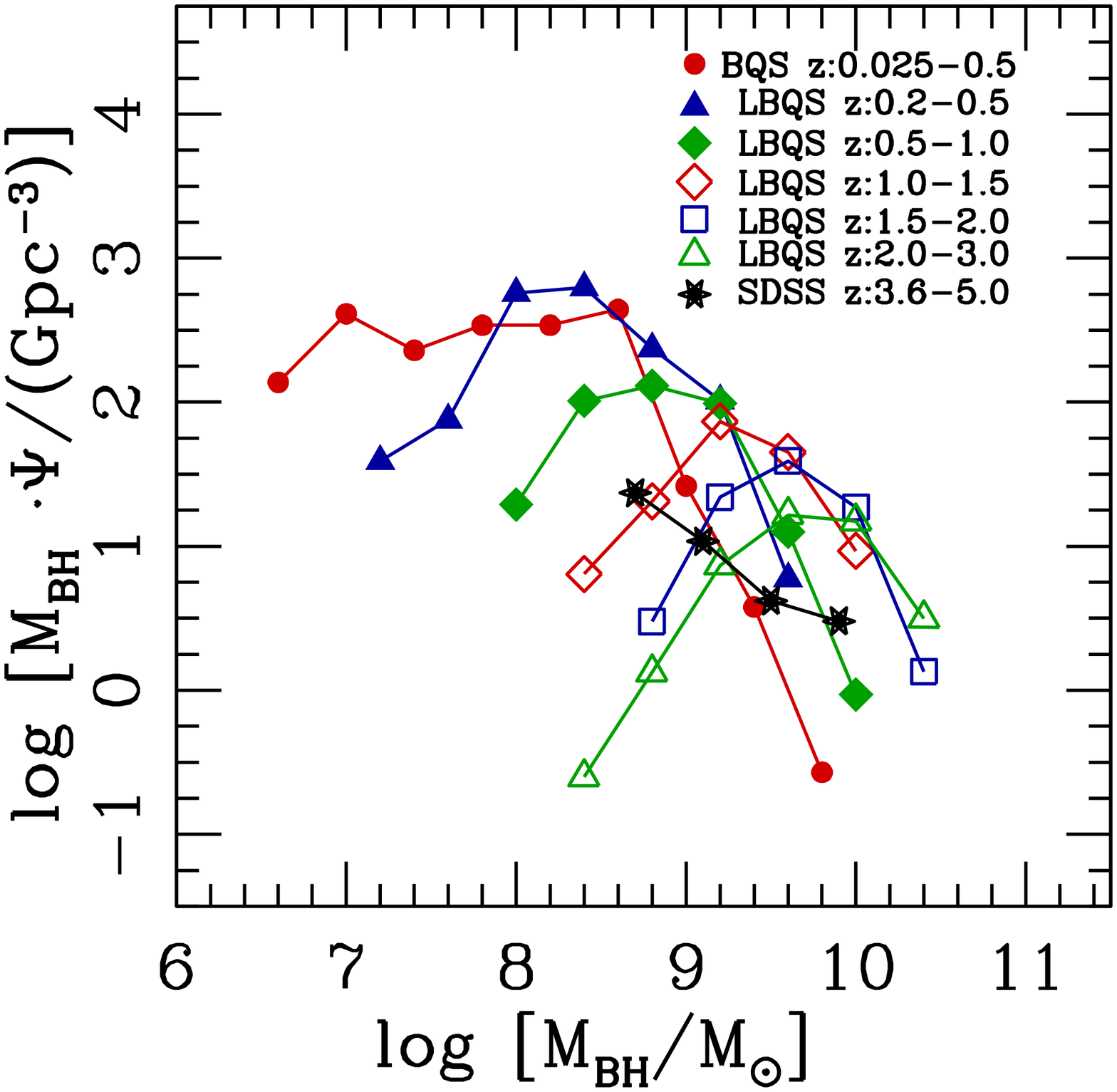}{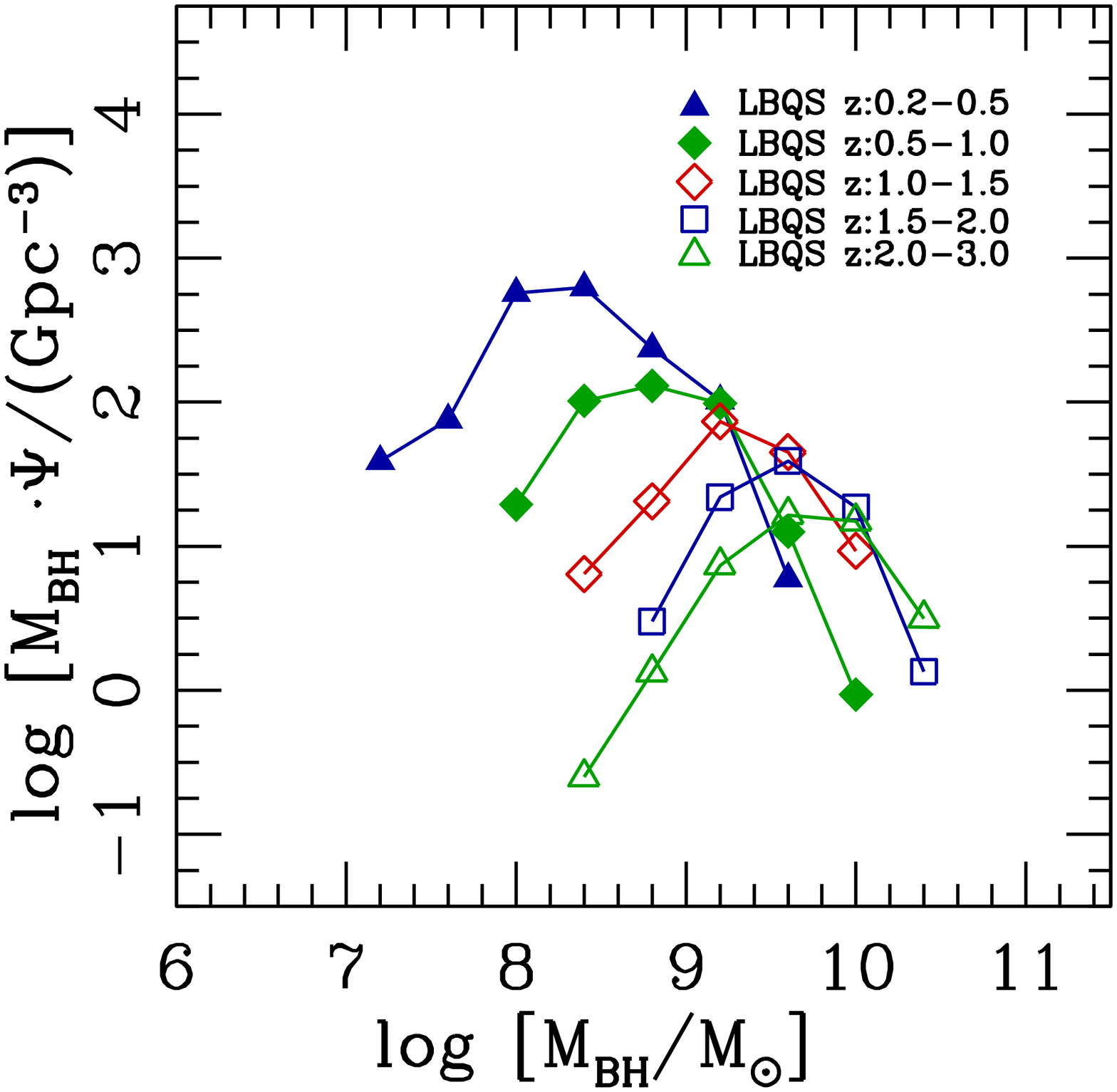}
\vspace{0.2cm}
\epsscale{0.95}
\plotone{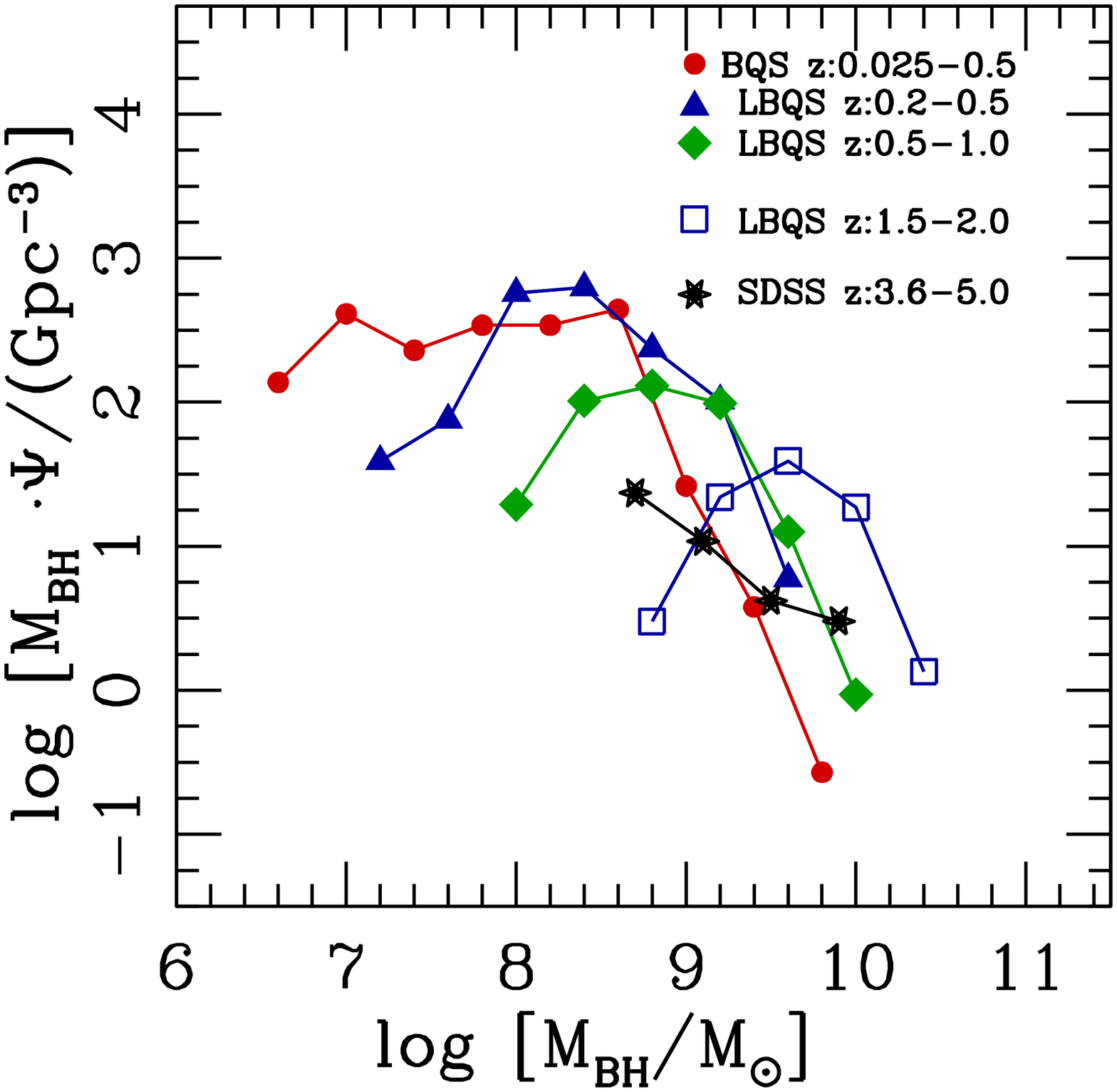}
\caption{Distributions of $\Psi \cdot \mbh$ as a function of black hole mass;
$\Psi$ is the black hole mass function.
{\it Top Panel}$-$ the LBQS (for all five redshift bins), the BQS, and the SDSS
color-selected samples;
{\it Middle Panel} $-$ the LBQS alone (for the five different redshift bins);
{\it Lower Panel} $-$ the BQS and the SDSS color-selected samples are shown with
selected redshift bins of the LBQS for ease of comparison.
\label{MFall2fig}}
\end{figure}

We determine the high mass end slopes, $\beta$, of each mass function in 
Fig.~\ref{MFallfig} ({\it top panel}) following the method of 
Vestergaard \et (2008) where $\Psi \propto M^{\beta}$.  We first 
discuss the mass functions located at redshifts below 3.5 and 
thereafter discuss the case at $z \geq$\,3.5.
For the LBQS mass functions a typical high end slope\footnote{Specifically, 
slopes of $-3.0, -3.4, -2.7, -2.9, -2.6$ were obtained for the redshift 
bins [0.2:0.5], [0.5:1.0], [1.0:1.5], [1.5:2.0], and [2.0:3.0], respectively,
for mass bins at or above 8.8\,dex, 9.2\,dex, 9.6\,dex, 9.6\,dex, and 
10.0\,dex, respectively.  The limits on the mass bins were adopted to ensure 
the mass bins are complete or nearly complete.  Note that given the 
uncertainties of between 0.4 to 1.4 there are no reasons to believe the 
apparent flattening of the mass functions toward higher redshifts is real. }
$\beta$ between about $-3.0$ to $-3.4$ was obtained with uncertainties ranging 
between 0.4 to 1.4.  For the BQS we find a slope of $\beta \approx -3.6\pm 1.0$ 
for mass bins at 8.6\,dex and above. These slopes are all consistent to within
the uncertainties. Also, they agree with the slope of $\beta \approx -3.3$ 
observed for the SDSS DR3 mass function below redshifts of about 3.8 
(Vestergaard \et 2008).
It is interesting to note that the high-end slopes of the SDSS DR3 luminosity
functions also have values of $\beta \approx -3.3$ (Richards \et 2006).

However, for the mass function above redshift 3.5, i.e., of the SDSS 
color-selected sample, we see a somewhat flatter slope of 
$\beta \approx -1.75\pm0.56$, determined for mass bins at 8.6\,dex and above. 
The reality of this flatter slope is also evident from Figure~\ref{MFallfig}.
Vestergaard \et{} also reported an apparent flattening of the DR3 mass function
slopes at $z \gsim 4$ but attributed this to the large errors and small number
statistics in those redshift bins.  Since the color-selected sample is a 
well-defined and complete sample within the survey area and flux limits, 
the result for this sample should be robust, suggesting that there is indeed 
a real trend toward the high-mass black hole distribution to flatten at the 
highest redshifts.

The constancy of the high mass end slope and of the amplitude of the mass 
function with redshift (at $z \lsim 3$) may appear surprising, especially 
considering the large changes occurring for the evolution of the luminosity 
function (\eg Boyle \et 2000; Richards \et 2006). For example, the amplitude
of the SDSS DR3 luminosity function drops more than two orders of magnitude
at an absolute magnitude of $-$27 between redshifts two and 0.5. In
comparison, the DR3 black hole mass function normalization changes only by 
a factor of a few and displays a similar constancy of the high mass end slope 
and amplitude at a wide range of redshifts. 
While this slope constancy may be real, we note the possibility that the 
statistical uncertainties in the black hole mass estimates of a factor of
a few prohibits us from detecting subtle differences or any cosmic evolution
of the high mass end slope.
One issue to keep in mind is that contrary to the quasar luminosity, the 
mass of black holes does not decrease with time. The only way for the
black holes to change the shape of the mass functions with cosmic time
is by growing in mass or by their activity to decrease sufficiently to 
drop out of the quasar survey owing to its lower flux limits.

The mass function and the cumulative mass density distribution both
display a rapid increase in amplitude at the highest masses by a 
factor of $\sim$5 and $\sim$10, respectively, over a period of a 
Giga-year from redshifts 4 (SDSS) to 2.5 (LBQS, 2\,$\leq z \leq$\,3).
This is an interesting feature, especially considering the slower 
amplitude decrease below $z \approx$ 1;  it is even more evident in 
the particular representation of the mass function shown in Figure~\ref{MFall2fig}. 
We are evidently directly seeing the population of massive black holes 
build up at these epochs as the density of actively accreting (massive) 
black holes rises rapidly toward lower redshift.  This is also consistent 
with the observed rise in the quasar space density at these redshifts 
(\eg Osmer 1982; Schmidt \et 1995; Warren \et 1994; Fan \et 2001b) and the 
relatively high Eddington luminosity ratios observed for the SDSS color-selected 
sample (Fig.~\ref{mlolzfig}).

We note that the space density information revealed by the luminosity
functions does not make the mass functions obsolete. The luminosity functions 
tell us the space density of active black holes radiating at a given luminosity
and redshift. The luminosity itself tells us how fast the black hole is accreting 
matter given the efficiency by which it converts matter to radiation but 
does not reveal whether we are observing a highly accreting, low mass black hole 
or a massive black hole accreting at moderate or low rates.  The mass estimates 
confirm the latter scenario. The dashed line in Fig.~\ref{mlolzfig} (top panel)
shows the SDSS flux limits folded with the cut-off in line widths of 1000\,\kms{} 
adopted for the SDSS quasars (Schneider \et 2003). If the former scenario was 
reality we would see the data points accumulate on and just above this dashed line.  
Instead, we see a much wider distribution well above this limit.  This is an 
important fact to keep in mind.
The mass estimates and the mass functions also serve to break the degeneracy of
the luminosity functions due to the loosely constrained radiative efficiency 
(\eg Wyithe \& Radmanabhan 2006), as noted in the introduction.

\begin{figure}
\epsscale{1.05}
\plotone{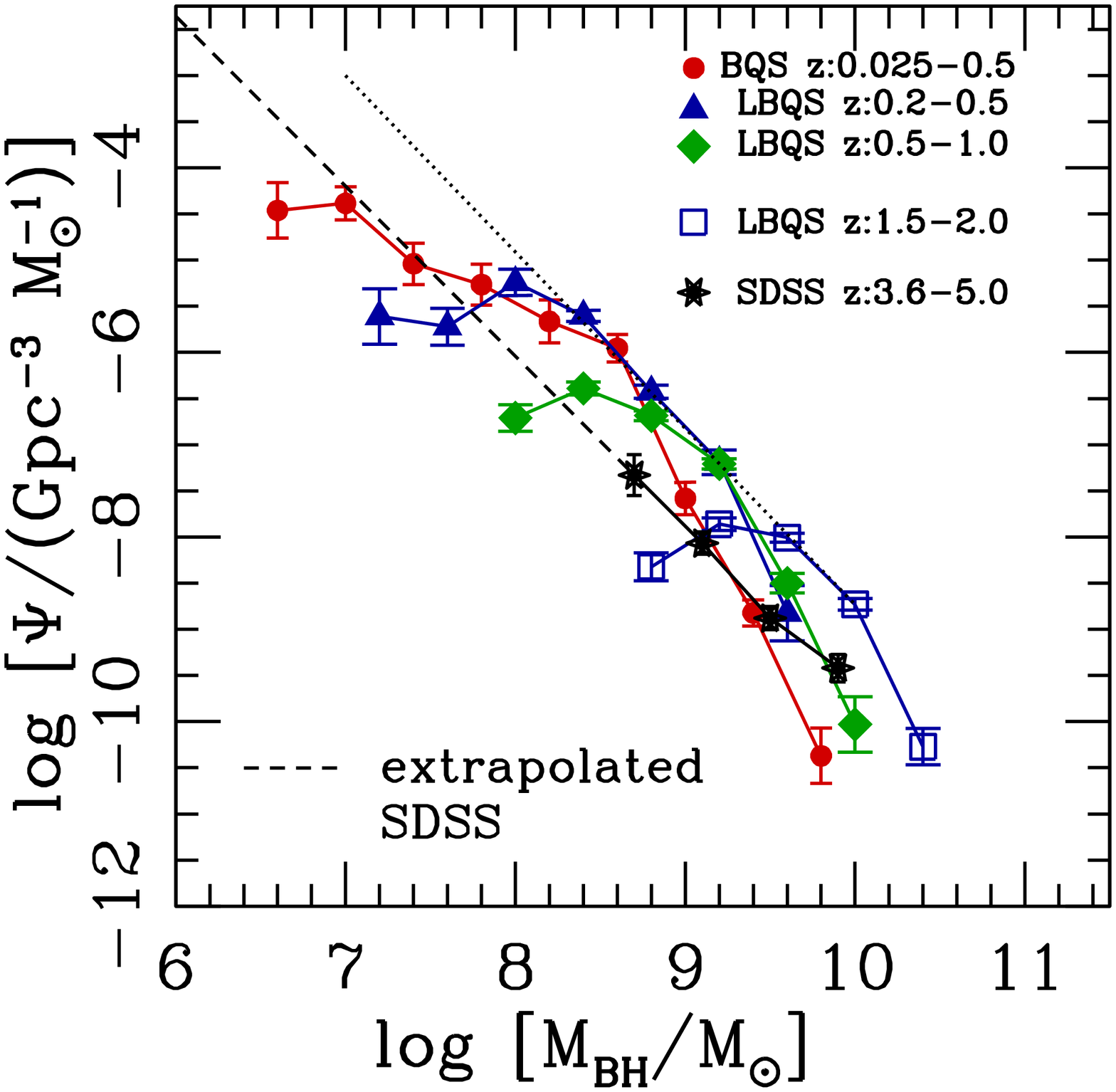}
\vspace{0.15cm}
\plotone{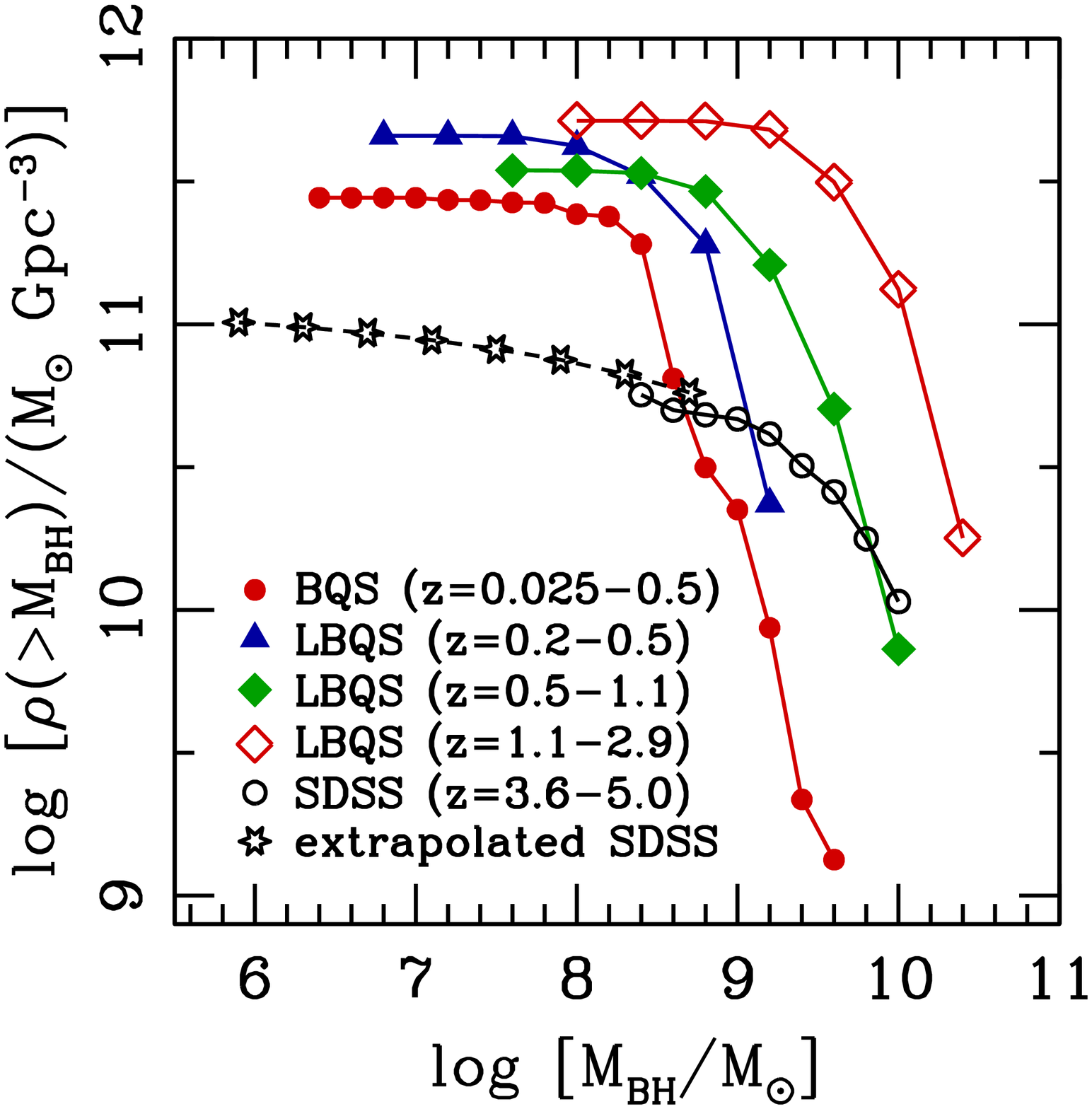}
\caption{Mass functions (top panel) and cumulative mass densities (lower panel) as a function
of black hole mass. These diagrams show the same distributions as Figure~\ref{MFallfig}
(bottom panel) and Figure~\ref{cumMdens.fig} (lower panel) with an extrapolation applied
to the SDSS distributions. In the top panel the SDSS mass function is extrapolated
(dashed line) to very low mass values based on the observed slope of the mass function.
This extrapolated mass function is used to compute the resulting cumulative mass density
for that sample, assuming we had no flux limits. The extrapolated cumulative mass density
is overplotted in the lower panel on the previously introduced mass densities from
Figure~\ref{cumMdens.fig}.
The dotted line in the top panel is a guideline to the maximum growth in the black holes
that the observations allow (see section~6 for discussion).
\label{extrapolfig}}
\end{figure}

The distinctly lower amplitudes of the mass function and the cumulative
mass density distribution for the SDSS sample at redshift four 
(Fig.~\ref{MFallfig}) show that 
there are somewhat fewer active black holes at a given mass and/or less 
massive active black holes at that epoch than at later cosmic times. In 
particular, high-$z$ quasars contribute very little to the mass density of 
active black holes. To examine this a little further and to test if this 
is merely a consequence of our inability to detect very low mass black holes 
at $z \approx$ 4, we extrapolated the SDSS mass function to very low masses 
based on the observed slope at the lower mass end ($\lsim 10^9$\,\Msol).
This extrapolation is shown in the top panel of Figure~\ref{extrapolfig}
(dashed line), allowing a comparison with the mass functions of the BQS 
and LBQS samples. We also show with a dotted line the general slope along 
which the LBQS mass functions at redshifts lower than two move with cosmic 
time. It is quite conceivable that, in the case of no survey flux limits, 
the LBQS mass functions would extend approximately along this line, 
since the SDSS DR3 mass functions (Vestergaard \et 2008) exhibit a similar 
behavior.  The dotted line is hence a guideline to the maximum growth that 
can occur from the earliest epochs according to observations.  
The extrapolated mass function 
at $z \approx$\,4 is clearly parallel to the dotted line, offset by about 
0.6\,$-$\,0.65\,dex in mass and about 1.25\,dex in space density. This 
parallel offset suggests that the SDSS quasars are either rarer than at 
lower epochs by a factor of $\sim$17 or they need to grow in mass by an 
average factor of 4 to 4.5 by redshift 2.5 (ignoring black hole mergers).  
Given the time passed between redshifts of four and 2.5 (of order 1 Gyr) 
and quasar life times of much less than a Giga year (\eg Martini 2004; 
Hopkins \& Hernquist 2008), the massive black holes that are contributing 
to the mass function at $z \approx$\,4 are not the same most massive black 
holes observed at $z \approx$\,2.5. The most massive black holes at 
$z \approx$\,4 would have ceased their activity and dropped out of our 
surveys by $z \approx$\,2.5.
Hence, the shift in the mass functions at the highest masses is not due to 
simple mass growth and a space density increase must dominate.  Since galaxy 
mergers (and hence black hole mergers) are expected to occur more frequently 
in the earlier universe than at present, we expect that 
the less massive black holes (than typically observed at redshift four)
undergo a combination of both minor and major mergers as 
well as growth by mass accretion between the epochs at redshifts four and two. 
This will cause these black holes to shift to higher masses and the mass 
function amplitude at the high mass end to shift upwards. 

In the lower panel of Figure~\ref{extrapolfig} we show the cumulative
mass density for the SDSS extrapolated mass function. Unless the distribution 
of low-mass black holes is significantly steeper than the extrapolated 
power-law discussed here then 
it is clear that the lack of detectable low-mass black holes at $z \approx$\,4 
is not the reason for the large amplitude discrepancies (both panels) between 
the quasar populations observed at very early ($z \approx$ 4) and at later 
($z<$2) epochs. A significant build-up of existing and future supermassive 
black holes is required to explain the quasar populations at later times.

We therefore conclude that the active black hole population at high 
redshift must be different since significantly fewer active black holes 
exist and with lower mass density at $z \approx$ 4 than at 
$z \approx$ 2.5 and $z \approx$ 0.5, respectively. 
An interesting theoretical exercise would be to place lower limits on
the distribution of 'seed' (\ie low mass) black holes at redshifts of
four required to explain the local distribution of (lower mass) black
holes (\mbh $\lsim 10^8$\Msol) for later observational tests of this
prediction, for example, similar to the work by Volonteri \et (2008). 

When comparing all the samples at redshifts below three the high-mass 
end of the mass functions show little mass growth in these black holes. 
In fact, the density of the highest mass black holes decreases with cosmic 
time at epochs later than $z \approx$\,1, consistent with a general decrease 
in activity of the most massive black holes. This is also and more clearly
seen in the representation of the mass function shown 
for different mass bins as a function of redshift, as we discuss next.
Although the LBQS only has about 1000 quasars 
distributed over a range of redshifts, it is still large enough to
allow a determination of how the mass function depends on redshift
for a given mass bin (Fig.~\ref{MFzfig}). 
(The value of the LBQS in this regard will become particularly apparent 
when discussed in relation to the mass function of the SDSS DR3 quasar 
sample which is statistically limited near a redshift of 3.)
This representation of the mass function directly shows at which epochs 
active black holes at a given mass are the most abundant. For example, 
the black holes of lower masses of 7.8 $-$ 8.6 dex peak in their 
comoving space density in the local universe, while the more massive
black hole of 10$^{10}$ \Msol{} are the most active at redshifts of
2 $-$ 3. Figure~\ref{MFzfig}f shows that the expected density of such 
black holes is more than two orders of magnitude lower in the local 
universe.  By comparing the mass functions for individual mass bins 
the cosmic downsizing (\eg Ueda \et 2003) of active black holes is
clearly evident: the most massive black holes are the most active at
high redshifts while the lowest mass black holes are the most active
at low redshift.

\section{Summary \label{Summary.sec}}
We present the mass functions of actively accreting black holes for
the Bright Quasar Survey, the Large Bright Quasar Survey, and the
SDSS color-selected sample in the Fall Equatorial Stripe and the
database from which these mass functions are derived. 
We find similar amplitudes and slopes (of about $\beta \approx -3.3$ 
with uncertainties between 0.4 and 1.4)
of the high mass end of the mass functions for different redshift bins 
and across quasar samples for redshifts below $\sim$3.5. This is similar 
to what was seen for the DR3 mass functions at a similar redshift range
(Vestergaard \et 2008). However, for the well-defined and complete SDSS 
color-selected sample of quasars at redshifts between 3.6 and 5 we find 
a somewhat flatter slope of $\beta = -1.75\pm0.56$.  Comparison of the mass 
functions and the cumulative mass density distributions for the different 
samples shows that the active black hole population at redshifts of about 
4 must be different than the black hole populations below a redshift of 
about 2.5.  In fact, we may be witnessing a fast build-up of the black 
hole population between redshifts of 4 and redshifts of $\sim$2.

The mass functions presented here will be discussed in further detail
in relation to other existing black hole mass functions 
and, in particular, the SDSS DR3 mass functions (Vestergaard \et 2008) 
in a future paper (M. Vestergaard \et 2009, in preparation). At that 
time we will also apply sophisticated statistical analysis (Kelly \et 
2008a, 2009) to investigate the nature of the actual underlying mass 
distribution free of measurement uncertainties and, to a certain extent, 
of survey flux limits.

\acknowledgments
We thank Dave Sanders for providing his tabulation of observed bolometric
luminosities of the BQS sample, Xiaohui Fan for comments on the manuscript, 
and Brandon Kelly for discussions.
MV thanks for their hospitality the astronomy departments at the Ohio State 
University and the University of Arizona where most of this work was 
performed.
We gratefully acknowledge financial support through HST grants 
HST-AR-10691, HST-GO-10417, and HST-GO-10833 from NASA through the 
Space Telescope Science Institute, which is operated by the Association
of Universities for Research in Astronomy, Inc., under NASA contract 
NAS5-26555.

{\it facilities:} \facility{HST (FOS,STIS)}, 
\facility{KPNO: 2.1m (Gold)}, \facility{SDSS}, 
\facility{MMT (blue channel spectrograph)}, \facility{Keck: I (LRIS)}, 
\facility{UKST (direct, objective prism)}.


\clearpage



\begin{deluxetable}{lccccllll}
\tablecaption{Line Widths and Nuclear Luminosities of the LBQS quasars.\label{LBQS_FWL.tab}}
\tablewidth{0pt}
\tablehead{
\colhead{} & \colhead{} & 
\colhead{FWHM(\hb)} & 
\colhead{FWHM(\mgii)} & 
\colhead{FWHM(\civ)} & 
\colhead{log[$L_{\lambda 1350}$} &  
\colhead{log[$L_{\lambda 2100}$} &  
\colhead{log[$L_{\lambda 3000}$} &  
\colhead{log[$L_{\lambda 5100}$} \\
\colhead{Name} & \colhead{$z$} & 
\colhead{(km s$^{-1}$)} & \colhead{(km s$^{-1}$)} & 
\colhead{(km s$^{-1}$)} &
\colhead{/erg s$^{-1}$]} &
\colhead{/erg s$^{-1}$]} &
\colhead{/erg s$^{-1}$]} &
\colhead{/erg s$^{-1}$]} \\
\colhead{(1)} &
\colhead{(2)} &
\colhead{(3)} &
\colhead{(4)} &
\colhead{(5)} &
\colhead{(6)} &
\colhead{(7)} &
\colhead{(8)} &
\colhead{(9)} 
}
\startdata
Q0000$+$0159 & 1.073 &  \nodata &  4000$^{+500}_{-450}$ & \nodata & 42.81$^{+0.18}_{-0.31}$ & 42.52$^{+0.25}_{-0.62}$ & 42.29$^{+0.32}_{-42.29}$ & 41.94$^{+0.45}_{-41.94}$ \\ 
Q0001$-$0050 & 1.459 & \nodata & 4000$^{+400}_{-400}$ & 5000$^{+350}_{-325}$ & 42.98$^{+0.20}_{-0.39}$ & 42.69$^{+0.28}_{-0.98}$ & 42.46$^{+0.36}_{-42.46}$ & 42.11$^{+0.50}_{-42.11}$ \\
Q0002$-$0243 &  0.432 & 3200$^{+325}_{-300}$ & 3111$^{+92}_{-124}$ & \nodata & 41.99$^{+0.14}_{-0.20}$ & 41.70$^{+0.19}_{-0.34}$ & 41.47$^{+0.24}_{-0.61}$ & 41.12$^{+0.35}_{-{41.12}}$ \\
\enddata
\tablecomments{FWHM measurements are based on Forster \et (2001) and the luminosities are computed 
from the $B_J$ survey magnitudes (Hewett \et 2001, and references therein). For cases where the 
relative luminosity error exceed 1.0 the negative error of the logarithm of the luminosity cannot 
be computed and this error is instead assigned the value of the luminosity itself.
This Table is listed in its entirety in the online journal. Only sample entries are listed here.}
\end{deluxetable}



\begin{deluxetable}{lccccc}
\tablecaption{LBQS Black Hole Mass Estimates and Luminosities\label{LBQS_ML.tab}}
\tablewidth{0pt}
\tablehead{
\colhead{Name} & \colhead{$z$} &  \colhead{$B_J$} & 
\colhead{log[\mbh/\Msol]} & 
\colhead{log[\lbol/erg s$^{-1}$]} & \colhead{log \lol} \\
\colhead{(1)} &
\colhead{(2)} &
\colhead{(3)} &
\colhead{(4)} &
\colhead{(5)} &
\colhead{(6)} 
}
\startdata
Q0000$+$0159 & 1.073 & 18.43 & 8.92$^{+0.16}_{-0.26}$ & 46.57$^{+0.26}_{-0.71}$&  $-$0.441$^{+0.284}_{-0.120}$ \\
Q0001$-$0050 & 1.459 & 18.74 & 9.13$^{+0.18}_{-0.18}$ & 46.74$^{+0.29}_{-1.17}$&  $-$0.483$^{+0.315}_{-0.127}$ \\
Q0002$-$0243 & 0.432 & 18.27 & 8.32$^{+0.23}_{-0.23}$ & 45.71$^{+0.25}_{-0.68}$&  $-$0.713$^{+0.311}_{-0.953}$
\enddata
\tablecomments{This Table is listed in its entirety in the online journal. Only sample entries are listed here.}
\end{deluxetable}



\begin{deluxetable}{lccc}
\tablecaption{Basic Properties of LBQS Sources without Black Hole Mass Estimates\label{LBQS_Lz_missingQ.tab}}
\tablewidth{0pt}
\tablehead{
\colhead{Name} & \colhead{$z$} & \colhead{$B_J$} & 
\colhead{log[\lbol/erg s$^{-1}$]} \\
\colhead{(1)} &
\colhead{(2)} &
\colhead{(3)} &
\colhead{(4)} 
}
\startdata
Q0004$+$0147 & 1.710 &  18.13 & 47.14 \\
Q0010$-$0012 & 2.154 &  18.46 & 47.22 \\
Q0013$-$0029 & 2.083 &  18.18 & 47.30 \\
Q0018$+$0047 & 1.835 &  17.82 & 47.32 \\
Q0018$-$0220 & 2.596 &  17.44 & 47.79 \\
\enddata
\tablecomments{This Table is listed in its entirety in the online journal. Only sample entries are listed here.}
\end{deluxetable}



\begin{deluxetable}{lcccr}
\tablecaption{Line Widths and Nuclear Luminosities of the SDSS Color-selected quasars.\label{SDSS_FWL.tab}}
\tablewidth{0pt}
\tablehead{
\colhead{} & \colhead{} & \colhead{} & 
\colhead{FWHM(\civ)} & 
\colhead{log[$L_{\lambda 1350}$} \\  
\colhead{Name} & \colhead{$z$} & \colhead{$p$\tablenotemark{a}} & 
\colhead{(km s$^{-1}$)} &
\colhead{/erg s$^{-1}$]} \\
\colhead{(1)} &
\colhead{(2)} &
\colhead{(3)} &
\colhead{(4)} &
\colhead{(5)} 
}
\startdata
J001950.06$-$004040.9  &  4.32 & 0.90  &  4706$^{+1396}_{-539}$ &  43.436$^{+0.020}_{-0.020}$ \\
J003525.29$+$004002.8  &  4.75 & 0.99  &  2150$^{+150}_{-100}$ &  43.366$^{+0.031}_{-0.033}$ \\
J005922.65$+$000301.4  &  4.16 & 0.78  &  3325$^{+425}_{-175}$ &  43.536$^{+0.012}_{-0.012}$ \\
\enddata
\tablenotetext{a}{Selection probability (Fan \et 2001)}
\tablecomments{FWHM and luminosity measurements are adopted from Vestergaard (2004a). This Table is listed in its entirety in the online journal. Only sample entries are listed here.}
\end{deluxetable}



\begin{deluxetable}{lcccr}
\tablecaption{SDSS Color$-$selected Sample Black Hole Mass Estimates and Luminosities\label{SDSS_ML.tab}}
\tablewidth{0pt}
\tablehead{
\colhead{Name} & \colhead{$z$} &  
\colhead{log[\mbh/\Msol]} & 
\colhead{log[\lbol/erg s$^{-1}$]} & \colhead{log \lol} \\
\colhead{(1)} &
\colhead{(2)} &
\colhead{(3)} &
\colhead{(4)} &
\colhead{(5)} 
}
\startdata
J001950.06$-$004040.9  & 4.32 & 9.345$^{+0.360}_{-0.771}$ &  47.230$^{+0.048}_{-0.054}$ &  $-$0.2146$^{+0.3602}_{-0.7742}$ \\
J003525.29$+$004002.8  & 4.75 & 8.627$^{+0.333}_{-0.823}$ &  47.160$^{+0.053}_{-0.060}$ &   0.4330$^{+0.3339}_{-0.8320}$ \\
J005922.65$+$000301.4  & 4.16 & 9.096$^{+0.338}_{-0.817}$ &  47.330$^{+0.046}_{-0.051}$ &   0.1342$^{+0.3376}_{-0.8187}$ \\
\enddata
\tablecomments{This Table is listed in its entirety in the online journal. Only sample entries are listed here.}
\end{deluxetable}



\begin{deluxetable}{cccccr}
\tablecaption{BQS Black Hole Mass Function \label{BQS_MF.tab}}
\tablewidth{0pt}
\tablehead{
\colhead{$<z>$} & \colhead{$\Delta z$\tablenotemark{a}} & \colhead{$<\mbh>$\tablenotemark{b}} & 
\colhead{$\Psi (M,z)$} & \colhead{$\sigma(\Psi)$~~~~} & \colhead{N} \\ 
\colhead{} & \colhead{} & \colhead{(\Msol)} & 
\colhead{(Gpc$^{-3}$\Msol$^{-1}$)} & \colhead{(Gpc$^{-3}$\Msol$^{-1}$)} & 
\colhead{} \\ 
\colhead{(1)} &
\colhead{(2)} &
\colhead{(3)} &
\colhead{(4)} &
\colhead{(5)} &
\colhead{(6)} 
}
\startdata
0.25& 0.25 & 6.6 & 3.448E-05 & 3.448E-05 & 1 \\
& &          7.0 & 4.112E-05 & 2.108E-05 & 6\\
& &          7.4 & 9.114E-06 & 6.162E-06 & 7\\
& &          7.8 & 5.419E-06 & 3.634E-06 & 11\\
& &          8.2 & 2.165E-06 & 1.529E-06 & 17\\
& &          8.6 & 1.107E-06 & 4.501E-07 & 22\\
& &          9.0 & 2.619E-08 & 1.297E-08 & 13\\
& &          9.4 & 1.506E-09 & 5.765E-10 & 9\\
& &          9.8 & 4.235E-11 & 4.235E-11 & 1\\

\enddata
\tablenotetext{a}{The range of the redshift bin is $<z> \pm \Delta z$.}
\tablenotetext{b}{The central mass in the bin. The mass bin size is 0.4 dex
and expands $\pm$0.2 dex relative to the central mass value.}
\end{deluxetable}



\begin{deluxetable}{cccccrc}
\tablecaption{LBQS Black Hole Mass Function \label{LBQS_MF.tab}}
\tablewidth{0pt}
\tablehead{
\colhead{$<z>$} & \colhead{$\Delta z$\tablenotemark{a}} & \colhead{$<\mbh>$\tablenotemark{b}} & 
\colhead{$\Psi (M,z)$} & \colhead{$\sigma(\Psi)$~~~~} & \colhead{N} & 
\colhead{$\rho(> \mbh)$\tablenotemark{c}} \\
\colhead{} & \colhead{} & \colhead{(\Msol)} & 
\colhead{(Gpc$^{-3}$\Msol$^{-1}$)} & \colhead{(Gpc$^{-3}$\Msol$^{-1}$)} & 
\colhead{} & \colhead{(\Msol Mpc$^{-3}$)} \\
\colhead{(1)} &
\colhead{(2)} &
\colhead{(3)} &
\colhead{(4)} &
\colhead{(5)} &
\colhead{(6)} &
\colhead{(7)} 
}
\startdata
  0.35  & 0.15 & 6.8 &    \nodata   &  \nodata    &    0  & 461.03 \\
  0.35  & 0.15 & 7.2 &   2.4518E-06 & 2.4518E-06  &    1  & 460.50 \\
  0.35  & 0.15 & 7.6 &   1.8975E-06 & 1.1120E-06  &    5  & 459.29 \\
  0.35  & 0.15 & 8.0 &   5.7041E-06 & 2.2265E-06  &   26  & 423.53 \\
  0.35  & 0.15 & 8.4 &   2.4266E-06 & 3.5559E-07  &   75  & 335.24 \\
  0.35  & 0.15 & 8.8 &   3.7082E-07 & 6.5793E-08  &   39  & 196.06 \\
  0.35  & 0.15 & 9.2 &   6.5978E-08 & 2.3601E-08  &   15  & 37.805 \\
  0.35  & 0.15 & 9.6 &   1.5007E-09 & 1.5007E-09  &    1  & \nodata\\[3pt]
\hline \\[-8pt]
  0.75  & 0.25 & 7.6 &    \nodata   &  \nodata    &    0  & 322.67 \\
  0.75  & 0.25 & 8.0 &   1.9470E-07 & 7.6028E-08  &    9  & 322.07 \\
  0.75  & 0.25 & 8.4 &   4.0175E-07 & 7.4011E-08  &   54  & 315.12 \\
  0.75  & 0.25 & 8.8 &   2.0598E-07 & 2.7534E-08  &   90  & 268.67 \\
  0.75  & 0.25 & 9.2 &   6.0881E-08 & 8.4790E-09  &   77  & 145.79 \\
  0.75  & 0.25 & 9.6 &   3.1654E-09 & 8.8983E-10  &   13  & 51.704 \\
  0.75  & 0.25 &10.0 &   3.6235E-10 & 1.8121E-10  &    4  & 8.9008 \\[3pt]
\tableline \\[-8pt]
  1.25  & 0.25 & 7.6 &    \nodata   &   \nodata   &    0  & 394.74 \\
  1.25  & 0.25 & 8.0 &   5.3376E-09 & 5.3376E-09  &    1  & 394.70 \\
  1.25  & 0.25 & 8.4 &   2.5527E-08 & 1.0412E-08  &    7  & 394.38 \\
  1.25  & 0.25 & 8.8 &   3.0814E-08 & 6.2483E-09  &   27  & 391.50 \\
  1.25  & 0.25 & 9.2 &   4.5757E-08 & 5.0341E-09  &  105  & 338.64 \\
  1.25  & 0.25 & 9.6 &   1.0996E-08 & 1.3039E-09  &   76  & 182.61 \\
  1.25  & 0.25 &10.0 &   1.2521E-09 & 3.0845E-10  &   20  & 63.610 \\
  1.25  & 0.25 &10.4 &   2.1249E-11 & 2.1249E-11  &    1  & \nodata\\[3pt]
\hline \\[-8pt]
  1.75  & 0.25 & 8.4 &     \nodata  &   \nodata   &    0  & 374.11 \\
  1.75  & 0.25 & 8.8 &   4.7534E-09 & 1.9897E-09  &    6  & 373.94 \\
  1.75  & 0.25 & 9.2 &   1.3410E-08 & 2.3147E-09  &   38  & 367.59 \\
  1.75  & 0.25 & 9.6 &   9.5608E-09 & 1.1030E-09  &   79  & 282.15 \\
  1.75  & 0.25 &10.0 &   1.9367E-09 & 3.0588E-10  &   41  & 98.379 \\
  1.75  & 0.25 &10.4 &   5.3655E-11 & 3.0978E-11  &    3  & \nodata\\[3pt]
\hline \\[-8pt]
  2.50  & 0.50 & 8.0 &     \nodata  &  \nodata    &    0  & 291.62 \\
  2.50  & 0.50 & 8.4 &   9.9367E-10 & 9.9367E-10  &    1  & 291.62 \\
  2.50  & 0.50 & 8.8 &   2.1144E-09 & 1.1199E-09  &    4  & 291.21 \\
  2.50  & 0.50 & 9.2 &   4.5198E-09 & 1.0667E-09  &   24  & 289.26 \\
  2.50  & 0.50 & 9.6 &   4.1379E-09 & 6.0390E-10  &   55  & 259.84 \\
  2.50  & 0.50 &10.0 &   1.4573E-09 & 2.1311E-10  &   52  & 150.49 \\
  2.50  & 0.50 &10.4 &   1.3643E-10 & 3.5502E-11  &   15  & 24.924 
\enddata
\tablenotetext{a}{The range of the redshift bin is $<z> \pm \Delta z$.}
\tablenotetext{b}{The central mass in the bin. The mass bin size is 0.4 dex
and expands $\pm$0.2 dex relative to the central mass value.}
\tablenotetext{c}{The lower mass limit is the central value of the mass bin
listed in column 3.}
\end{deluxetable}



\begin{deluxetable}{cccccr}
\tablecaption{The LBQS Redshift Dependent Black Hole Mass Function \label{LBQS_MF_zdep.tab}}
\tablewidth{0pt}
\tablehead{
\colhead{$<\mbh>$\tablenotemark{a}} & \colhead{$<z>$} & \colhead{$\Delta z$\tablenotemark{b}} & 
\colhead{$\Psi (M,z)$} & \colhead{$\sigma(\Psi)$~~~~} & \colhead{N} \\ 
\colhead{} & \colhead{(\Msol)} & &
\colhead{(Gpc$^{-3}$\Msol$^{-1}$)} & \colhead{(Gpc$^{-3}$\Msol$^{-1}$)} \\ 
\colhead{(1)} &
\colhead{(2)} &
\colhead{(3)} &
\colhead{(4)} &
\colhead{(5)} &
\colhead{(6)} \\
}
\startdata
   8.00 &  0.35 & 0.15 & 5.7292E-06 &2.2287E-06  &   25 \\
   8.00 &  0.75 & 0.25 & 1.9470E-07 &7.6028E-08  &    9 \\
   8.00 &  1.25 & 0.25 &   \nodata  &  \nodata   &    0 \\
   8.00 &  1.75 & 0.25 &   \nodata  &  \nodata   &    0 \\
   8.00 &  2.50 & 0.50 &   \nodata  &  \nodata   &    0 \\[3pt]
\tableline \\[-8pt]
   8.40 &  0.35 & 0.15 & 2.4872E-06 &3.5733E-07  &   78 \\
   8.40 &  0.75 & 0.25 & 4.0640E-07 &7.4157E-08  &   55 \\
   8.40 &  1.25 & 0.25 & 2.5527E-08 &1.0412E-08  &    7 \\
   8.40 &  1.75 & 0.25 &   \nodata  &  \nodata   &    0 \\
   8.40 &  2.50 & 0.50 & 9.9367E-10 &9.9367E-10  &    1 \\[3pt]
\tableline \\[-8pt]
   8.80 &  0.35 & 0.15 & 3.7392E-07 &6.6183E-08  &   40 \\
   8.80 &  0.75 & 0.25 & 2.0598E-07 &2.7534E-08  &   90 \\
   8.80 &  1.25 & 0.25 & 3.2505E-08 &6.3618E-09  &   29 \\
   8.80 &  1.75 & 0.25 & 4.7534E-09 &1.9897E-09  &    6 \\
   8.80 &  2.50 & 0.50 & 2.1144E-09 &1.1199E-09  &    4 \\[3pt]
\tableline \\[-8pt]
   9.20 &  0.35 & 0.15 & 6.4744E-08 &2.3427E-08  &   15 \\
   9.20 &  0.75 & 0.25 & 6.2003E-08 &8.5161E-09  &   79 \\
   9.20 &  1.25 & 0.25 & 4.6456E-08 &5.0355E-09  &  108 \\
   9.20 &  1.75 & 0.25 & 1.3817E-08 &2.3514E-09  &   39 \\
   9.20 &  2.50 & 0.50 & 4.6557E-09 &1.0753E-09  &   25 \\[3pt]
\tableline \\[-8pt]
   9.60 &  0.35 & 0.15 & 1.5007E-09 &1.5007E-09  &    1 \\
   9.60 &  0.75 & 0.25 & 3.1654E-09 &8.8983E-10  &   13 \\
   9.60 &  1.25 & 0.25 & 1.1309E-08 &1.3366E-09  &   77 \\
   9.60 &  1.75 & 0.25 & 9.8252E-09 &1.1188E-09  &   81 \\
   9.60 &  2.50 & 0.50 & 4.1379E-09 &6.0390E-10  &   55 \\[3pt]
\tableline \\[-8pt]
  10.00 &  0.35 & 0.15 &   \nodata  &  \nodata   &    0 \\
  10.00 &  0.75 & 0.25 & 9.3214E-11 &9.3214E-11  &    1 \\
  10.00 &  1.25 & 0.25 & 9.2919E-10 &2.7036E-10  &   15 \\
  10.00 &  1.75 & 0.25 & 1.8614E-09 &2.9647E-10  &   40 \\
  10.00 &  2.50 & 0.50 & 1.4888E-09 &2.1542E-10  &   53 \\[3pt]
\tableline \\[-8pt]
  10.40 &  0.35 & 0.15 &   \nodata  &  \nodata   &    0 \\
  10.40 &  0.75 & 0.25 &   \nodata  &  \nodata   &    0 \\
  10.40 &  1.25 & 0.25 &   \nodata  &  \nodata   &    0 \\
  10.40 &  1.75 & 0.25 & 5.3655E-11 &3.0978E-11  &    3 \\
  10.40 &  2.50 & 0.50 & 1.2549E-10 &3.3775E-11  &   14 
\enddata
\tablenotetext{a}{The central mass in the bin. The mass bin size is 0.4 dex
and expands $\pm$0.2 dex relative to the central mass value.}
\tablenotetext{b}{The range of the redshift bin is $<z> \pm \Delta z$.}
\end{deluxetable}



\begin{deluxetable}{cccccr}
\tablecaption{Black Hole Mass Function of the SDSS Color-selected Sample \label{SDSS_MF.tab}}
\tablecolumns{6}
\tablehead{
\colhead{$<z>$} & \colhead{$\Delta z$\tablenotemark{a}} & \colhead{$<\mbh>$\tablenotemark{b}} & 
\colhead{$\Psi (M,z)$} & \colhead{$\sigma(\Psi)$~~~~} & \colhead{N} \\ 
\colhead{} & \colhead{} & \colhead{(\Msol)} & 
\colhead{(Gpc$^{-3}$\Msol$^{-1}$)} & \colhead{(Gpc$^{-3}$\Msol$^{-1}$)} & 
\colhead{} \\ 
\colhead{(1)} &
\colhead{(2)} &
\colhead{(3)} &
\colhead{(4)} &
\colhead{(5)} &
\colhead{(6)} 
}
\startdata
\cutinhead{Measured \mbh{} Values Only (Original Mass Function)}
4.3 & 0.7 &  8.7  & 4.690E-08 & 3.162E-08 & 6 \\
4.3 & 0.7 &  9.1  & 8.607E-09 & 2.800E-09 & 13 \\
4.3 & 0.7 &  9.5  & 1.325E-09 & 4.462E-10 &  9\\
4.3 & 0.7 &  9.9  & 3.798E-10 & 1.569E-10 &  6\\
\cutinhead{Measured \mbh{} and Assigned Median \mbh{} Values}
4.3 & 0.7 &  8.7 & 4.690E-08 & 3.162E-08 &  6\\
4.3 & 0.7 &  9.1 & 1.072E-08 & 3.073E-09 &  17\\
4.3 & 0.7 &  9.5 & 1.325E-09 & 4.462E-10 &  9\\
4.3 & 0.7 &  9.9 & 3.798E-10 & 1.569E-10 &  6\\
\cutinhead{Measured \mbh{} and Assigned Median \mbh \,$+ 1\sigma$ Values}
4.3 & 0.7 & 8.7 & 4.690E-08 & 3.162E-08  &  6\\
4.3 & 0.7 & 9.1 & 8.607E-09 & 2.800E-09  &  13\\
4.3 & 0.7 & 9.5 & 1.715E-09 & 4.997E-10  &  12\\
4.3 & 0.7 & 9.9 & 5.597E-10 & 2.386E-10  &  7\\
\cutinhead{Measured \mbh{} and Assigned Median \mbh \,$- 1\sigma$ Values}
4.3 & 0.7 & 8.7 & 4.975E-08 & 3.175E-08  &  7\\
4.3 & 0.7 & 9.1 & 9.585E-09 & 2.856E-09  &  16\\
4.3 & 0.7 & 9.5 & 1.325E-09 & 4.462E-10  &  9\\
4.3 & 0.7 & 9.9 & 3.798E-10 & 1.569E-10  &  6\\
\enddata
\tablenotetext{a}{The range of the redshift bin is $<z> \pm \Delta z$.}
\tablenotetext{b}{The central mass in the bin. The mass bin size is 0.4 dex
and expands $\pm$0.2 dex relative to the central mass value.}
\end{deluxetable}

\clearpage


\LongTables
\begin{deluxetable}{cccccrc}
\tablecaption{Corrected LBQS Black Hole Mass Function \label{LBQS_MFcor.tab}}
\tablewidth{400pt}
\tablecolumns{7}
\tablehead{
\colhead{$<z>$} & \colhead{$\Delta z$\tablenotemark{a}} & \colhead{$<\mbh>$\tablenotemark{b}} &
\colhead{$\Psi (M,z)$} & \colhead{$\sigma(\Psi)$~~~~} & \colhead{N} &
\colhead{$\rho(> \mbh)$\tablenotemark{c}} \\
\colhead{} & \colhead{} & \colhead{(\Msol)} &
\colhead{(Gpc$^{-3}$\Msol$^{-1}$)} & \colhead{(Gpc$^{-3}$\Msol$^{-1}$)} &
\colhead{} & \colhead{(\Msol Mpc$^{-3}$)} \\
\colhead{(1)} &
\colhead{(2)} &
\colhead{(3)} &
\colhead{(4)} &
\colhead{(5)} &
\colhead{(6)} &
\colhead{(7)}
}
\startdata
\cutinhead{Measured \mbh{} and Assigned Median \mbh{} Values} 
0.35&  0.15&  6.8 &    \nodata  &  \nodata    &  0  & 475.05 \\
0.35&  0.15&  7.2 &  2.4518E-06 & 2.4518E-06  &  1  & 474.52 \\
0.35&  0.15&  7.6 &  1.8975E-06 & 1.1120E-06  &  5  & 473.31 \\
0.35&  0.15&  8.0 &  5.7292E-06 & 2.2287E-06  &  25 & 437.86 \\
0.35&  0.15&  8.4 &  2.7017E-06 & 3.7761E-07  &  82 & 345.18 \\
0.35&  0.15&  8.8 &  3.8550E-07 & 6.7189E-08  &  41 & 190.48 \\
0.35&  0.15&  9.2 &  6.4744E-08 & 2.3427E-08  &  15 &  23.53 \\
0.35&  0.15&  9.6 &  1.5007E-09 & 1.5007E-09  &  1  & \nodata\\
\hline
0.75&  0.25&  7.6 &   \nodata   &   \nodata  & 0  &  318.57 \\
0.75&  0.25&  8.0 &  1.9470E-07 & 7.6028E-08 & 9  &  317.97 \\
0.75&  0.25&  8.4 &  4.3218E-07 & 7.8509E-08 & 56 &  311.02 \\
0.75&  0.25&  8.8 &  2.2154E-07 & 2.8136E-08 & 98 &  259.42 \\
0.75&  0.25&  9.2 &  6.5052E-08 & 8.6269E-09 & 84 &  130.77 \\
0.75&  0.25&  9.6 &  3.1654E-09 & 8.8983E-10 & 13 &   29.34 \\
0.75&  0.25& 10.0 &  9.3214E-11 & 9.3214E-11 & 1  &  \nodata \\
\hline
1.25 & 0.25 & 8.000 &  \nodata   &  \nodata   &  0 & 389.25 \\
1.25 & 0.25 & 8.400 & 2.5527E-08 & 1.0412E-08 &  7 & 388.93 \\
1.25 & 0.25 & 8.800 & 3.2505E-08 & 6.3618E-09 & 29 & 385.78 \\
1.25 & 0.25 & 9.200 & 5.0448E-08 & 5.2050E-09 & 118& 332.50 \\
1.25 & 0.25 & 9.600 & 1.2247E-08 & 1.3828E-09 & 84 & 157.13 \\
1.25 & 0.25 &10.000 & 9.2919E-10 & 2.7036E-10 & 15 & 35.06 \\
\hline
1.75 & 0.25 & 8.400 &  \nodata   &  \nodata   & 0  & 420.21 \\
1.75 & 0.25 & 8.800 & 4.7534E-09 & 1.9897E-09 & 6  & 420.04 \\
1.75 & 0.25 & 9.200 & 1.5414E-08 & 2.4894E-09 & 43 & 412.79 \\
1.75 & 0.25 & 9.600 & 1.1405E-08 & 1.1958E-09 & 95 & 313.36 \\
1.75 & 0.25 &10.000 & 2.1402E-09 & 3.1767E-10 & 46 & 90.84 \\
1.75 & 0.25 &10.400 & 5.3655E-11 & 3.0978E-11 & 3  & \nodata \\
\hline
2.50 & 0.50 & 8.000 &  \nodata   &  \nodata   & 0  & 338.39 \\
2.50 & 0.50 & 8.400 & 9.9367E-10 & 9.9367E-10 & 1  & 338.39 \\
2.50 & 0.50 & 8.800 & 2.1144E-09 & 1.1199E-09 & 4  & 337.98 \\
2.50 & 0.50 & 9.200 & 4.6557E-09 & 1.0753E-09 & 25 & 336.03 \\
2.50 & 0.50 & 9.600 & 5.2545E-09 & 6.7097E-10 & 71 & 306.28 \\
2.50 & 0.50 &10.000 & 1.8430E-09 & 2.3892E-10 & 66 & 147.96 \\
2.50 & 0.50 &10.400 & 1.2549E-10 & 3.3775E-11 & 14 & 24.92 \\
\cutinhead{Measured \mbh{} and Assigned Median \mbh{}\,$+ 1\sigma$ Values} 
0.35 & 0.15&  6.8 &   \nodata    & \nodata    &  0 &  502.79 \\
0.35 & 0.15&  7.2 &   2.4518E-06 & 2.4518E-06 &  1 &  502.25 \\
0.35 & 0.15&  7.6 &   1.8975E-06 & 1.1120E-06 &  5 &  501.05 \\
0.35 & 0.15&  8.0 &   5.7292E-06 & 2.2287E-06 & 25 &  465.59 \\
0.35 & 0.15&  8.4 &   2.4872E-06 & 3.5733E-07 & 78 &  377.61 \\
0.35 & 0.15&  8.8 &   4.5933E-07 & 8.2115E-08 & 44 &  224.05 \\
0.35 & 0.15&  9.2 &   6.9354E-08 & 2.3876E-08 & 16 &  35.714 \\
0.35 & 0.15&  9.6 &   1.5007E-09 & 1.5007E-09 &  1 &  \nodata\\
\hline
0.75 & 0.25&  7.6 &      \nodata &    \nodata &  0 &  335.75 \\
0.75 & 0.25&  8.0 &   1.9470E-07 & 7.6028E-08 &  9 &  335.15 \\
0.75 & 0.25&  8.4 &   4.0640E-07 & 7.4157E-08 & 55 &  328.20 \\
0.75 & 0.25&  8.8 &   2.1765E-07 & 2.9418E-08 & 92 &  281.43 \\
0.75 & 0.25&  9.2 &   6.7637E-08 & 8.8044E-09 & 86 &  146.66 \\
0.75 & 0.25&  9.6 &   4.3793E-09 & 1.0455E-09 & 18 &  32.929 \\
0.75 & 0.25& 10.0 &   9.3214E-11 & 9.3214E-11 &  1 &  \nodata\\
\hline
1.25 & 0.25&  8.0 &      \nodata &    \nodata &  0 &  409.27 \\
1.25 & 0.25&  8.4 &   2.5527E-08 & 1.0412E-08 &  7 &  408.95 \\
1.25 & 0.25&  8.8 &   3.2505E-08 & 6.3618E-09 & 29 &  405.79 \\
1.25 & 0.25&  9.2 &   4.6456E-08 & 5.0355E-09 &108 &  353.76 \\
1.25 & 0.25&  9.6 &   1.3702E-08 & 1.4729E-09 & 93 &  176.43 \\
1.25 & 0.25& 10.0 &   9.8256E-10 & 2.7558E-10 & 16 &  35.057 \\
\hline
1.75 & 0.25&  8.4 &      \nodata &    \nodata &  0 &  473.23 \\
1.75 & 0.25&  8.8 &   4.7534E-09 & 1.9897E-09 &  6 &  473.06 \\
1.75 & 0.25&  9.2 &   1.3817E-08 & 2.3514E-09 & 39 &  465.81 \\
1.75 & 0.25&  9.6 &   1.0912E-08 & 1.1868E-09 & 89 &  376.75 \\
1.75 & 0.25& 10.0 &   2.5353E-09 & 3.4376E-10 & 55 &  162.89 \\
1.75 & 0.25& 10.4 &   7.5198E-11 & 3.7732E-11 &  4 &  14.133 \\
\hline
2.50 & 0.50&  8.0 &      \nodata &    \nodata &  0 &  402.11 \\
2.50 & 0.50&  8.4 &   9.9367E-10 & 9.9367E-10 &  1 &  402.11 \\
2.50 & 0.50&  8.8 &   2.1144E-09 & 1.1199E-09 &  4 &  401.69 \\
2.50 & 0.50&  9.2 &   4.6557E-09 & 1.0753E-09 & 25 &  399.74 \\
2.50 & 0.50&  9.6 &   4.1379E-09 & 6.0390E-10 & 55 &  369.99 \\
2.50 & 0.50& 10.0 &   1.9763E-09 & 2.4675E-10 & 71 &  255.72 \\
2.50 & 0.50& 10.4 &   2.4082E-10 & 5.1114E-11 & 24 &   34.871 \\
2.50 & 0.50& 10.8 &   3.4121E-12 & 3.4121E-12 &  1 &  \nodata\\
\cutinhead{Measured \mbh{} and Assigned Median \mbh \,$- 1\sigma$ Values}
0.35 & 0.15&  6.8 &   \nodata  &   \nodata  &  0  & 464.72 \\
0.35 & 0.15&  7.2 & 2.4518E-06 & 2.4518E-06 &  1  & 464.19 \\
0.35 & 0.15&  7.6 & 1.8975E-06 & 1.1120E-06 &  5  & 462.98 \\
0.35 & 0.15&  8.0 & 6.2681E-06 & 2.2497E-06 & 29  & 427.52 \\
0.35 & 0.15&  8.4 & 2.5162E-06 & 3.5851E-07 & 79  & 333.11 \\
0.35 & 0.15&  8.8 & 3.7392E-07 & 6.6183E-08 & 40  & 190.48 \\
0.35 & 0.15&  9.2 & 6.4744E-08 & 2.3427E-08 & 15  & 23.53 \\
0.35 & 0.15&  9.6 & 1.5007E-09 & 1.5007E-09 &  1  & \nodata \\
\hline
0.75 & 0.25&  7.6 &   \nodata  &   \nodata  &  0  & 310.60 \\
0.75 & 0.25&  8.0 & 1.9470E-07 & 7.6028E-08 &  9  & 310.00 \\
0.75 & 0.25&  8.4 & 4.7127E-07 & 7.9844E-08 & 64  & 301.02 \\
0.75 & 0.25&  8.8 & 2.1364E-07 & 2.7751E-08 & 95  & 252.17 \\
0.75 & 0.25&  9.2 & 6.2003E-08 & 8.5161E-09 & 79  & 126.47 \\
0.75 & 0.25&  9.6 & 3.1654E-09 & 8.8983E-10 & 13  & 29.34 \\
0.75 & 0.25& 10.0 & 9.3214E-11 & 9.3214E-11 &  1  & \nodata \\
\hline
1.25 & 0.25&  8.0 &   \nodata  &   \nodata  &  0  & 378.83 \\
1.25 & 0.25&  8.4 & 2.5527E-08 & 1.0412E-08 &  7  & 378.52 \\
1.25 & 0.25&  8.8 & 3.5906E-08 & 6.6951E-09 & 32  & 374.77 \\
1.25 & 0.25&  9.2 & 5.0778E-08 & 5.1932E-09 &120  & 315.32 \\
1.25 & 0.25&  9.6 & 1.1577E-08 & 1.3500E-09 & 79  & 153.21 \\
1.25 & 0.25& 10.0 & 9.2919E-10 & 2.7036E-10 & 15  & 35.06 \\
\hline
1.75 & 0.25&  8.4 &   \nodata  &   \nodata  &  0  & 395.31 \\
1.75 & 0.25&  8.8 & 4.7534E-09 & 1.9897E-09 &  6  & 395.14 \\
1.75 & 0.25&  9.2 & 1.7965E-08 & 2.6306E-09 & 52  & 385.18 \\
1.75 & 0.25&  9.6 & 1.1090E-08 & 1.1821E-09 & 92  & 274.61 \\
1.75 & 0.25& 10.0 & 1.8614E-09 & 2.9647E-10 & 40  & 90.84 \\
1.75 & 0.25& 10.4 & 5.3655E-11 & 3.0978E-11 &  3  & \nodata \\
\hline
2.50 & 0.50&  8.0 &   \nodata  &   \nodata  &  0  & 310.87 \\
2.50 & 0.50&  8.4 & 9.9367E-10 & 9.9367E-10 &  1  & 310.87 \\
2.50 & 0.50&  8.8 & 2.1144E-09 & 1.1199E-09 &  4  & 310.45 \\
2.50 & 0.50&  9.2 & 5.7111E-09 & 1.1510E-09 & 32  & 308.50 \\
2.50 & 0.50&  9.6 & 5.7243E-09 & 7.0062E-10 & 77  & 257.20 \\
2.50 & 0.50& 10.0 & 1.4888E-09 & 2.1542E-10 & 53  & 144.92 \\
2.50 & 0.50& 10.4 & 1.2549E-10 & 3.3775E-11 & 14  & 24.92 \\
\enddata
\tablenotetext{a}{The range of the redshift bin is $<z> \pm \Delta z$.}
\tablenotetext{b}{The central mass in the bin. The mass bin size is 0.4 dex
and expands $\pm$0.2 dex relative to the central mass value.}
\tablenotetext{c}{The lower mass limit is the central value of the mass bin
listed in column 3.}
\end{deluxetable}

\clearpage


\begin{deluxetable}{cccccc}
\tablecaption{Cumulative Mass Densities of the BQS and SDSS Color-selected Samples \label{BQS_SDSS_cumMdens.tab}}
\tablewidth{0pt}
\tablecolumns{6}
\tablehead{
\colhead{} & \colhead{\underline{\bf ~~~~BQS~~~~}} & \multicolumn{4}{c}{\underline{\bf ~~~~~~~~~~~~~~~~~~~~~~~~~~~~SDSS Color-selected Sample~~~~~~~~~~~~~~~~~~~~~~~~~~~~~~~~~}} \\[5pt]
\colhead{$<\mbh>$\tablenotemark{a}} & 
\colhead{$\rho(> \mbh)$\tablenotemark{b}} &
\colhead{$\rho(> \mbh)$\tablenotemark{b,c}} &
\colhead{~$\rho_{\rm corr, Med}(> \mbh)$\tablenotemark{b,d}~~} &
\colhead{~$\rho_{\rm corr, Med+ \sigma}(> \mbh)$\tablenotemark{b,e}~~} &
\colhead{~$\rho_{\rm corr, Med- \sigma}(> \mbh)$\tablenotemark{b.f}~~} \\
\colhead{(\Msol)} & 
\colhead{(\Msol Mpc$^{-3}$)} &
\colhead{(\Msol Mpc$^{-3}$)} &
\colhead{(\Msol Mpc$^{-3}$)} & 
\colhead{(\Msol Mpc$^{-3}$)} & 
\colhead{(\Msol Mpc$^{-3}$)} \\
\colhead{(1)} &
\colhead{(2)} &
\colhead{(3)} &
\colhead{(4)} &
\colhead{(5)} &
\colhead{(6)}
}
\startdata
6.4 &  278.2  &  \nodata & \nodata & \nodata & \nodata \\
6.6 &  277.8  &  \nodata & \nodata & \nodata & \nodata \\
6.8 &  277.8  &  \nodata & \nodata & \nodata & \nodata \\
7.0 &  277.5  &  \nodata & \nodata & \nodata & \nodata \\
7.2 &  272.9  &  \nodata & \nodata & \nodata & \nodata \\
7.4 &  272.3  &  \nodata & \nodata & \nodata & \nodata \\
7.6 &  267.4  &  \nodata & \nodata & \nodata & \nodata \\
7.8 &  265.8  &  \nodata & \nodata & \nodata & \nodata \\
8.0 &  243.3  &  \nodata & \nodata & \nodata & \nodata \\
8.2 &  238.9  &  \nodata & \nodata & \nodata & \nodata \\
8.4 &  190.8  &   56.56  & 61.07 & 67.89 & 58.41\\ 
8.6 &  64.72  &   49.96  & 54.46 & 61.29 & 51.80\\ 
8.8 &  31.52  &   48.25  & 52.76 & 59.58 & 50.09\\ 
9.0 &  22.43  &   46.69  & 51.20 & 58.02 & 46.69\\ 
9.2 &  8.656  &   41.33  & 45.84 & 52.67 & 41.33\\ 
9.4 &  2.168  &   31.99  & 31.99 & 43.32 & 31.99\\ 
9.6 &  1.334  &   25.95  & 25.95 & 33.10 & 25.95\\ 
9.8 &  \nodata &  17.73  & 17.73 & 17.73 & 17.73\\ 
10.0 & \nodata &  10.69  & 10.69 & 10.69 & 10.69\\  

\enddata
\tablenotetext{a}{The central mass in the bin. The mass bin size is 0.4 dex
and expands $\pm$0.2 dex relative to the central mass value.}
\tablenotetext{b}{
The lower mass limit is the central value of the mass bin listed in column 1.}
\tablenotetext{c}{Cumulative mass density for the original mass function based on measured 
spectral measurements only.}
\tablenotetext{d}{
Cumulative mass density for the full sample based on measured spectral measurements and 
assigned mass values (median \mbh{}) for sources without a suitable spectrum.}
\tablenotetext{e}{
Cumulative mass density for the full sample based on measured spectral measurements and 
assigned mass values (median \mbh{} $+ 1\sigma$) for sources without a suitable spectrum.} 
\tablenotetext{f}{
Cumulative mass density for the full sample based on measured spectral measurements and 
assigned mass values (median \mbh{} $- 1\sigma$) for sources without a suitable spectrum.} 

\end{deluxetable}

\clearpage




\begin{thebibliography}{}
\bibitem[Avni \& Bahcall(1980)]{avni80}  Avni, Y., \& Bahcall, J.~N.\ 1980, \apj, 235, 694
\bibitem[Blandford \& McKee(1982)]{1982ApJ...255..419B} Blandford, R.~D., \& 
	McKee, C.~F.\ 1982, \apj, 255, 419
\bibitem[Boyle et al.(2000)]{2000MNRAS.317.1014B} Boyle, B.~J., Shanks, T., 
	Croom, S.~M., Smith, R.~J., Miller, L., Loaring, N., \& Heymans, C.\ 
	2000, \mnras, 317, 1014
\bibitem[Boroson \& Green(1992)]{bg92} Boroson, T. \& Green, R. 1992, \apj, 80, 109
\bibitem[Brotherton(1996)]{1996ApJS..102....1B} Brotherton, M.~S.\ 1996, \apjs, 102, 1
\bibitem[Collin et al. 2006]{} Collin, S., Kawaguchi, T., Peterson, B.M., 
	Vestergaard, M. 2006, \aap, 456, 75
\bibitem[Crampton et al.(1987)]{1987ApJ...314..129C} Crampton, D., Cowley, A.~P., \& 
	Hartwick, F.~D.~A.\ 1987, \apj, 314, 129 
\bibitem[Denney et al. 2008]{} Denney, K., Peterson, B.M., Dietrich, M., 
	Vestergaard, M., Bentz, M.C. 2009, \apj, 692, 246
\bibitem[Dietrich and Hamann 2004]{} Dietrich, M. \& Hamann, F. 2004, \apj, 611, 761
\bibitem[Fan(2006)]{2006NewAR..50..665F} Fan, X.\ 2006, New Astronomy Review, 50, 665
\bibitem[Fan et al. (2001a)]{} Fan, X., et al.\ 2001a, \aj, 121, 31
\bibitem[Fan et al. (2001b)]{} Fan, X., et al.\ 2001b, \aj, 122, 2833
\bibitem[Ferrarese (2003)]{2003ASPC..291..196F} Ferrarese, L.\ 2003, Hubble's Science 
	Legacy: Future Optical/Ultraviolet Astronomy from Space, 291, 196 
\bibitem[Foltz et al.(1987)]{1987AJ.....94.1423F} Foltz, C.~B., Chaffee, F.~H., Jr., 
	Hewett, P.~C., MacAlpine, G.~M., Turnshek, D.~A., Weymann, R.~J., \& 
	Anderson, S.~F.\ 1987, \aj, 94, 1423 
\bibitem[Forster et al.(2001)]{forster01} Forster, K., Green, P.J., Aldcroft, T.L., Vestergaard, M.,
	Foltz, C.B., \& Hewett, P.C., 2001, \apjs, 134, 35
\bibitem[Granato et al.(2004)]{2004ApJ...600..580G} Granato, G.~L., De
        Zotti, G., Silva, L., Bressan, A., \& Danese, L.\ 2004, \apj, 600, 580
\bibitem[Grier et al.(2008)]{} Grier, C.~J., et al.\ 2008, \apj, 688, 837
\bibitem[Green et al.(1986)]{1986ApJS...61..305G} Green, R.~F., Schmidt, M., 
	\& Liebert, J.\ 1986, \apjs, 61, 305
\bibitem[Hewett et al.(1995)]{1995AJ....109.1498H} Hewett, P.~C., Foltz, C.~B., \& Chaffee, F.~H.\ 1995, \aj, 109, 1498 
\bibitem[Hewett et al. (2001)]{} Hewett, P.C., Foltz, C.B., Chaffee, F.H. 2001, \aj, 122, 518 
\bibitem[Huchra \& Burg(1992)]{1992ApJ...393...90H} Huchra, J., \& Burg, R.\ 1992, \apj, 393, 90
\bibitem[Hogg(1999)]{1999astro.ph..5116H} Hogg, D.~W.\ 1999, ArXiv Astrophysics e-prints, arXiv:astro-ph/9905116
\bibitem[Hopkins \& Hernquist(2008)]{2008arXiv0809.3789H} Hopkins, P.~F., \& Hernquist, L.\ 2008, arXiv:0809.3789
\bibitem[Jester et al.(2005)]{jester05} Jester, S., et al.\ 2005, \aj, 130, 873
\bibitem[Kelly et al.(2008a)]{2008ApJ...682..874K} Kelly, B.~C., Fan, X., \& Vestergaard, M.\ 2008a, \apj, 682, 874
\bibitem[Kelly et al.(2008b)]{2008ApJS..176..355K} Kelly, B.~C., Bechtold, J., Trump, J.~R., Vestergaard, M., \& Siemiginowska, A.\ 2008b, \apjs, 176, 355
\bibitem[Kelly et al. (2009)]{} Kelly, B., Vestergaard, M., \& Fan, X. 2009, \apj, 692, 1388
\bibitem[Kibblewhite et al.(1984)]{1984amd..conf..277K} Kibblewhite, E.~J., Bridgeland, M.~T., Bunclark, 
	P.~S., \& Irwin, M.~J.\ 1984, Astronomical Microdensitometry Conference, 277
\bibitem[Kollmeier et al.(2006)]{2006ApJ...648..128K} Kollmeier, J.~A., et al.\ 2006, \apj, 648, 128
\bibitem[Koo \& Kron(1982)]{1982A&A...105..107K} Koo, D.~C., \& Kron, R.~G.\ 1982, \aap, 105, 107
\bibitem[Krolik(2001)]{2001ApJ...551...72K} Krolik, J.~H.\ 2001, \apj, 551, 72
\bibitem[Kuhn et al.(2001)]{2001ApJS..136..225K} Kuhn, O., Elvis, M., Bechtold, J., \& Elston, R.\ 2001, \apjs, 136, 225 
\bibitem[Magorrian et al.(1998)]{1998AJ....115.2285M} Magorrian, J., et al.\ 1998, \aj, 115, 2285 
\bibitem[Marconi et al.(2008)]{2008ApJ...678..693M} Marconi, A., Axon, D.~J., 
	Maiolino, R., Nagao, T., Pastorini, G., Pietrini, P., Robinson, A., \& 
	Torricelli, G.\ 2008, \apj, 678, 693 
\bibitem[McLure \& Dunlop(2004)]{2004MNRAS.352.1390M} McLure, R.~J., \& Dunlop, J.~S.\ 
	2004, \mnras, 352, 1390
\bibitem[McLure \& Jarvis(2002)]{2002MNRAS.337..109M} McLure, R.~J., \& Jarvis, M.~J.\ 2002, \mnras, 337, 109 
\bibitem[McNamara \& Nulsen(2007)]{2007ARA&A..45..117M} McNamara, B.~R., \& Nulsen, 
	P.~E.~J.\ 2007, \araa, 45, 117
\bibitem[Martini(2004)]{2004cbhg.symp..169M} Martini, P.\ 2004, Coevolution 
of Black Holes and Galaxies, 169 
\bibitem[Morris et al.(1991)]{1991AJ....102.1627M} Morris, S.~L., Weymann, R.~J., 
	Anderson, S.~F., Hewett, P.~C., Francis, P.~J., Foltz, C.~B., Chaffee, F.~H., 
	\& MacAlpine, G.~M.\ 1991, \aj, 102, 1627
\bibitem[Netzer \& Trakhtenbrot(2007)]{2007ApJ...654..754N} Netzer, H., \& 
	Trakhtenbrot, B.\ 2007, \apj, 654, 754
\bibitem[Onken et al.(2004)]{onken04} Onken, C.A., et al. 2004, \apj,   615, 645
\bibitem[Osmer(1982)]{osmer82} Osmer, P.~S.\ 1982, \apj, 253, 28
\bibitem[]{} Peterson, B.M., 1993, \pasp, 105, 247
\bibitem[Peterson et al. (2004)]{peterson04} Peterson, B.M. et al. 2004, \apj,   613, 682
\bibitem[Richards et al.(2006)]{2006AJ....131.2766R} Richards, G.~T., et
        al.\ 2006, \aj, 131, 2766 
\bibitem[Richards et al.(2002)]{2002AJ....124....1R} Richards, G.~T., 
	Vanden Berk, D.~E., Reichard, T.~A., Hall, P.~B., Schneider, D.~P., 
	SubbaRao, M., Thakar, A.~R., \& York, D.~G.\ 2002, \aj, 124, 1 
\bibitem[Sanders et al. (1989)]{sanders89} Sanders, D.\,B., Phinney, E.\,S., 
	Neugebauer, G., Soifer, B.\,T., Matthews, K.\ 1989, \apj, 347, 29
\bibitem[Shen et al.(2008)]{2008ApJ...680..169S} Shen, Y., Greene, J.~E., 
	Strauss, M.~A., Richards, G.~T., \& Schneider, D.~P.\ 2008, \apj, 680, 169
\bibitem[]{} Shemmer, O., \et 2004, \apj, 614, 547
\bibitem[Schlegel et al.(1998)]{schlegel98} Schlegel, D.~J., Finkbeiner, D.~P., \& 
	Davis, M.\ 1998, \apj, 500, 525
\bibitem[Schneider et al. 2001]{} Schneider, D.\ et al. 2001, \aj,  130, 367
\bibitem[Schneider et al. 2003]{dr1qcat} Schneider, D.\ et al. 2003, \aj,  126, 2579
\bibitem[Schneider et al. 2005]{dr3qcat} Schneider, D.\ et al. 2005, \aj,  130, 367
\bibitem[Schmidt \& Green(1983)]{sg83} Schmidt, M., \& Green, R.~F.\ 1983, \apj, 269, 352
\bibitem[Schmidt, Schneider, \& Gunn (1995)]{SSG95} Schmidt, M., Schneider, D.~P., 
	\& Gunn, J. E. 1995, \aj, 110, 68
\bibitem[Smith et al.(2005)]{2005MNRAS.359...57S} Smith, R.~J., Croom, S.~M., Boyle, B.~J., 
	Shanks, T., Miller, L., \& Loaring, N.~S.\ 2005, \mnras, 359, 57 
\bibitem[Somerville et al.(2008)]{2008MNRAS.391..481S} Somerville, R.~S., Hopkins, P.~F., Cox, T.~J., Robertson, B.~E., \& Hernquist, L.\ 2008, \mnras, 391, 481
\bibitem[Springel et al.(2005)]{2005MNRAS.361..776S} Springel, V., Di 
	Matteo, T., \& Hernquist, L.\ 2005, \mnras, 361, 776 
\bibitem[Tremaine et al.(2002)]{2002ApJ...574..740T} Tremaine, S., et al.\ 2002, \apj, 574, 740 
\bibitem[Ueda et al.(2003)]{2003ApJ...598..886U} Ueda, Y., Akiyama, M., Ohta, K., \& 
	Miyaji, T.\ 2003, \apj, 598, 886 
\bibitem[V{\'e}ron-Cetty et al.(2004)]{2004A&A...417..515V}
        V{\'e}ron-Cetty, M.-P., Joly, M., \& V{\'e}ron, P.\ 2004, \aap, 417, 515 
\bibitem[Vestergaard(2002)]{V02} Vestergaard, M.\ 2002, \apj, 571, 733
\bibitem[Vestergaard(2004a)]{V04} Vestergaard, M.\ 2004a, \apj, 601, 676
\bibitem[Vestergaard 2004b]{} Vestergaard, M. 2004b, in AGN Physics with the Sloan Digital Sky 
	Survey, \apj, ASP Conference Series, 311, 69 
\bibitem[Vestergaard (2009)]{V09} Vestergaard, M.\ 2009, in the STScI Spring Symposium
        2007 on 'Black Holes', ed. A. Koekemoer, Cambridge University Press, in press (astro-ph: arXiv:0904.2615)
\bibitem[Vestergaard \& Peterson(2006)]{VP06} Vestergaard, M., \& Peterson, B.~M.\ 2006, \apj, 641, 689
\bibitem[Vestergaard \et (2008)]{Vea08} Vestergaard, M., Fan, X., 
	Tremonti, C.\,A., Osmer, P.\,O., Richards, G.\,T.\ 2008, ApJ, 674, L1
\bibitem[Vestergaard \& Wilkes(2001)]{VW01} Vestergaard, M., \& Wilkes, B.~J.\ 2001, \apjs, 134, 1
\bibitem[Vestergaard, Wilkes, \& Barthel(2000)]{2000ApJ...538L.103V}
        Vestergaard, M., Wilkes, B.\ J., \& Barthel, P.\ D.\ 2000, \apjl, 538, L103
\bibitem[Volonteri et al.(2008)]{2008MNRAS.383.1079V} Volonteri, M., 
Lodato, G., \& Natarajan, P.\ 2008, \mnras, 383, 1079
\bibitem[Wandel et al. (1999)]{} Wandel, A., Peterson, B.M., Malkan, M. 1999, \apj, 526, 579 
\bibitem[Warner et al.(2003)]{2003ApJ...596...72W} Warner, C., 
         Hamann, F., \& Dietrich, M.\ 2003, \apj, 596, 72 
\bibitem[Warren et al.(1994)]{who94} Warren, S.~J., Hewett, P.~C.,
        \& Osmer, P.~S.\ 1994, \apj, 421, 412
\bibitem[Wills and Browne 1986]{} Wills, B., Browne, I. 1986, \apj, 302, 56
\bibitem[Wyithe \& Padmanabhan(2006)]{2006MNRAS.372.1681W} Wyithe,
J.~S.~B., \& Padmanabhan, T.\ 2006, \mnras, 372, 1681
\bibitem[York et al.(2000)]{2000AJ....120.1579Y} York, D.~G., et al.\ 2000, \aj, 120, 1579 
\end{thebibliography}
\end{document}